\documentclass[reprint,aps,prb,longbibliography]{revtex4-1}
\usepackage{mathtools}
\usepackage{esint}
\usepackage{graphicx}
\usepackage{graphics}
\usepackage{multirow}
\usepackage{bm}
\usepackage{listings}
\usepackage{color}
\usepackage[colorlinks]{hyperref}
\usepackage{cancel}
\usepackage{braket}
\usepackage{relsize}
\usepackage{bm}
\usepackage{amsmath}
\usepackage{amsbsy}
\usepackage{amssymb}
\usepackage{amsfonts}
\usepackage{mathrsfs}
\usepackage{sidecap}

\usepackage{subfigure}

\makeatletter
\newcommand*{\rom}[1]{\expandafter\@slowromancap\romannumeral #1@}
\makeatother

\def \bG{\mathscr{G}}
\def \tTh{\tilde{\Theta}}
\def \Gam{\Gamma_{\zeta,\mu}}
\def \Gaminv{\Gamma^{-1}_{\zeta,\mu}}
\def \bk{\mathbf{k}}
\def \bx{\mathbf{x}}
\def \bR{\mathbf{R}}
\def \bg{\mathbf{g}}
\def \nue{\nu_\mathbf{k}}
\def \etae{\eta_\mathbf{k}}
\def \nup{\nu_\mathbf{q}}
\def \etap{\eta_\mathbf{q}}
\def \bU{\mathcal{U}}
\def \bq{\mathbf{q}}

\def \RIalp{R_{I'_\alpha}}
\def \RJbet{R_{J'_\beta}}
\def \Ialp{I'_\alpha}
\def \Jbet{J'_\beta}
\def \Lkappa{\mathlarger{\mathlarger{\kappa}}}
\def \psiIalpha{\psi^{(1)}_{n,\bk,\bq,I_\alpha}}
\def \psiJbeta{\psi^{(1)}_{n,\bk,\bq,J_\beta}}

\def \HJbeta{\mathcal{H}^{(1)}_{\bk,\bq,J_\beta}}
\def \gIalpha{g^{(1)}_{n,\bk,-\bq,I_\alpha}}
\def \gJbeta{g^{(1)}_{n,\bk,\bq,J_\beta}}
\def \lambdaIalpha{\lambda^{(1)}_{n,\bk,-\bq,I_\alpha}}
\def \lambdaJbeta{\lambda^{(1)}_{n,\bk,\bq,J_\beta}}
\def \lambdaFIalpha{{\mu}^{(1)}_{-\bq,I_\alpha}}
\def \lambdaFJbeta{{\mu}^{(1)}_{\bq,J_\beta}}
\def \VxcJbeta{{\mathcal{V}_{xc}}^{(1)}_{\bq,J_\beta}}
\def \phiJbeta{\phi^{(1)}_{\bq,J_\beta}}
\def \rhoJbeta{\rho^{(1)}_{\bq,J_\beta}}
\def \bIalpha{b^{(1)}_{\bq,I_\alpha}}
\def \bJbeta{b^{(1)}_{\bq,J_\beta}}
\def \bItalpha{\tilde{b}^{(1)}_{\bq,I_\alpha}}
\def \bJtbeta{\tilde{b}^{(1)}_{\bq,J_\beta}}
\def \VIalpha{V^{(1)}_{\bq,I_\alpha}}
\def \VJbeta{V^{(1)}_{\bq,J_\beta}}
\def \VItalpha{\tilde{V}^{(1)}_{\bq,I_\alpha}}
\def \VJtbeta{\tilde{V}^{(1)}_{\bq,J_\beta}}
\def \VnlJbeta{{\mathcal{V}_{nl}}^{(1)}_{\bk,\bq,J_\beta}}

\newcommand{\rev}[1]{\textcolor{black}{#1}}

\begin{document}

\title{Cyclic- and helical-symmetry-adapted phonon formalism within density functional perturbation theory}
\author{Abhiraj Sharma}
\affiliation{Department of Civil Engineering, Indian Institute of Technology Roorkee, Uttarakhand, $247667$, India}
\author{Phanish Suryanarayana}
\email[Email:]{ phanish.suryanarayana@ce.gatech.edu}
\affiliation{College of Engineering, Georgia Institute of Technology, Atlanta, GA $30332$, USA}
\affiliation{College of Computing, Georgia Institute of Technology, Atlanta, GA $30332$, USA}
\begin{abstract}
We present a first-principles framework for the calculation of phonons in nanostructures with cyclic and/or helical symmetry. In particular, we derive a cyclic- and helical-symmetry-adapted representation of the dynamical matrix at arbitrary phonon wavevectors within a variationally formulated, symmetry-adapted density functional perturbation theory framework. In so doing, we also derive the acoustic sum rules for cylindrical geometries, which include a rigid-body rotational mode in addition to the three translational modes. We implement the cyclic- and helical-symmetry-adapted formalism within a high-order finite-difference discretization. Using carbon nanotubes as representative systems, we demonstrate the accuracy of the framework through excellent agreement with periodic plane-wave results. We further apply the framework to compute the Young’s and shear moduli of carbon nanotubes, as well as the scaling laws governing the dependence of ring and radial breathing mode phonon frequencies on nanotube diameter. The elastic moduli are found to be in agreement with previous density functional theory and experimental results, while the phonon scaling laws show qualitative agreement with previous atomistic simulations.
\end{abstract}

\maketitle
\allowdisplaybreaks

\section{Introduction}
Over the past few decades, Kohn--Sham density functional theory (DFT) \cite{Hohenberg, Kohn1965} has firmly established itself as a cornerstone of materials and chemical sciences research. The widespread popularity of Kohn--Sham DFT stems from its generality, simplicity, and favorable accuracy-to-cost ratio relative to other such \emph{ab initio}  methods. Despite substantial advances in numerical algorithms and high-performance computing implementations \cite{gavini2022roadmap}, the computational cost associated with solving the Kohn--Sham problem remains significant, scaling cubically with the number of atoms. As a result, the range of systems accessible to such a rigorous first-principles investigation  remains severely constrained.

Low-dimensional materials have attracted increasing attention over recent decades due to their distinctive mechanical, electronic, vibrational, and thermal properties \cite{xia2003one}. These material systems are not limited to engineered nanostructures such as nanosheets, nanotubes, nanorods, nanowires, nanoribbons, nanodots, and nanoclusters, but also occur naturally in the form of viruses, RNA, and proteins. Non-translational symmetries are ubiquitous in such systems, with cyclic and helical symmetries being among the most prevalent \cite{james2006objective, allen2007nanocrystalline1}. Even when absent in the undeformed configurations, mechanical deformations such as bending and twisting can induce cyclic and helical symmetries \cite{james2006objective, Koskinen2010RPBC}.  Exploiting these symmetries leads to substantial computational savings and simplifications in the analysis of physical properties, motivating the development of cyclic- and helical-symmetry-adapted formulations in atomistic force-field methods \cite{james2006objective, DumitricaobjectiveMD, DayalnonequilibriumMD, aghaei2011symmetry, aghaei2013symmetry}, tight-binding approaches \cite{mintmire1993symmetries, allen2007nanocrystalline3, allen2007nanocrystalline2, popov2004carbon, popov2006radius, gunlycke2008lattice, Zhang2009CNT, Zhang2009dislocation, Koskinen2010RPBC, zhang2017inhomogeneous}, machine-learned force fields (MLFFs) \cite{sharma2025cyclic}, and Kohn--Sham DFT \cite{saito1992electronic, OnoHir2005, OnoHir2005gold, Banerjee2016cyclic, ghosh2019symmetry, banerjee2021ab, sharma2021real}. In particular, the symmetry-adapted Kohn--Sham framework has enabled diverse applications, including bending moduli of 2D materials \cite{Kumar_2022, Kumar_2020}, elastic moduli of nanotubes \cite{Bhardwaj_2021, bhardwaj2022elastic}, the flexoelectric effect in 2D materials \cite{codony_2021, kumar_codony_2021}, and  impact of mechanical deformations on the electronic \cite{bhardwaj2023ab, bhardwaj2022strain} and spintronic \cite{Bhardwaj_2024} properties of nanotubes. However, these studies are restricted to static behavior and response.

Phonons characterize the dynamic response of crystalline materials to small ionic displacements about their equilibrium, zero-force positions. In particular, they correspond to the normal modes of lattice vibrations, with the phonon frequencies and their associated mode shapes at a given wavevector obtained from the eigenvalues and eigenvectors of the corresponding dynamical matrix, respectively. Phonons play a central role in determining a wide range of material properties and response behaviors that cannot be captured by static models, including structural stability \cite{clatterbuck2003phonon, liu2007ab}, thermal conductivity \cite{savrasov1996electron}, elastic moduli \cite{wu2005trends, karki2000high}, heat capacity \cite{lee1995ab, nie2007ab}, coefficients of thermal expansion \cite{fleszar1990first, togo2010first}, and superconductivity \cite{savrasov1996electron}. 

Phonons are generally computed within DFT using one of three approaches: the frozen-phonon method \cite{yin1982calculation}, molecular dynamics simulations \cite{car1985unified}, and density functional perturbation theory (DFPT) \cite{baroni2001phonons, baroni1987green, gonze1989density, zein1984density}. In the frozen-phonon approach, the elements of the dynamical matrix are evaluated using finite-difference approximations to atomic displacement  derivatives.  While conceptually straightforward, this method is computationally expensive, as it requires large supercells to accurately capture low-frequency modes. Alternatively, vibrational properties may be extracted from molecular dynamics simulations, where time-averaged correlations of atomic trajectories are used to compute phonon spectra. This approach is likewise computationally demanding, requiring not only large supercells to access long-wavelength modes, but also a large number of time steps to achieve adequate statistical convergence. These limitations have motivated the development of DFPT-based approaches, including periodic plane-wave formulations \cite{dal1993ab, dal1997density, savrasov1992linear, yu1994linear, baroni1987green, gonze1997dynamical, verstraete2008density, refson2006variational, ABINIT, Espresso, CASTEP, rivano2024density,  sohier2017density} and, more recently, real-space methods \cite{shang2018all, sharma2023calculation} that naturally accommodate both periodic and Dirichlet boundary conditions. By solving linear-response equations, these approaches enable the computation of phonons with arbitrary wavelengths within the fundamental domain, thereby avoiding the need for large supercells. As a result, such approaches are significantly more efficient than both the frozen-phonon and molecular dynamics schemes. Even so, phonon calculations remain expensive, with computational costs scaling quartically with the number of atoms. Moreover, they are only able to exploit translational/periodic symmetry. This restricts the range of systems that can be investigated, especially low-dimensional materials and their responses to mechanical deformations such as bending and torsion, thereby motivating the present effort.

In this work, we present a first-principles framework for phonon calculations in nanostructures exhibiting cyclic and/or helical symmetry. Specifically, we derive a symmetry-adapted dynamical matrix at arbitrary phonon wavevectors within a variationally formulated, symmetry-adapted DFPT framework, together with the corresponding acoustic sum rules for cylindrical geometries. The formulation is implemented using a high-order finite-difference scheme and validated for carbon nanotubes through excellent agreement with periodic plane-wave results.  The framework is further applied to compute the Young’s and shear moduli, as well as phonon scaling laws for carbon nanotubes, obtaining results consistent with prior theoretical and experimental studies.

The remainder of this manuscript is organized as follows. In Sec.~\ref{Sec:Symmetry}, we introduce the mathematical preliminaries for cyclic and helical symmetry. In Sec.~\ref{Sec:Symmetry-adapted DFT}, we summarize the symmetry-adapted formulation of Kohn--Sham DFT. In Sec.~\ref{Sec:Symmetry-adapted DFPT}, we present the cyclic- and helical-symmetry-adapted formalism for phonon calculations, and discuss its implementation in Sec.~\ref{Sec:Implementation}. In Sec.~\ref{Sec:Results}, we apply the developed framework to the study of phonons in carbon nanotubes. Finally, we provide concluding remarks in Sec.~\ref{Sec:Conclusions}.


\begin{widetext}
\section{Cyclic and helical symmetry} \label{Sec:Symmetry}
\begin{figure*}[htbp]
\subfigure[Cyclic]{\includegraphics[keepaspectratio=true,width=0.23\textwidth]{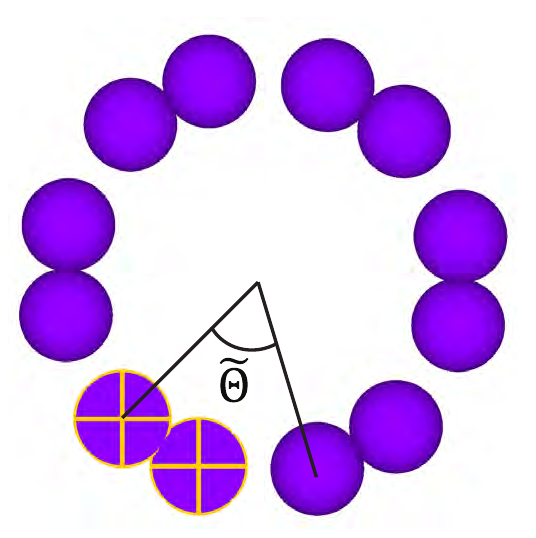}  \label{fig:Cyclicsym}} \hspace{5mm}
\subfigure[Helical]{\includegraphics[keepaspectratio=true,width=0.19\textwidth]{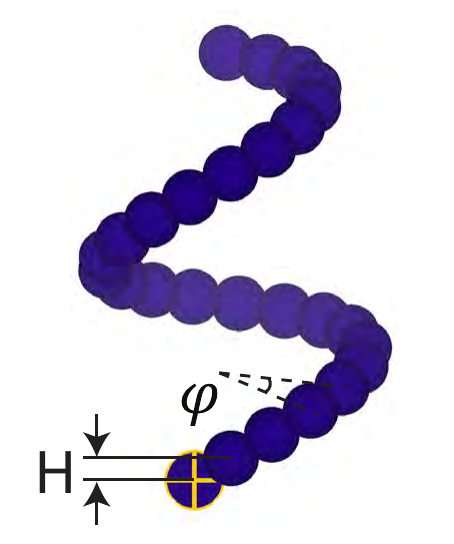}  \label{fig:helicalsym}} \hspace{10mm}
\subfigure[Cyclic and helical]{ \includegraphics[keepaspectratio=true,width=0.19\textwidth]{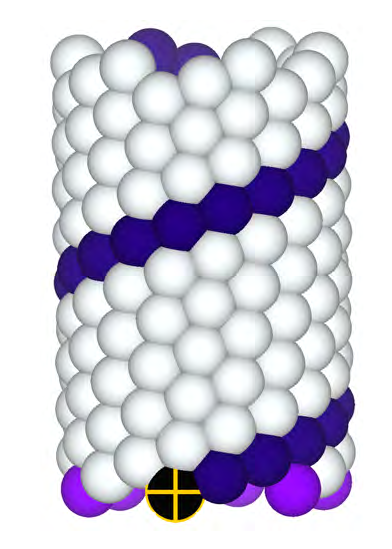} \label{fig:cyclixsym}}
    \caption{Illustration of nanostructures exhibiting (a) cyclic, (b) helical, and (c) combined cyclic and helical symmetries. Fundamental atoms are indicated by plus markers, while their cyclic and helical images are shown in magenta and purple, respectively.}
    \label{Fig:Nanostructure}
\end{figure*}

Consider a quasi-one-dimensional nanostructure comprising of $\hat{N}_a$ fundamental atoms  and exhibiting cyclic and/or helical symmetry, as illustrated in Fig.~\ref{Fig:Nanostructure}. Let \rev{$\tilde{\Theta}$} denote the angle between an atom and its nearest cyclic image \rev{(Fig.~\ref{fig:Cyclicsym})}, and let $\varphi$ denote the circumferential-plane angle between an atom and its nearest helical image, which is separated by a distance $H$ along the longitudinal direction \rev{(Fig.~\ref{fig:helicalsym})}. The symmetry group of the nanostructure can be represented as a set of isometries obtained from the direct product (denoted by $\times$) of a cyclic symmetry group $\mathfrak{C}$ and a helical symmetry group $\mathfrak{H}$ \cite{mcweeny2002symmetry}:
\begin{equation}
\bG = \mathfrak{C} \times \mathfrak{H} = \left\{\Gam = (\mathcal{Q}_{\zeta \tTh}|\mathcal{T}_0) (\mathcal{Q}_{\mu \varphi}|\mathcal{T}_{\mu H}): \zeta = 0,1,2,\cdots,\mathfrak{N}-1, \mu \in \mathbb{Z} \right\} \,,
\end{equation}
where $\mathfrak{N} = 2 \pi / \tTh$ is the order of $\mathfrak{C}$, and $\mathcal{Q}_{\zeta \tTh}$ and $\mathcal{T}_{\mu H}$ are the rotation and translation operators whose action (denoted by $\circ$) on an atom rotate and translate it by $\zeta \tTh$ and $\mu H$, respectively.  The Cartesian coordinates $\bR_{I'}$ of any atom $I' \in \mathbb{Z}$ in the nanostructure can therefore be generated by the action of an element $\Gamma_{\zeta_{I'},\mu_{I'}} \in \bG$ on the Cartesian coordinates $\bR_{I}$ of the corresponding fundamental atom $I\in \{1,2,\cdots,\hat{N}_a\}$ as:
\begin{eqnarray}
\bR_{I'} = \Gamma_{\zeta_{I'},\mu_{I'}} \circ \bR_I = \mathbf{Q}_{\zeta_{I'} \tTh} \mathbf{Q}_{\mu_{I'} \varphi} \bR_I + \mathbf{t}_{\mu_{I'} H} \,,
\end{eqnarray}
where $\zeta_{I'} \in \{0,1,2, \cdots, \mathfrak{N}-1\}$ and $\mu_{I'} \in \mathbb{Z}$ are chosen such that
\begin{align}
 {I'}  = \Gamma_{\zeta_{I'},\mu_{I'}} \circ I 
     = \begin{cases}
 I + \hat{N}_a \zeta_{I'} + \hat{N}_a \mathfrak{N} \mu_{I'}\,, & \mu_{I'} \geq 0 \\
 -I - \hat{N}_a \zeta_{I'} +\hat{N}_a \mathfrak{N} (\mu_{I'}+1)\,, & \mu_{I'} < 0
 \end{cases} \,,
\end{align}
with
\begin{align}
\mathbf{Q}_{\zeta_{I'}\tTh}  =
\begin{bmatrix}
\cos (\zeta_{I'}\tTh)   & -\sin (\zeta_{I'}\tTh) & 0 \vspace{0.05in}\\
\sin (\zeta_{I'}\tTh) & \cos (\zeta_{I'}\tTh) & 0 \vspace{0.05in} \\
0 & 0 & 1
\end{bmatrix} \,,  \quad 
\mathbf{Q}_{\mu_{I'}\phi}  =
\begin{bmatrix}
\cos (\mu_{I'}\varphi)   & -\sin (\mu_{I'}\varphi) & 0 \vspace{0.05in}\\
\sin (\mu_{I'}\varphi) & \cos (\mu_{I'}\varphi) & 0 \vspace{0.05in} \\
0 & 0 & 1
\end{bmatrix} \,, \quad 
\mathbf{t}_{\mu_{I'}H}  = 
\begin{bmatrix}
0 \vspace{0.05in}\\
0 \vspace{0.05in} \\
\mu_{I'}H 
\end{bmatrix} \,.
\end{align}
Similarly, the action of these symmetry operators on a function rototranslate it as:
\begin{equation}
\Gam \circ f(\bx) = f(\Gamma^{-1}_{\zeta,\mu} \circ \bx)\,, \quad \bx \in \mathscr{N} \,,
\end{equation}
where $\mathscr{N}$ is the radially compact region around the nanostructure and $\Gamma^{-1}_{\zeta,\mu} \circ \bx = \mathbf{Q}^{T}_{\zeta \tTh} \mathbf{Q}^{T}_{\mu \varphi} \bx - \mathbf{t}_{\mu H}$.

The geometry of the nanostructure renders it commensurate with the helical coordinate system defined by the coordinates $(r,\tilde{\theta},z)$, which are related to the Cartesian coordinates $\mathbf{x} = [x_1,x_2,x_3]$ through the transformation:
\begin{align}
\begin{pmatrix}
r\\\tilde{\theta}\\z
\end{pmatrix}  = \begin{bmatrix}
\sqrt{x_1^2 +x_2^2}\\
\tan^{-1}\left(\frac{x_2}{x_1}\right) -\frac{\varphi}{H} x_3 +k\pi \\
x_3
\end{bmatrix}
 \,,  \quad 
  k =\begin{cases}
0, \quad  x_1,x_2 \geq 0 \\
1, \quad x_1 <0 \\
2, \quad \text{otherwise} \,.
\end{cases}
\end{align}
Indeed, the nanostructure is periodic along both the $\tilde{\theta}$ and $z$ directions in this helical coordinate system. Therefore, the helical coordinates of the image atom $\mathbf{\tilde{R}}_{I'}$ are related to those of the corresponding fundamental atom $\mathbf{\tilde{R}}_{I}$ by a translation vector, i.e., $\mathbf{\tilde{R}}_{I'} =\mathbf{\tilde{R}}_{I} + (0 \, \, \zeta_{I'} \tTh \, \, \mu_{I'} H)^T$.  Moreover, the 2D reciprocal space associated with the nanostructure in this coordinate system is discrete-continuous and periodic, with the first Brillouin zone (denoted by $\bU$) defined by the set of $|\bG|$ points $(\nu,\eta)$ such that, $\nu \in \{0,1,2,\cdots, \mathfrak{N}-1\}$ and $\eta \in [-\pi/H, \pi/H)$ \cite{sharma2021real}. These Brillouin zone points are commonly referred to as wavevectors. In what follows, the electron and phonon wavevectors are denoted by $\bk=(0,\nu_\bk,\eta_\bk)$ and $\bq=(0,\nu_\bq,\eta_\bq)$, respectively.


\section{Symmetry-adapted DFT} \label{Sec:Symmetry-adapted DFT}
Neglecting spin and adopting the frozen-core pseudopotential approximation \cite{Martin2004}, together with the local density approximation (LDA) for exchange--correlation \cite{Kohn1965}, the cyclic- and helical-symmetry-adapted real-space Kohn--Sham energy functional takes the form \cite{sharma2021real}:
\begin{align}
E[\Psi, \bg, \phi, \hat{\bR}, \bG] = - \frac{1}{N_k} \sum_{\bk \in \bU} \sum_{n=1}^{\hat{N}_s} g_{n,\bk} {\langle \psi_{n,\bk}| \nabla^2| \psi_{n,\bk} \rangle} + {\langle \varepsilon_{xc}(\rho)| \rho \rangle} -\frac{1}{8\pi} {\langle \nabla \phi|\nabla \phi \rangle} + {\langle \rho+b| \phi \rangle} - \frac{1}{2} \sum_{I=1}^{\hat{N}_a} \sum_{I'\in \bG \circ I} \langle \tilde{b}_{I'}|\tilde{V}_{I'} \rangle \nonumber \\
+ \frac{1}{2} {\langle \tilde{b}+b|V_c \rangle} + \frac{2}{N_k} \sum_{\bk \in \bU} \sum_{n=1}^{\hat{N}_s} g_{n,\bk} \sum_{I=1}^{\hat{N}_a} \sum_{p=1}^{\mathcal{P}_I} \gamma_{I,p} \left|{\langle \psi_{n,\bk}|\tilde{\chi}_{I,p,\bk}\rangle} \right|^2 + \frac{2 \sigma}{N_k} \sum_{\bk \in \bU} \sum_{n=1}^{\hat{N}_s} \left(g_{n,\bk} \log g_{n,\bk} + (1-g_{n,\bk}) \log(1-g_{n,\bk}) \right)  \,, \label{Eq:Energy}
\end{align}
where $\Psi$ is the set of symmetry-adapted Kohn-Sham orbitals $\psi_{n,\bk}$ that have  compact support in the radial direction; $\bg$ is the set of occupation numbers $g_{n,\bk}$; $\phi$ is the symmetry-adapted electrostatic potential that has compact support in the radial direction; $\hat{\bR}$ is the set of fundamental atoms in the nanostructure; $N_{k}$ is the number of electron wavevectors in the discretized $\bU$; $\hat{N}_s$ is the number of Kohn-Sham orbitals considered at  each electron wavevector; $\langle \cdot|\cdot \rangle$ denotes the inner product over the fundamental domain of the nanostructure; $\varepsilon_{xc}$ is the exchange-correlation energy per electron; $\rho$ is the electron density:
\begin{align} \label{Eq:ElectronDensity}
\rho = \frac{2}{N_k}  \sum_{\bk \in \bU} \sum_{n=1}^{\hat{N}_s} g_{n,\bk} |\psi_{n,\bk}|^2 \,;
\end{align}
$b = \sum_{I=1}^{\hat{N}_a} \sum_{I'\in \bG \circ I} b_{I'}$ is the total ionic pseudocharge density, where $b_{I'} = -\frac{1}{4\pi} \nabla^2 V_{I'}$ is the spherically symmetric and localized ionic pseudocharge density corresponding to the pseudopotential $V_{I'}$; $\tilde{b} = \sum_{I=1}^{\hat{N}_a} \sum_{I'\in \bG \circ I} \tilde{b}_{I'}$ is the total reference ionic pseudocharge density, where $\tilde{b}_{I'} = -\frac{1}{4\pi} \nabla^2 \tilde{V}_{I'}$ is the spherically symmetric and localized reference ionic pseudocharge density corresponding to the reference ionic potential $\tilde{V}_{I'}$; $V_c = \sum_{I=1}^{\hat{N}_a} \sum_{I'\in \bG \circ I} (\tilde{V}_{I'} - V_{I'})$; $\mathcal{P}_I$ is the number of projectors associated with the $I^\text{th}$ atom; $\tilde{\chi}_{I,p,\bk}$ is the symmetry-adapted nonlocal pseudopotential projector corresponding to $I^\text{th}$ atom and its images:
\begin{align}
\tilde{\chi}_{I,p,\bk} = \sum_{I' \in \bG \circ I} e^{-i m_p (\zeta_{I'} \tTh +\mu_{I'} \varphi)} e^{i \bk \cdot (\tilde{\bR}_{I'}- \tilde{\bR}_{I})}\chi_{I',p} \,,
\end{align}
where $i = \sqrt{-1}$, $m_p$ is the magnetic quantum number and $\chi_{I,p}$ is the nonlocal pseudopotential projector within the Kleinman-Bylander representation \cite{kleinman1982efficacious} having a  normalization constant $\gamma_{I,p}$; and $\sigma$ is the electronic smearing.

The electronic ground state  of the nanostructure is the solution to the optimization problem:
\begin{align}
\min_{\substack{\Psi,\bg}} \max_\phi E[\Psi,\bg,\phi,\hat{\bR},\bG] \quad \text{s.t.} \quad \langle \psi_{n,\bk}|\psi_{n,\bk} \rangle = 1 \quad \text{and} \quad \frac{2}{N_k} \sum_{\bk \in \bU}\sum_{n=1}^{\hat{N}_s} g_{n,\bk} = \hat{N}_e \,,
\end{align}
where the first constraint is on the normality of the orbitals and the second constraint is on the total number of electrons in the fundamental domain. The corresponding Lagrangian can be defined as:
\begin{align}
\mathscr{L}[\Psi,\bg,\phi,\Lambda,\mu,\hat{\bR},\bG] = & E[\Psi,\bg,\phi,\hat{\bR},\bG] - C[\Psi,\bg,\Lambda,\mu,\hat{\bR},\bG] \nonumber \\
= & E[\Psi,\bg,\phi,\hat{\bR},\bG] - \frac{2}{N_k} \sum_{\bk \in \bU} \sum_{n=1}^{\hat{N}_s} g_{n,\bk} \lambda_{n,\bk} \left(\langle \psi_{n,\bk}|\psi_{n,\bk} \rangle - 1 \right) - \mu \left(\frac{2}{N_k} \sum_{\bk \in \bU}\sum_{n=1}^{\hat{N}_s} g_{n,\bk} - \hat{N}_e \right) \,,
\end{align}
where $\Lambda$ is a set of Lagrange multipliers $\lambda_{n,\bk}$ for enforcing the normality  of the orbitals, and $\mu$ is the Lagrange multiplier for enforcing the constraint on the number of electrons in the fundamental domain, commonly referred to as the Fermi level. It follows from the stationarity of the Lagrangian that:
\begin{subequations} \label{Eqn:Euler-Lagrange}
\begin{align}
\left( \mathcal{H}_\bk := -\frac{1}{2} \nabla^2 + \mathcal{V}_{xc} + \phi + {\mathcal{V}_{nl}}_\bk \right) \psi_{n,\bk} &=\lambda_{n,\bk} \psi_{n,\bk} \,, \quad \Gaminv \circ \psi_{n,\bk} = e^{i \left(\nue \zeta \tTh+ \etae \mu H \right)} \psi_{n,\bk} \,, \label{Eq:EL:Eigen} \\
-\frac{1}{4 \pi} \nabla^2 \phi &= \rho + b \,; \quad \Gaminv \circ \phi = \phi \,, \label{Eq:EL:Poisson}\\
g_{n,\bk} &= \left(1+e^{ \frac{\lambda_{n,\bk} - \mu}{\sigma}}\right)^{-1} \,, \quad \mu \,\,\, \text{is s.t.} \,\,\, \frac{2}{N_k}  \sum_{\bk \in \bU} \sum_{n=1}^{\hat{N}_s} g_{n,\bk} = \hat{N}_e \,, \label{Eq:EL:Occ}
\end{align} 
\end{subequations} 
where $\mathcal{H}_\bk$ is the wavevector-dependent Hamiltonian operator, with the exchange-correlation and nonlocal pseudopotential operators taking the form:
\begin{subequations} 
\begin{align}
\mathcal{V}_{xc} &=  \rho \frac{d \varepsilon_{xc}}{d \rho} +\varepsilon_{xc} \,, \label{Eq:EL:Vxc}\\
{\mathcal{V}_{nl}}_\bk &= \sum_{I=1}^{\hat{N}_a} \sum_{p=1}^{\mathcal{P}_I} \gamma_{I,p} |\tilde{\chi}_{I,p,\bk}\rangle \langle \tilde{\chi}_{I,p,\bk}| \,. \label{Eq:EL:Vnl}
\end{align}
\end{subequations}
\rev{The boundary condition satisfied by the symmetry-adapted Kohn--Sham orbitals in Eq.~\ref{Eq:EL:Eigen} corresponds to Bloch's theorem in the helical coordinate system \cite{sharma2021real}. As noted in the previous section, the corresponding Brillouin zone consists of the points $(\nu,\eta)$ such that $\nu \in \{0,1,2,\cdots,\mathfrak{N}-1\}$ and $\eta \in [-\pi/H,\pi/H)$.} The Hamiltonian depends on the electron density, which in turn depends on the orbitals, and therefore the governing equations must be solved \rev{self-consistently. A common approach is the self-consistent field (SCF) method, i.e., a fixed-point iteration in which the eigenproblem in Eq.~\ref{Eq:EL:Eigen} is solved to obtain the orbitals, from which, together with the occupations in Eq.~\ref{Eq:EL:Occ}, the electron density is computed using Eq.~\ref{Eq:ElectronDensity}. The electrostatic potential is then obtained by solving the Poisson problem in Eq.~\ref{Eq:EL:Poisson}; together with the exchange-correlation potential in Eq.~\ref{Eq:EL:Vxc}, it is used to update the eigenproblem, and the process is repeated until self-consistency is achieved.} The resulting electronic ground-state quantities can then be used to evaluate the energy and its first-order derivatives,  which includes the atomic forces and stress tensor \cite{sharma2021real}. 

\rev{The cyclic- and helical-symmetry-adapted formalism presented above is for spin-unpolarized calculations within the pseudopotential approximation and at the LDA level of exchange and correlation. As in standard DFT, the pseudopotential approximation is adopted to reduce computational cost. The extension to all-electron calculations follows the standard DFT treatment \cite{suryanarayana2011mesh, Phanish2010}: the nonlocal pseudopotential energy term in Eq.~\ref{Eq:Energy} and the corresponding operator in the Hamiltonian in Eq.~\ref{Eq:EL:Eigen} are set to zero, and the ionic pseudopotential is replaced by the corresponding Coulomb potential. The framework can also be generalized to collinear spin-polarized calculations by solving separate eigenproblems for the two spin channels, which are coupled through the spin-resolved and total electron densities, following standard DFT. It can further be extended to semilocal exchange-correlation functionals, as demonstrated in earlier cyclic- and helical-symmetry-adapted work \cite{sharma2021real}.}

\section{Symmetry-adapted phonons}\label{Sec:Symmetry-adapted DFPT}

The evaluation of second-order energy derivatives, from which quantities such as phonons can be obtained, requires the first-order corrections to the zeroth-order Kohn--Sham solutions, as dictated by the \(2n+1\) theorem \cite{gonze1989density}. We now derive the governing equations for the first-order correction terms describing the response of the nanostructure to atomic perturbations, thereby establishing a cyclic- and helical-symmetry-adapted formulation of DFPT, which provides the quantities required for the evaluation of the symmetry-adapted dynamical matrix, from which the phonons can be calculated.

\subsection{Perturbative expansion}
The periodicities of the nanostructure in the helical coordinate system allows for the atomic perturbations to be expressed in terms of plane-waves as:
\begin{eqnarray}
\Delta \tilde{R}_{\Ialp} = \sum_{\bq \in \bU} e^{i \bq \cdot \tilde{\bR}_{I'} } \tilde{u}_{\bq,I_\alpha}\,\quad \forall \, I' \in \bG \circ I, \,\, I=1,2,\cdots \hat{N}_a, \,\, \alpha = 1,2,3 \,, \label{Eqn:DFT_helical}
\end{eqnarray}
where $\Delta \tilde{R}_{I'_\alpha}$ denotes the perturbation  of the atom at $\mathbf{\tilde{R}}_{I'}$ in the $\alpha^\text{th}$ direction, and $\tilde{u}_{\bq,I_\alpha}$ is the discrete Fourier transform of the distribution associated with the displacment of the $I^\text{th}$ atom and its images in the $\alpha^\text{th}$ direction:
\begin{eqnarray}
\tilde{u}_{\bq,I_\alpha} = \frac{1}{|\bG|} \sum_{I'\in \bG \circ I}  e^{-i \bq \cdot \tilde{\bR}_{I'} } \Delta \tilde{R}_{I'_\alpha} \,. \label{Eqn:IDFT_helical}
\end{eqnarray}
\rev{The} atomic displacement in the Cartesian coordinate system can be obtained using the transformation:
\begin{eqnarray}
\begin{bmatrix}
\Delta R_{I'_1} \\
\Delta R_{I'_2} \\
\Delta R_{I'_3} 
\end{bmatrix} = \mathbf{Q}_{\tilde{R}_{I'_2}} \mathbf{Q}_{\frac{\varphi}{H}\tilde{R}_{I'_3}}
\begin{pmatrix}
\Delta \tilde{R}_{I'_1} \\
\tilde{R}_{I_1}\left(\Delta \tilde{R}_{I'_2} + \frac{\varphi}{H} \Delta \tilde{R}_{I'_3}\right)\\
\Delta \tilde{R}_{I'_3}
\end{pmatrix} \,. \label{Eqn:Helical2Cart}
\end{eqnarray}
The Kohn-Sham variables, namely orbitals, occupations, eigenvalues, and Fermi level, associated with the perturbed nanostructure can therefore be expressed as:
\begin{eqnarray}
\Lkappa_s &=& \Lkappa^{(0)}_s + \sum_{I=1}^{\hat{N}_a}  \sum_{I'\in\bG \circ I}\sum_{\alpha=1}^3 \Lkappa^{(1)}_{s,\Ialp} \Delta \RIalp + \text{higher order terms} \nonumber \\
&=& \Lkappa^{(0)}_s + \sum_{\bq \in \bU} \sum_{I=1}^{\hat{N}_a} \sum_{\alpha=1}^3 \Lkappa^{(1)}_{s,\bq,I_\alpha} u_{\bq,I_\alpha} + \text{higher order terms} \,,
\end{eqnarray}
where $\Lkappa_s$ represents any of the aforementioned Kohn-Sham variables, $\Lkappa^{(0)}_s$ denotes the unperturbed solution, $\Lkappa^{(1)}_s$ denotes the first-order correction term, and 
\begin{subequations}
\begin{eqnarray}
\begin{bmatrix}
u_{\bq,I_1} \\
u_{\bq,I_2} \\
u_{\bq,I_3}
\end{bmatrix}  &=& e^{i \bq \cdot\tilde{\bR}_{I}} \mathbf{Q}_{\tilde{R}_{I_2}} \mathbf{Q}_{\frac{\varphi}{H} \tilde{R}_{I_3}}
\begin{pmatrix}
\tilde{u}_{\bq,I_1} \\
\tilde{R}_{I_1} \left(\tilde{u}_{\bq,I_2} + \frac{\varphi}{H} \tilde{u}_{\bq,I_3} \right)\\
\tilde{u}_{\bq,I_3} \label{Eqn:IDFT_coef_cart}
\end{pmatrix} \,, \\
\begin{bmatrix}
\Lkappa^{(1)}_{s,\bq,I_1} \\
\Lkappa^{(1)}_{s,\bq,I_2} \\
\Lkappa^{(1)}_{s,\bq,I_3}
\end{bmatrix} &=& \sum_{I' \in \bG \circ I} e^{i \bq \cdot (\tilde{\bR}_{I'}- \tilde{\bR}_{I})}  \mathbf{Q}^T_{\zeta_{I'} \tTh} \mathbf{Q}^T_{\mu_{I'} \varphi}
\begin{bmatrix}
\Lkappa^{(1)}_{s,I'_1} \\
\Lkappa^{(1)}_{s,I'_2} \\
\Lkappa^{(1)}_{s,I'_3}
\end{bmatrix} \,.
\end{eqnarray}
\end{subequations}
\subsection{Sternheimer equation}
The Lagrangian associated with the perturbed nanostructure can be written within the adiabatic harmonic approximation as:
\begin{align}
&\mathscr{L}[\Lkappa^{(1)},\Lkappa^{(2)},\bR + \Delta \bR] = \bar{E}[\hat{\bR},\bG] + \frac{1}{|\bG|}\sum_{I=1}^{\hat{N}_a}  \sum_{I'\in\bG \circ I}\sum_{\alpha=1}^3 \left(\frac{\partial E}{\partial \RIalp} + \sum_{s=1}^{|\Lkappa|} \left \{\left\langle\frac{\delta E}{\delta \Lkappa_s}-\frac{\delta C}{\delta \Lkappa_s}\bigg| \Lkappa^{(1)}_{s,\Ialp} \right\rangle_{\mathscr{N}} \right \}\right) \Delta \RIalp + \frac{1}{2 |\bG|} \sum_{I=1}^{\hat{N}_a}  \nonumber \\
&\sum_{I' \in \bG \circ I}  \sum_{\alpha=1}^3  \sum_{J=1}^{\hat{N}_a}\sum_{J' \in \bG \circ J} \sum_{\beta=1}^3 \left(\frac{\partial^2 E}{\partial \RIalp \partial \RJbet} + \sum_{s=1}^{|\Lkappa|} \left \{ \sum_{t=1}^{|\Lkappa|} \left \langle \Lkappa^{(1)}_{s,\Ialp} \bigg|\frac{\delta^2 E}{\delta \Lkappa_s \delta \Lkappa_t} - \frac{\delta^2 C}{\delta \Lkappa_s \delta \Lkappa_t} \bigg| \Lkappa^{(1)}_{t,\Jbet} \right \rangle_{\mathscr{N}} + \left \langle \frac{\delta \partial E}{\delta \Lkappa_s \partial \RIalp}\bigg|\Lkappa^{(1)}_{s,\Jbet} \right \rangle_{\mathscr{N}}  \right. \right.   \nonumber \\
&\left. \left.  + \left \langle \Lkappa^{(1)}_{s,\Ialp} \bigg| \frac{\partial \delta E}{\partial \RJbet \delta \Lkappa_s} \right \rangle_{\mathscr{N}} + \left \langle \frac{\delta E}{\delta \Lkappa_s}-\frac{\delta C}{\delta \Lkappa_s} \bigg| \Lkappa^{(2)}_{s,\Ialp,\Jbet} \right \rangle_{\mathscr{N}} \right\} \right)\Delta R^*_{I'_\alpha} \Delta \RJbet \,,
\end{align}
where $\Lkappa$ is a set with cardinality $|\Lkappa|$ consisting of all aforementioned Kohn-Sham variables, $\bR$ is the set of all atoms in the nanostructure, $E$ is the energy of the unperturbed nanostructure, and $\langle \cdot| \cdot \rangle_{\mathscr{N}}$ denotes the inner product over the entire nanostructure. Note that the Lagrangian has been normalized by the symmetry group order $|\bG|$. Utilizing the perturbative expansions of the Kohn-Sham variables, the Lagrangian takes the following form about the structural ground state of the unperturbed nanostructure:
\begin{align}
&\mathscr{L}[\Psi^{(1)},\bg^{(1)}, \mathbf{\Lambda}^{(1)},\mu^{(1)},\bR + \Delta \bR] = \bar{E}[\hat{\bR},\bG] + \frac{1}{2} \sum_{\bq \in \bU} \sum_{I=1}^{\hat{N}_a} \sum_{J=1}^{\hat{N}_a} \sum_{\alpha=1}^3 \sum_{\beta=1}^3  \Bigg(\frac{1}{|\bG|} \sum_{I'\in\bG \circ I} \sum_{J'\in\bG \circ J} e^{-i \bq \cdot (\tilde{\bR}_{I'} - \tilde{\bR}_{I} - \tilde{\bR}_{J'} + \tilde{\bR}_{J}) } \nonumber \\
& \left[\mathbf{Q}^T_{\zeta_{I'} \tTh + \mu_{I'} \varphi} \frac{\partial^2 E}{\partial \bR_{I'} \partial \bR_{J'}} \mathbf{Q}_{\zeta_{J'} \tTh + \mu_{J'} \varphi} \right]_{\alpha \beta} + \left \langle \bIalpha \Big|\phiJbeta \right\rangle + \frac{2}{N_k}\sum_{\bk \in \bU^-} \sum_{n=1}^{\hat{N}_s} g_{n,\bk} \left\{\left \langle \psiIalpha\right|\mathcal{H}_{\bk+\bq} - \lambda_{n,\bk}\left|\psiJbeta \right \rangle + \right. \nonumber \\
& \left. \left \langle \psiIalpha \Big| \HJbeta  - \delta_{\bq \mathbf{0}}\lambdaJbeta \Big|   \psi_{n,\bk} \right \rangle + \left \langle \psi_{n,\bk} \Big| {\mathcal{V}_{nl}}^{(1)}_{\bk+\bq,-\bq,I_\alpha} -\delta_{\bq \mathbf{0}}\lambdaIalpha \Big|\psiJbeta \right \rangle \right\} + \frac{2}{N_k} \sum_{\bk \in \bU} \sum_{n=1}^{\hat{N}_s} \bigg[\gIalpha \nonumber \\
& \bigg\{ \sigma\frac{\gJbeta}{g_{n,\bk} \left(1-g_{n,\bk} \right)} +  \delta_{\bq \mathbf{0}} \left \langle \psi_{n,\bk} \Big| \HJbeta\Big|\psi_{n,\bk} \right \rangle - \lambdaFJbeta \bigg\} + \gJbeta \left\{\delta_{\bq \mathbf{0}}\left \langle \psi_{n,\bk} \Big| {\mathcal{V}_{nl}}^{(1)}_{\bk,-\bq,I_\alpha} \Big| \psi_{n,\bk} \right \rangle - \lambdaFIalpha \right\} \bigg]\Bigg) u^*_{\bq,I_\alpha}u_{\bq,J_\beta} \,, \label{Eqn:Lagrangian_perturbed}
\end{align}
where $\delta_{ij}$ is the Kronecker delta function, $\bU^{-} = \bU \cup -\bU$, and
\begin{eqnarray}
\HJbeta &:=& \VxcJbeta + \phiJbeta + \VnlJbeta\,; \nonumber \\
 \VxcJbeta & =& \frac{d \mathcal{V}_{xc}}{d \rho} \rhoJbeta = \left(\rho \frac{d^2 \varepsilon_{xc}}{d \rho^2} + 2 \frac{d \varepsilon_{xc}}{d \rho} \right) \rhoJbeta \,, \nonumber\\
\phiJbeta &=& \langle G|\rhoJbeta + \bJbeta \rangle  \,, \nonumber \\
\VnlJbeta & =& \sum_{p=1}^{\mathcal{P}_J} \gamma_{J,p} \left(|\tilde{\chi}^{(1)}_{J_\beta,p,\bk,\bq}\rangle \langle \tilde{\chi}_{J,p,\bk}| + |\tilde{\chi}_{J,p,\bk+\bq}\rangle \langle \tilde{\chi}^{(1)}_{J_\beta,p,\bk,\mathbf{0}}| \right) \,. \label{Eq:PerturbationsQuant}
\end{eqnarray}
In addition,
\begin{subequations}
\begin{eqnarray}
\rhoJbeta &=& \frac{2}{N_k} \sum_{\bk \in \bU} \sum_{n=1}^{\hat{N}_s}\left(\gJbeta  |\psi_{n,\bk}|^2 + g_{n,\bk} \psi^{(1)}_{n,-\bk,\bq,J_\beta} \psi_{n,\bk} + g_{n,\bk} \psi^*_{n,\bk} \psiJbeta \right) \,,\\
\bJbeta &=& -\sum_{J' \in \bG \circ J} e^{i \bq \cdot (\tilde{\bR}_{J'}- \tilde{\bR}_{J}) } \left[\mathbf{Q}^T_{\zeta_{J'} \tTh} \mathbf{Q}^T_{\mu_{J'} \varphi} \nabla b_{J'}\right]_\beta \,, \\
\tilde{\chi}^{(1)}_{J_\beta,p,\bk,\bq} &=& -\sum_{J'\in \bG \circ J} e^{-i m_p (\zeta_{J'} \tTh +\mu_{J'} \varphi)} e^{i (\bk+\bq) \cdot (\tilde{\bR}_{J'}- \tilde{\bR}_{J}) } \left[\mathbf{Q}^T_{\zeta_{J'} \tTh} \mathbf{Q}^T_{\mu_{J'} \varphi} \nabla \chi_{J',p}\right]_\beta \,.
\end{eqnarray}
\end{subequations}
Also, $G$ is the Green's function corresponding to the Laplacian operator, i.e., Coulomb kernel,  whereby $\phiJbeta$ can be written as the solution to the Poisson problem:
\begin{equation}
-\frac{1}{4 \pi} \nabla^2 \phiJbeta = \rhoJbeta + \bJbeta \,, \quad \Gaminv \circ \phiJbeta = e^{i \left(\nup \zeta \tTh+ \etap \mu H \right) } \phiJbeta \,.
\end{equation}
To obtain Eq.~\ref{Eqn:Lagrangian_perturbed}, in addition to the governing equations associated with the unperturbed Kohn--Sham variables (Eq.~\ref{Eqn:Euler-Lagrange}), we have employed several key mathematical identities and transformation relations.  First, we have used the block-circulant nature of the dynamical matrix in the helical coordinate system to decouple different phonon wavevectors. Second, we exploit the spherical symmetry of the ionic pseudocharge densities and the nonlocal pseudopotential projectors to transform derivatives with respect to atomic positions into derivatives with respect to spatial coordinates. Third, we have used the commutation of the Hamiltonian operator with the cyclic and helical symmetry group operators, periodic boundary conditions satisfied by the electron density and the electrostatic potential, and Bloch-periodic boundary conditions satisfied by the Kohn-Sham orbitals, $\psiJbeta$, $\rho^{(1)}_{\bq,J_\beta}$, and $\phi^{(1)}_{\bq,J_\beta}$ to reduce the inner products defined originally over the entire nanostructure to the ones defined over the fundamental domain of the unperturbed nanostructure. Lastly, we have used the conjugate symmetry property of inner product, self-adjoint property of the Hamiltonian operator, and time reversal symmetry: $\psi_{n,\bk,-\bq,J_\beta}^* = \psi_{n,-\bk,\bq,J_\beta}$, to simplify the terms in the Lagrangian.

It follows from the stationarity of the Lagrangian for the perturbed nanostructure in Eq.~\ref{Eqn:Lagrangian_perturbed} that:
\begin{subequations}
\begin{align}
&\left(\mathcal{H}_{\bk+\bq} - \lambda_{n,\bk} \right)\psiJbeta = \left(\delta_{\bq \mathbf{0}}\lambdaJbeta - \HJbeta\right) \psi_{n,\bk} ,\, \, \Gaminv \circ \psiJbeta = e^{i \left( (\nue+\nup) \zeta \tTh+ (\etae + \etap) \mu H \right) } \psiJbeta , \label{Eq:Sternheimer}\\
& \quad \quad \quad \quad \quad \quad \, \, \, \, \, \gJbeta = - \frac{g_{n,\bk} (1-g_{n,\bk})}{\sigma} \left(\delta_{\bq \mathbf{0}}\left \langle \psi_{n,\bk} \Big| \HJbeta\Big|\psi_{n,\bk} \right \rangle - \lambdaFJbeta \right) \,, \label{Eq:gJbeta} \\
& \quad \quad \quad \, \, \, \, \, \langle\psi_{n,\bk}|\psi^{(1)}_{n,\bk,\mathbf{0},J_\beta}\rangle = 0 \,, \label{Eq:Constrained1}\\
&\quad \quad \quad \, \, \sum_{\bk \in \bU} \sum_{n=1}^{N_s} \gJbeta = 0 \,, \label{Eq:Constrained2}
\end{align}
\end{subequations}
where the last two equations provide constraints on $\psiJbeta$ and $\gJbeta$, respectively. Multiplying Eq.~\ref{Eq:Sternheimer} by $\langle\psi_{n,\bk}|$, we obtain the relation for $\lambda^{(1)}_{n,\bk,\mathbf{0},J_\beta}$ as:
\begin{equation}
\lambda^{(1)}_{n,\bk,\mathbf{0},J_\beta} = \left \langle \psi_{n,\bk}\Big|\mathcal{H}^{(1)}_{\bk,\mathbf{0},J_\beta}\Big|\psi_{n,\bk}\right\rangle \,.
\end{equation}
Similarly, utilizing Eqs.~\ref{Eq:gJbeta} \&~\ref{Eq:Constrained2} we obtain the expression of $\lambdaFJbeta$ as:
\begin{equation}
\lambdaFJbeta = \delta_{\bq \mathbf{0}}\frac{\sum_{\bk \in \bU} \sum_{n=1}^{\hat{N}_s} g_{n,\bk} (1-g_{n,\bk}) \lambdaJbeta}{\sum_{\bk \in \bU} \sum_{n=1}^{\hat{N}_s} g_{n,\bk} (1-g_{n,\bk})} \,.
\end{equation}
Above, Eq.~\ref{Eq:Sternheimer} represents the cyclic- and helical-symmetry-adapted analogue of the  Sternheimer equation. The operator for this equation has singularity at $\bq = \mathbf{0}$ and is poorly conditioned when the eigenvalues at the wavevectors  $\bk$ and $\bk+\bq$ are close to each other. To overcome the numerical challenges associated with solving this ill-conditioned problem, we modify the Sternheimer equation as \cite{sharma2023calculation}:
\begin{subequations}
\begin{align}
&\left(\mathcal{H}_{\bk+\bq} + \mathcal{W}_{\bk+\bq} - \lambda_{n,\bk} \right)\psiJbeta = \left(\mathcal{P}_{n,\bk+\bq} - \mathcal{I}_{\bk+\bq} \right)\HJbeta \psi_{n,\bk} \,, \\
&\mathcal{W}_{\bk+\bq}  = \sum_{m=1}^{N_s} \xi_{m,\bk+\bq} |\psi_{m,\bk+\bq}\rangle \langle \psi_{m,\bk+\bq}| \,, \quad \mathcal{P}_{n,\bk+\bq} = \sum_{m=1}^{N_s} \zeta_{n,m,\bk,\bq} |\psi_{m,\bk+\bq}\rangle \langle \psi_{m,\bk+\bq}| \,, \\
&\zeta_{n,m,\bk,\bq} =  \begin{cases}
    \delta_{\bq \mathbf{0}} + (1-\delta_{\bq \mathbf{0}}) \left[1- \xi_{m, \bk+\bq
    } \left(\frac{1-g_{n,\bk}}{2\sigma}\right) \right],& \text{if }  m=n \,\&\, \lambda_{n,\bk} = \lambda_{m,\bk+\bq} \,, \\
    1- \xi_{m, \bk+\bq} \left( \frac{1-g_{n,\bk}}{2\sigma} \right),& \text{if }  m \neq n \,\&\, \lambda_{n,\bk} = \lambda_{m,\bk+\bq} \,, \\
    \frac{\xi_{m,\bk+\bq}}{\lambda_{n,\bk} - \lambda_{m,\bk+\bq}},& \text{otherwise} \,. 
\end{cases}
\end{align}
\end{subequations}
where $\mathcal{I}$ denotes the identity operator, and the coefficients $\xi_{m,\mathbf{k}+\mathbf{q}}$ are chosen to remove the singularity and improve the conditioning of the equation, thereby ensuring robust and faster convergence to the solution.

\rev{The cyclic- and helical-symmetry-adapted DFPT formalism presented above can be extended to all-electron calculations by replacing the ionic pseudopotential with the Coulomb potential and setting both the nonlocal pseudopotential operator in the Hamiltonian $\mathcal{H}_{\bk+\bq}$ and its perturbation $\VnlJbeta$ in Eq.~\ref{Eq:PerturbationsQuant} to zero. It can also be generalized to spin-polarized systems by solving the Sternheimer equations separately for the orbital perturbations in each spin channel, with the two channels coupled through the perturbations in the spin-resolved and total electron densities, following standard DFPT \cite{dal2001density}. The extension to semilocal exchange-correlation functionals requires a corresponding modification of $\VxcJbeta$ in Eq.~\ref{Eq:PerturbationsQuant}, as in standard DFPT \cite{dal2000ab}.}

\subsection{Dynamical matrix}
The harmonic term in the ground-state energy of the perturbed nanostructure, which forms part of the Lagrangian in Eq.~\ref{Eqn:Lagrangian_perturbed}, yields the  phonon wavevector dependent symmetry-adapted dynamical matrix $\mathbf{D_q}$. In particular, the $(I_\alpha, J_\beta)^\text{th}$ element of $\mathbf{D_q}$ takes the form:
\begin{align}
&[\mathbf{D_q}]_{I_\alpha J_\beta}  = \frac{1}{\sqrt{M_I M_J}} \Bigg\{\frac{1}{2} \left \langle \bIalpha +\bItalpha\Big|\VJtbeta- \VJbeta\right\rangle  + \frac{1}{2} \left \langle \VItalpha -\VIalpha \Big|\bJbeta + \bJtbeta \right \rangle+\left \langle \bIalpha \Big|\phiJbeta \right\rangle -\frac{\delta_{IJ}}{2} \sum_{J' \in \bG \circ J}  \nonumber \\
& \bigg[\mathbf{Q}^T_{\zeta_{J'} \tTh} \mathbf{Q}^T_{\mu_{J'} \varphi} \left(\big \langle \nabla (b_{J'}+\tilde{b}_{J'}) \big| \nabla V_c \big \rangle+ 2 \big\langle \nabla b_{J'}\big| \nabla \phi \big \rangle + \big \langle \nabla(\tilde{V}_{J'} - V_{J'}) \big | \nabla( b+\tilde{b}) \big \rangle \right) \mathbf{Q}_{\zeta_{J'} \tTh}\mathbf{Q}_{\mu_{J'} \varphi} \bigg]_{\alpha \beta} +  \frac{1}{N_k} \sum_{\bk \in \bU^-} \sum_{n=1}^{\hat{N}_s} \bigg(g^{(1)}_{n,\bk,\mathbf{0},J_\beta} \nonumber \\
& \left \langle \psi_{n,\bk}\Big|{V_{nl}}^{(1)}_{\bk,\mathbf{0},I_\alpha}\Big|\psi_{n,\bk} \right \rangle  + g_{n,\bk} \left \langle \psi_{n,\bk} \Big| {V_{nl}}^{(2)}_{\bk,I_\alpha,J_\beta}\Big|\psi_{n,\bk} \right\rangle + 2 g_{n,\bk} \left \langle \psi_{n,\bk}\Big|{V_{nl}}^{(1)}_{\bk+\bq,-\bq,I_\alpha}\Big|\psiJbeta \right\rangle \bigg) \Bigg\} \,, \label{Eq:Complete:Dq}
\end{align}
where $M_I$ is the mass of the $I^\text{th}$ fundamental atom, and 
\begin{equation}
{V_{nl}}^{(2)}_{\bk,I_\alpha,J_\beta} = 2\delta_{IJ} \sum_{p=1}^{\mathcal{P}_J} \gamma_{J,p} \Re  \bigg\{ | \tilde{\chi}^{(2)}_{J_{\alpha \beta},p,\bk} \rangle \langle \tilde{\chi}_{J,p,\bk}|  +  | \tilde{\chi}^{(1)}_{J_\alpha,p,\bk,\mathbf{0}} \rangle \langle \tilde{\chi}^{(1)}_{J_\beta,p,\bk,\mathbf{0}}| \bigg\} \,,
\end{equation}
with $\Re[.]$ denoting the real part, and 
\begin{equation}
\tilde{\chi}^{(2)}_{J_{\alpha \beta},p,\bk} = \sum_{J'\in \bG \circ J} e^{i \bk \cdot (\tilde{\bR}_{J'}- \tilde{\bR}_{J}) }  \left[\mathbf{Q}^T_{\zeta_{J'} \tTh} \mathbf{Q}^T_{\mu_{J'} \varphi} \nabla \left(\mathbf{Q}^T_{\zeta_{J'} \tTh} \mathbf{Q}^T_{\mu_{J'} \varphi} \nabla \chi_{J,p} \right)^T \right]_{\alpha \beta} \,.
\end{equation}
In the derivation of the dynamical matrix expression, we have used the spherically symmetric nature of the ionic psudopotentials, ionic pseudocharge densities, and nonlocal pseudopotential projectors to transform the derivatives with respect to atomic positions into derivatives with respect to spatial coordinates. In addition, we employ integration by parts and the Gauss divergence theorem to rewrite electrostatic terms involving second derivatives in terms of first derivatives, yielding expressions that are both more accurate and computationally more efficient.

The phonons associated with the wavevector $\bq$ can be written as the solution to the generalized  eigenproblem:
\begin{eqnarray}
\mathbf{D}_\bq \mathbf{v}_\bq = \omega^2_\bq \mathbf{v}_\bq \,,
\end{eqnarray}
where  $\omega_{\mathbf{q}}$ and $\mathbf{v}_{\mathbf{q}}$ are the phonon frequencies and mode shapes, respectively. The cylindrical geometry of the aforedescribed nanostructure permits four independent rigid body perturbations, namely, one translation along the axis, two translations perpendicular to the axis and one rotation about the axis. Correspondingly, there exist four zero-frequency phonon modes, also known as acoustic modes, for which the atomic forces on all atoms of the perturbed nanostructure vanish, i.e.,
\begin{align}
&\mathbf{f}_{K} = -\frac{d \bar{E}[\bR+\Delta \bR]}{d(\bR_{K}+\Delta \bR_{K})} = -\frac{1}{2} \sum_{\bq \in \bU} \sum_{I=1}^{\hat{N}_a} \sum_{J=1}^{\hat{N}_a} \sum_{\alpha=1}^3 \sum_{\beta=1}^3 \frac{d \left([\mathbf{v}^*_\bq]_{I_\alpha}[\mathbf{D_q}]_{I_\alpha J_\beta} [\mathbf{v}_\bq]_{J_\beta} \right)}{d\Delta \bR_{K}}  = \mathbf{0} \, \quad \forall \, K \in {1,2, \cdots, \hat{N}_a} \nonumber \\
&=> \Re \left( \sum_{\bq \in \bU} \sum_{I=1}^{\hat{N}_a} \sum_{I'\in\bG \circ I} \sum_{\alpha=1}^3  \sqrt{M_I} e^{i \bq \cdot (\tilde{\bR}_{I'}- \tilde{\bR}_{I})} \left[\mathbf{Q}^T_{\zeta_{I'} \tTh} \mathbf{Q}^T_{\mu_{I'}\varphi} \Delta \bR_{I'}  \right]_\alpha [\mathbf{D_q}]_{I_\alpha K_\beta}  \right) = \mathbf{0} \,\, \forall \, K \in {1,2, \cdots, \hat{N}_a}, \beta = 1,2,3 \,.
\end{align}
In arriving at the above relation, we have used Eqs.~\ref{Eqn:IDFT_helical},~\ref{Eqn:Helical2Cart},~\ref{Eqn:IDFT_coef_cart},~\ref{Eqn:Lagrangian_perturbed}, and~\ref{Eq:Complete:Dq} as well as the Hermitian nature of the symmetry-adapted dynamical matrix for any phonon wavevector. Inserting the perturbation vector for each of the rigid body motions in the above equation, we obtain the following symmetry-adapted acoustic sum rules:
\begin{subequations}
\begin{eqnarray}
\text{x-translation: }&& \sum_{I=1}^{\hat{N}_a} \sqrt{M_I} \Re\left([\mathbf{D}_{(0,-1,-\frac{\varphi}{H})}]_{I_1 K_\beta} + i [\mathbf{D}_{(0,-1,-\frac{\varphi}{H})}]_{I_2 K_\beta} \right) = 0  \,\, \forall \, K \in {1,2, \cdots, \hat{N}_a}, \beta = 1,2,3 \,,\\
\text{y-translation: }&& \sum_{I=1}^{\hat{N}_a} \sqrt{M_I} \Re\left(-i [\mathbf{D}_{(0,1,\frac{\varphi}{H})}]_{I_1 K_\beta} + [\mathbf{D}_{(0,1,\frac{\varphi}{H})}]_{I_2 K_\beta} \right) = 0 \,\, \forall \, K \in {1,2, \cdots, \hat{N}_a}, \beta = 1,2,3 \,,\\
\text{z-translation: }&& \sum_{I=1}^{\hat{N}_a} \sqrt{M_I}[\mathbf{D}_{\mathbf{0}}]_{I_3 K_\beta} = 0 \,\, \forall \, K \in {1,2, \cdots, \hat{N}_a}, \beta = 1,2,3 \,,\\
\text{Rotation: }&& \sum_{I=1}^{\hat{N}_a} \sqrt{M_I}\left(-\sin(\theta_I) [\mathbf{D}_{\mathbf{0}}]_{I_1 K_\beta} + \cos(\theta_I) [\mathbf{D}_{\mathbf{0}}]_{I_2 K_\beta} \right) = 0 \,\, \forall \, K \in {1,2, \cdots, \hat{N}_a}, \beta = 1,2,3 \,,
\end{eqnarray}
\end{subequations}
where $\theta_I=\left(\tilde{R}_{I_2}+ \frac{\varphi} {H} \tilde{R}_{I_3} \right)$. Note that the nanostructures with cyclic group order $\mathfrak{N} = 1$ demonstrate all the rigid body modes at $\bq = (0,0,0)$ and obey the same acoustic sum rules as discussed above.

\rev{The cyclic- and helical-symmetry-adapted expression for the dynamical matrix presented above is directly applicable to semilocal exchange-correlation functionals. Within the pseudopotential approximation, its extension to spin-polarized systems requires summation over the two spin channels for the nonlocal pseudopotential contributions, namely, the last three terms in Eq.~\ref{Eq:Complete:Dq}. As noted earlier, the all-electron formulation is obtained by replacing the ionic pseudopotential with the Coulomb potential and setting the nonlocal pseudopotential contributions to zero.}

\end{widetext}
\section{Implementation} \label{Sec:Implementation}
We have implemented the cyclic- and helical-symmetry-adapted Kohn-Sham DFT phonon framework  within the Cyclix-DFT \cite{sharma2021real} feature of the real-space code M-SPARC \cite{xu2020m, zhang2023version}, which is a \texttt{Matlab}-based variant of the large-scale electronic-structure code SPARC \cite{xu2021sparc, zhang2024sparc}. In particular, all quantities in the Kohn-Sham formalism are represented on a uniform grid in the aforementioned helical coordinate system. The differential operators are approximated using high order centered finite-differences and spatial integrations are approximated using the trapezoidal rule. The Brillouin zone is discretized at the points $(\nu, \eta_{MP} + \frac{\varphi}{H} \nu)$, where $\eta_{MP}$ is generated using the Monkhorst-Pack \cite{monkhorst1976special} scheme. The \rev{Perdew}-Zunger \cite{PhysRevB.23.5048} variant of the LDA \cite{Kohn1965}  exchange-correlation, optimized norm-conserving Vanderbilt pseudopotentials \cite{hamann2013optimized}, and Fermi-Dirac electronic smearing are employed.

The Sternheimer equations for $\psiJbeta$ are solved via a fixed-point iteration with respect to $\rhoJbeta$, using the $\beta^{\texttt{th}}$ component of the gradient of the $J^{\texttt{th}}$ atom's non-interacting electron density as the initial guess, and a restarted version \cite{pratapa2015restarted} of the periodic Pulay mixing scheme \cite{banerjee2016PeriodicPulay} to accelerate convergence. In each iteration, the linear system for $\psiJbeta$ is solved using the stabilized biconjugate gradient \cite{van1992bi} method, while the linear system for $\phiJbeta$ is solved using the alternating Anderson-Richardson (AAR) method \cite{suryanarayana2019alternating, pratapa2016anderson}, with an incomplete LU factorization of the discrete Laplacian matrix employed as the preconditioner and the solution from the previous iteration used as the initial guess. In so doing, the discrete Laplacian matrix-vector products are performed using the Kronecker product method \cite{sharma2018real}, imposing symmetry-adapted Bloch-periodic boundary conditions in the cyclic and helical directions and zero Dirichlet boundary conditions in the radial direction. Note that, due to the non-Hermitian nature of the finite-difference Laplacian in the helical coordinate system, the linear systems for both $\psiJbeta$ and $\phiJbeta$ are non-Hermitian, which necessitates the use of linear solvers capable of handling such systems. To avoid storing the memory-intensive $\psiJbeta$ for each perturbation, their contributions to the dynamical matrix are computed immediately after solving the corresponding Sternheimer equation, and the same memory space is reused to store $\psiJbeta$ for subsequent atomic perturbations. After the self-consistent iterations have converged for all atomic perturbations, the dynamical matrix at any given phonon wavevector is assembled, acoustic sum rules are applied, and it is eigendecomposed to obtain the corresponding phonon frequencies and mode shapes.

The calculations are parallelized at two levels: first, independent phonon eigenproblems corresponding to distinct phonon wavevectors are distributed as simultaneous cluster jobs; second, within each job, the computations over electron wavevectors are parallelized using \texttt{Matlab}'s \texttt{parfor} construct.


\section{Results and discussion} \label{Sec:Results}

\begin{figure}[htbp]
\subfigure[Zigzag]{\includegraphics[keepaspectratio=true,width=0.1\textwidth]{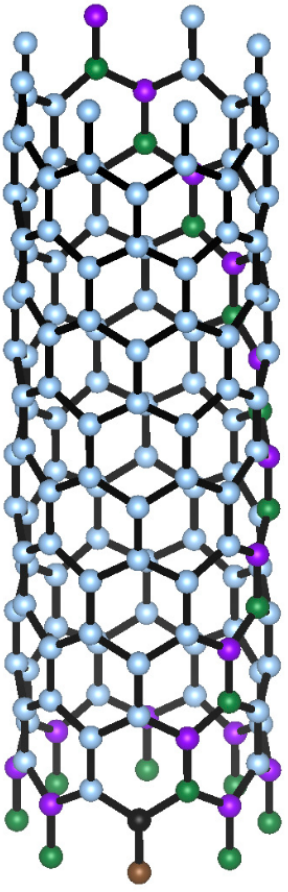}  \label{fig:zigzag}} \hspace{5mm}
\subfigure[Armchair]{\includegraphics[keepaspectratio=true,width=0.1\textwidth]{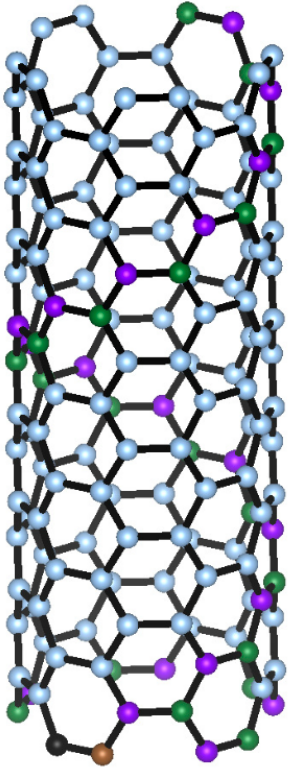}  \label{fig:armchair}} \hspace{5mm}
\subfigure[Chiral]{ \includegraphics[keepaspectratio=true,width=0.1\textwidth]{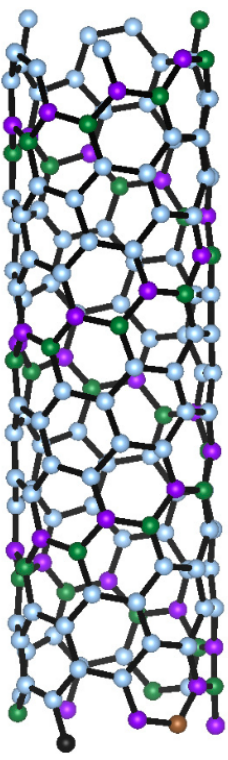} \label{fig:chiral}}
    \caption{Illustration of (a) zigzag, (b) armchair, and (c) chiral single-walled carbon nanotubes. The two fundamental carbon atoms are shown in black and brown, with their cyclic and helical images depicted in violet and green, respectively.}
    \label{Fig:CNT_chirality}
\end{figure}

We now apply the cyclic- and helical-symmetry-adapted framework to study phonons in single-walled carbon nanotubes \cite{ReviewCNT}. These structures exhibit both cyclic and helical symmetry, and can be generated by the action of the corresponding cyclic and helical symmetry groups on two fundamental carbon atoms selected from the monolayer graphene lattice. Depending on the choice of symmetry group and fundamental atoms, these nanotubes can be classified into three categories: (a) zigzag $(n,0)$, (b) armchair $(n,n)$, and (c) chiral $(n,m)$ with $n \neq m$, as illustrated in Fig.~\ref{Fig:CNT_chirality}. The $(n,m)$ nomenclature characterizes the chirality of a nanotube, with zigzag and armchair nanotubes classified as achiral and all remaining configurations classified as chiral. The unrelaxed radius of a carbon nanotube can be determined from its chirality using the relation: $r = \frac{a_0}{2\pi}\sqrt{3(n^2 + m^2 + nm)}$, 
where $a_0$ is the carbon--carbon bond length in graphene.

In all the symmetry-adapted simulations, we employ a 2-atom unit cell and a 12-order accurate finite-difference discretization with a mesh spacing of $0.11$~bohr. We discretize the Brillouin zone by including all $\nu$-points and $6\,\eta$-points per bohr of axial length. We use a $20$~bohr-thick annular region in the radial direction, which translates to $10$ bohr vacuum on either side.  These and other numerical parameters are chosen such that the phonon frequencies are converged to within $1~\mathrm{cm}^{-1}$.

\begin{figure}[htbp]
        \centering
        \includegraphics[keepaspectratio=true,width=0.36\textwidth]{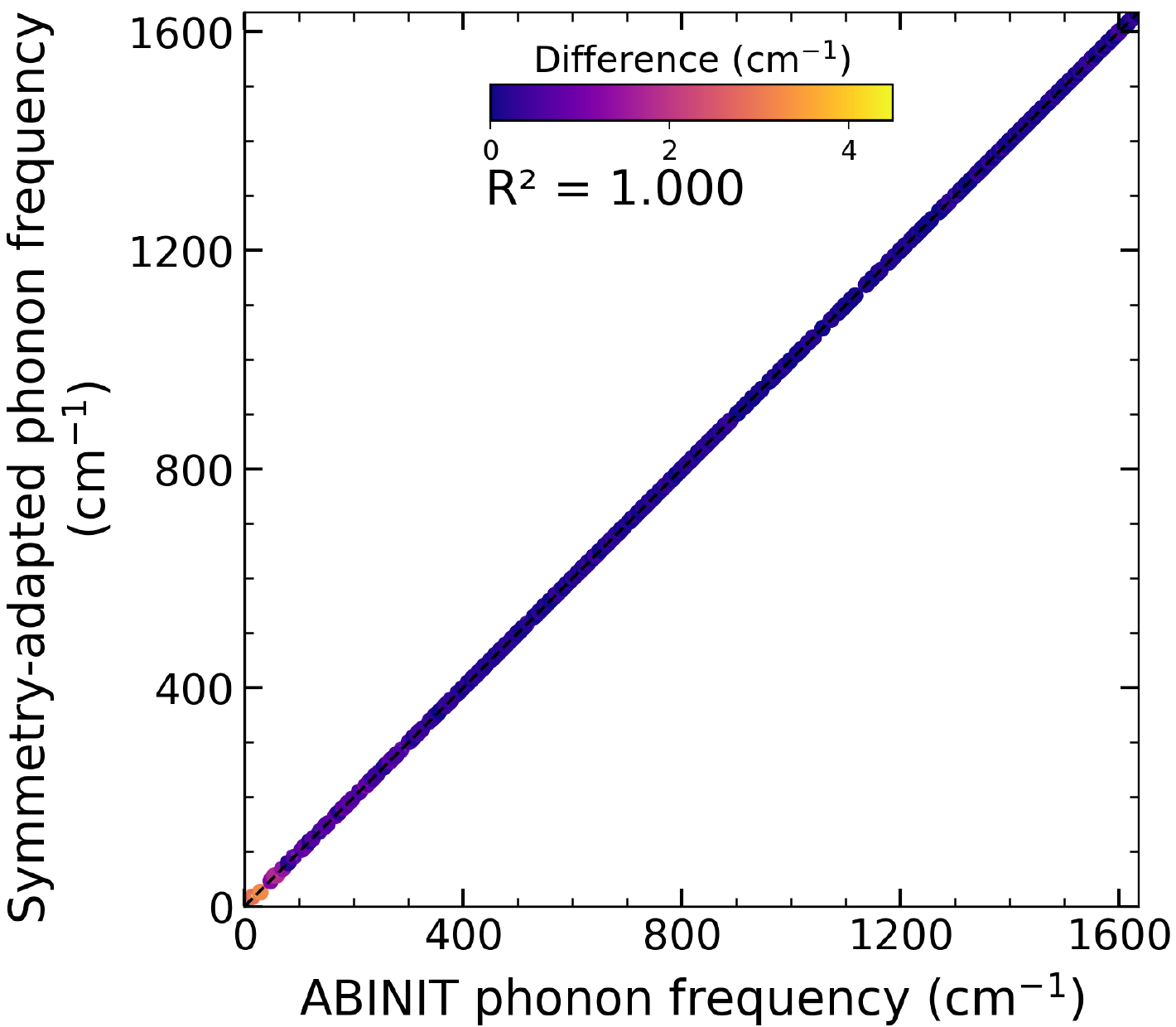}
        \caption{Comparison of the phonon frequencies computed using the cyclic- and helical-symmetry-adapted framework and the planewave code ABINIT.}
        \label{fig:abinit_cyclix_cmp}
 \end{figure}

First, we verify the accuracy of the symmetry-adapted framework by comparing the phonon spectra for a $(16,0)$ carbon nanotube with those obtained using the established plane-wave code ABINIT \cite{gonze2009abinit}. In ABINIT, we employ a 64-atom unit cell, planewave cutoff of $80$~Ha, and $12$ Brillouin-zone points along the axial direction. Indeed, since ABINIT is restricted to periodic boundary conditions, this is the smallest system that can be used to represent the $(16,0)$ carbon nanotube. As shown in Fig.~\ref{fig:abinit_cyclix_cmp}, the phonon frequencies obtained using the symmetry-adapted framework are in excellent agreement with those computed using ABINIT, with a maximum difference of approximately $4~\mathrm{cm}^{-1}$, occurring at the lowest nonzero phonon frequency. This difference can be further reduced by increasing the planewave cutoff in  ABINIT; however, doing so becomes computationally prohibitive. Indeed, phonon calculations are associated with a large computational cost, scaling quartically with system size, which renders simulations employing periodic boundary conditions intractable for chiral nanotubes and large-diameter nanotubes due to the substantially increased number of atoms in their periodic unit cells. In contrast, the unit cell in the symmetry-adapted calculations contains only the fundamental atoms, two in the case of carbon nanotubes, independent of nanotube diameter and chirality, thereby enabling accurate and efficient phonon calculations for such systems. The advantages of the symmetry-adapted framework become particularly pronounced for mechanical deformations such as torsion, where the number of atoms in the periodic unit cell increases dramatically, while remaining unchanged within the symmetry-adapted framework.

\begin{figure}[htbp]
        \centering
        \includegraphics[keepaspectratio=true,width=0.4\textwidth]{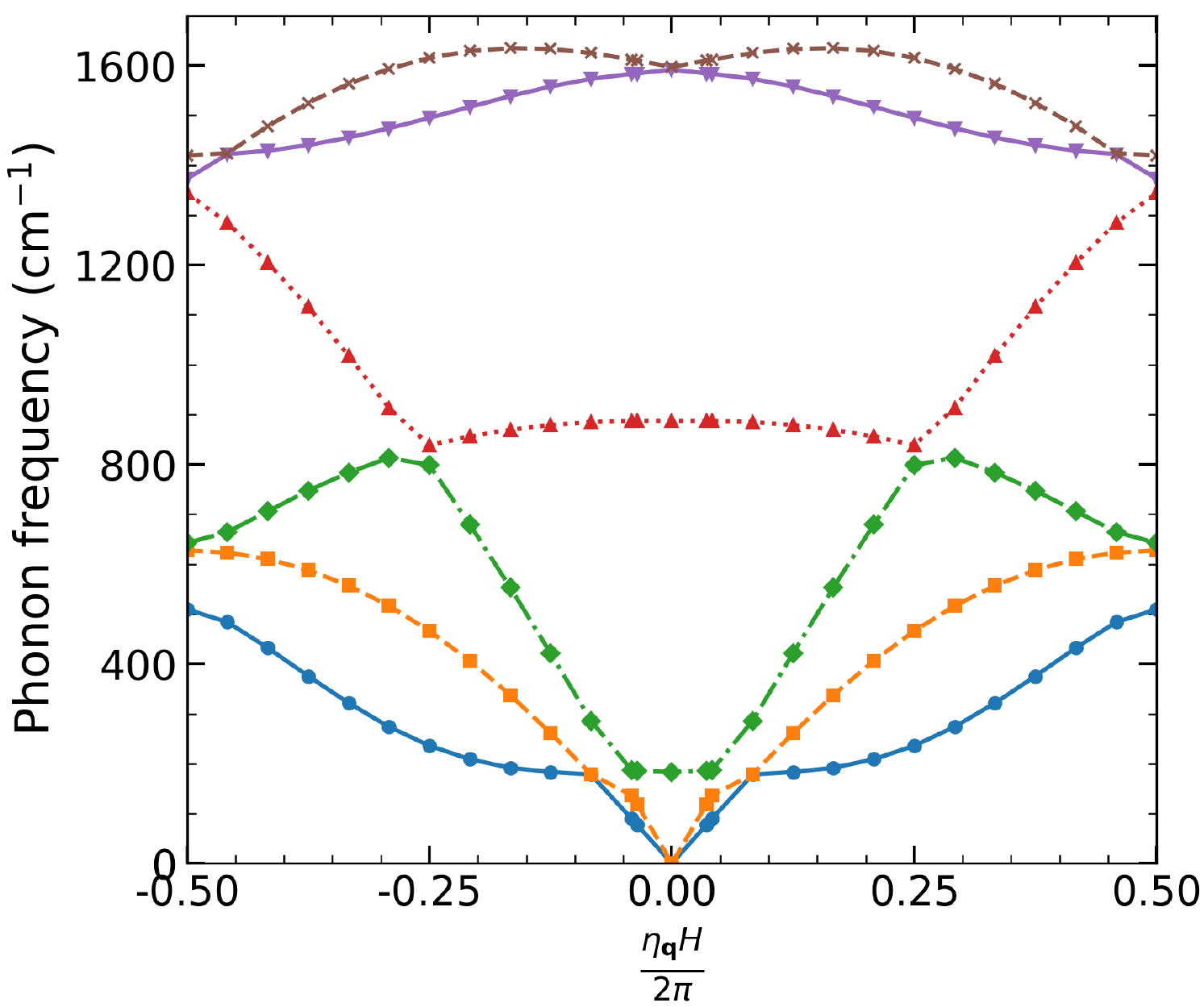}
        \caption{Cyclic- and helical-symmetry-adapted phonon band structure for the (16,0) carbon nanotube at $\nu_\mathbf{q} = 0$.}
        \label{fig:bandstructure_nu0}
 \end{figure}

The symmetry-adapted formulation offers advantages that go beyond accuracy and computational efficiency, providing a physically meaningful representation of vibrational properties. By exploiting the inherent cyclic and helical symmetries of the system, the phonon band structure can be resolved distinctly across different $\nu_{\bq}$ points. Such a representation simplifies the visualization and provides clearer physical insight into the vibrational modes, as illustrated in Fig.~\ref{fig:bandstructure_nu0}, which shows the phonon band structure at $\nu_{\bq}=0$ for the $(16,0)$ carbon nanotube. The bands exhibit reflection symmetry about the $\Gamma$-point across all phonon branches, consistent with time-reversal symmetry in the absence of an external magnetic field. In addition, the lowest two phonon branches display linear dispersion near the $\Gamma$-point. The slopes of these branches can be used to compute the Young’s and shear moduli using the relation $C=\rho_m H^2 v^2$, where $C$ denotes the modulus, $\rho_m$ is the mass density of the nanotube, and $v$ is the slope of the corresponding linear dispersion at $\bq=\mathbf{0}$. Using $\rho_m=2.26~\mathrm{g/cc}$ and the equilibrium value $H=4.001$~bohr, we obtain Young’s and shear moduli of $1.00$~TPa and $0.43$~TPa, respectively, in very good agreement with previous DFT results \cite{Zhang2009CNT, sharma2021real} and experimental measurements \cite{Hall2006CNT, treacy1996exceptionally}.

\begin{figure}[htbp]
\subfigure{\includegraphics[keepaspectratio=true,width=0.15\textwidth]{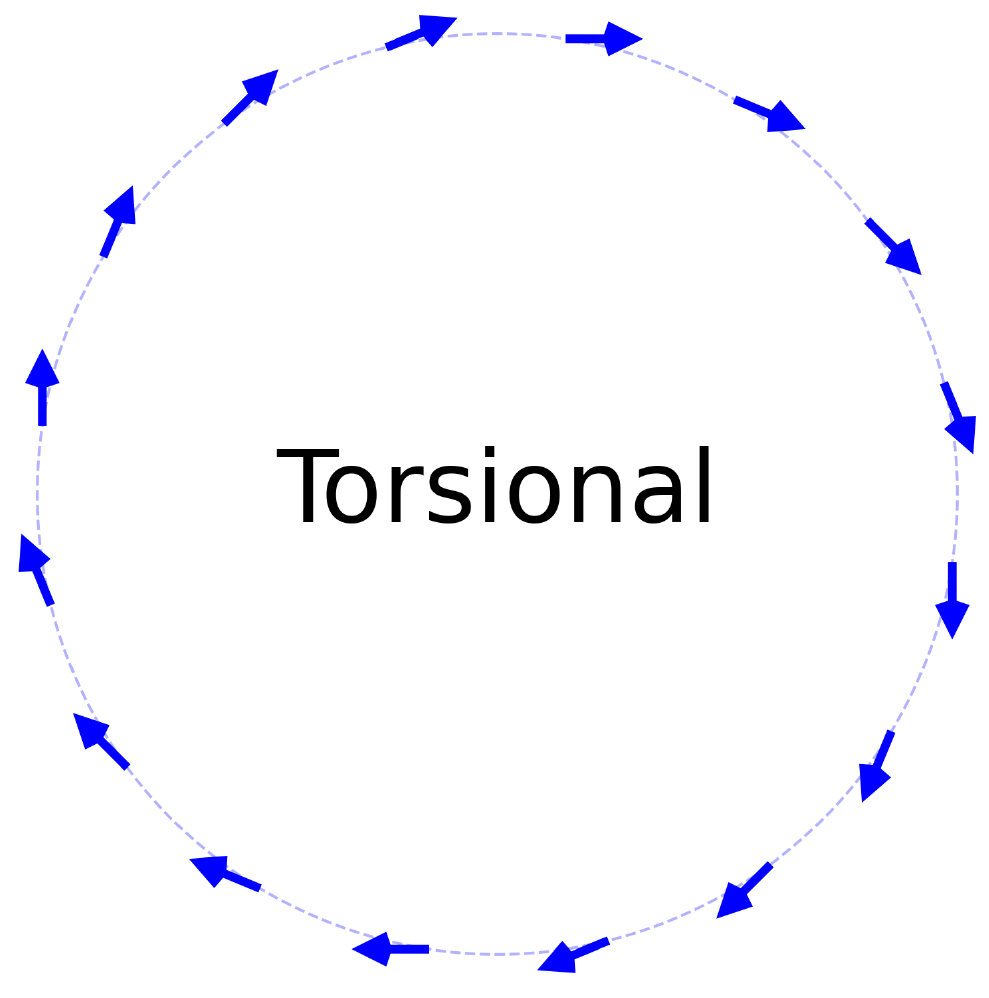}} 
\subfigure{\includegraphics[keepaspectratio=true,width=0.15\textwidth]{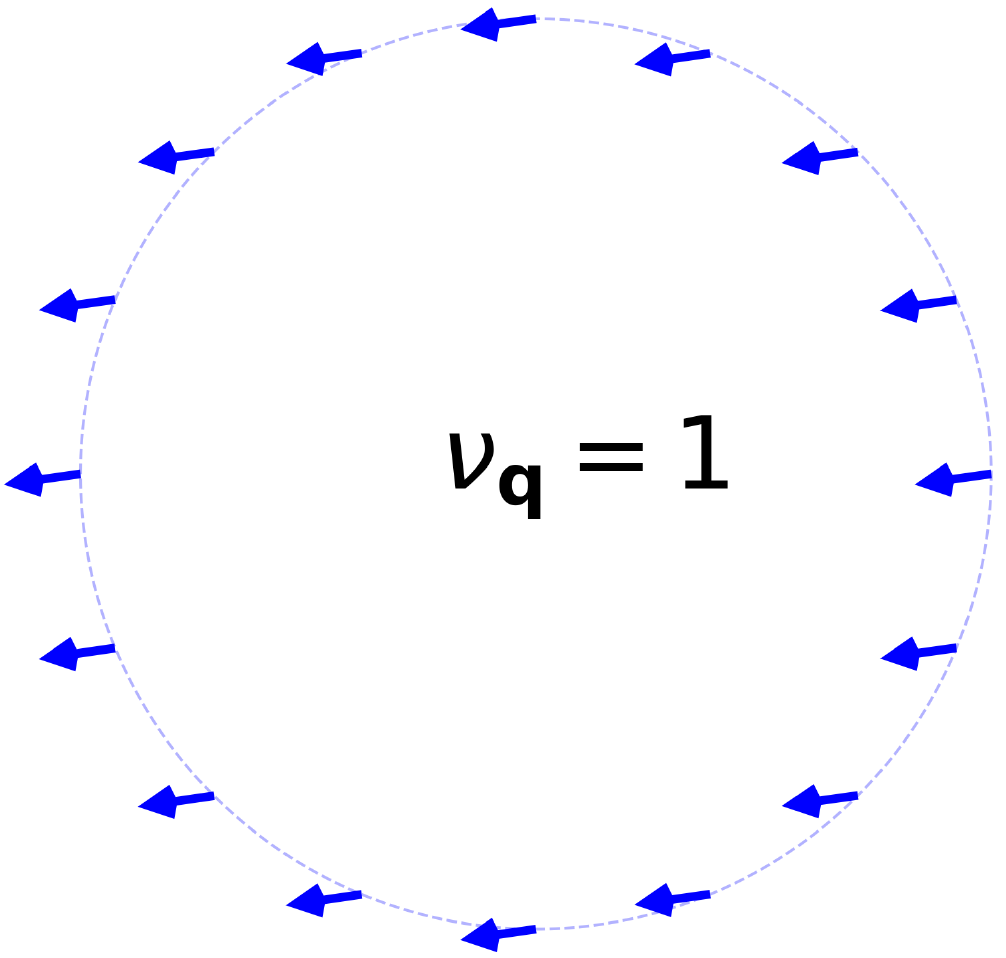}} 
\subfigure{\includegraphics[keepaspectratio=true,width=0.11\textwidth]{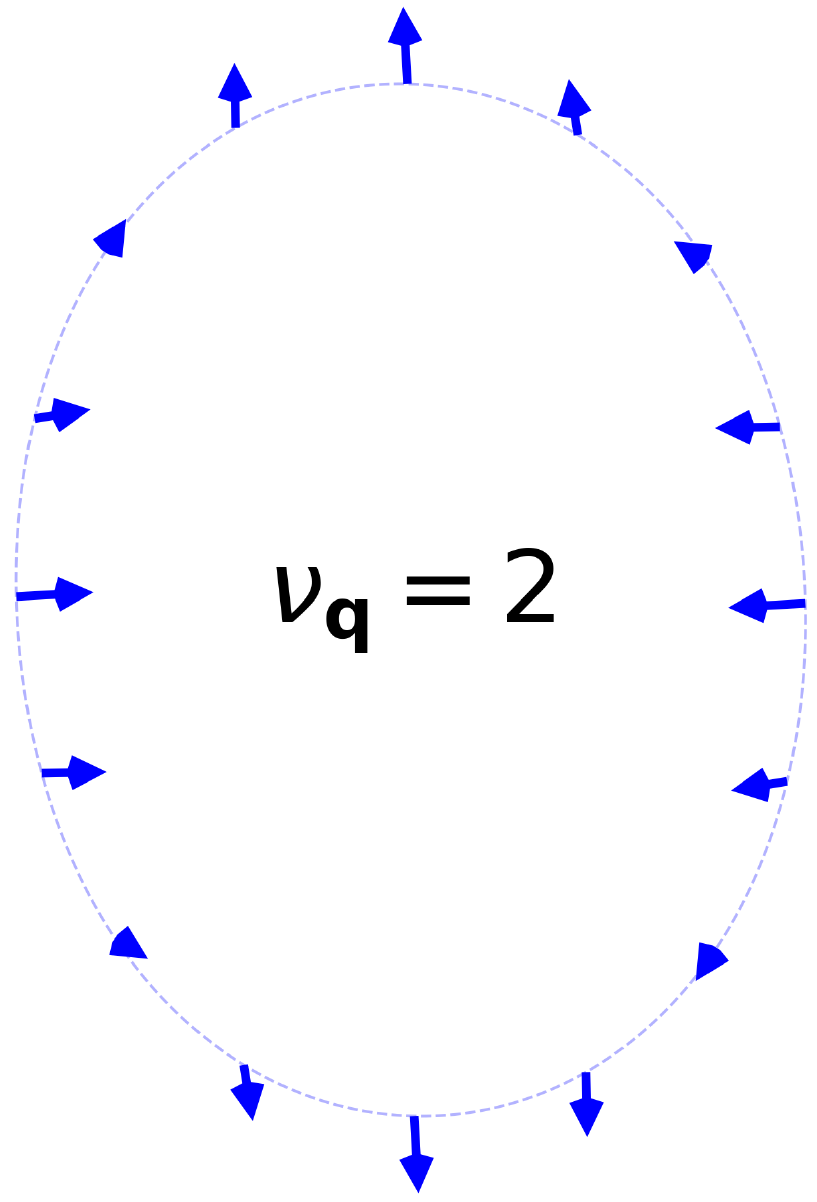}}
\subfigure{\includegraphics[keepaspectratio=true,width=0.15\textwidth]{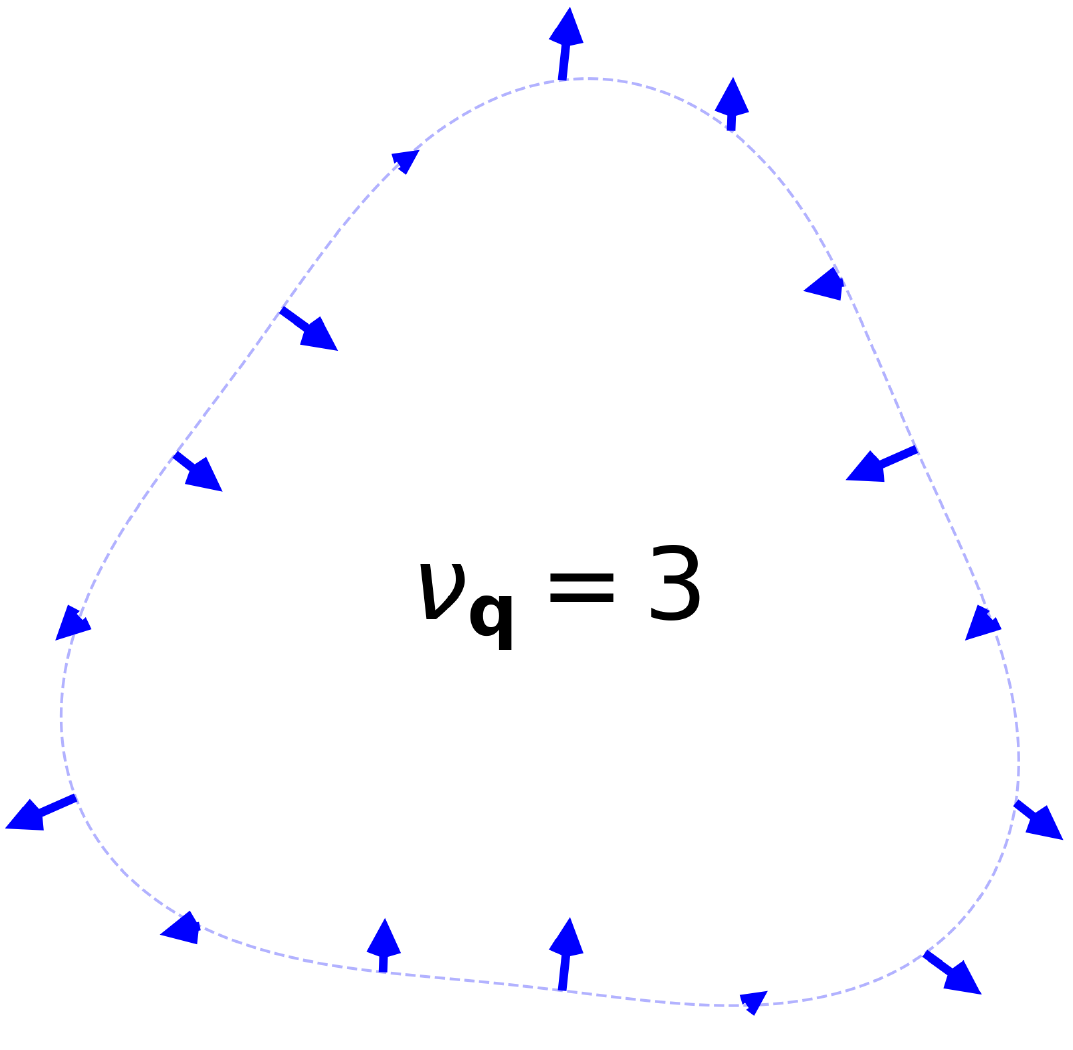}}
\subfigure{\includegraphics[keepaspectratio=true,width=0.15\textwidth]{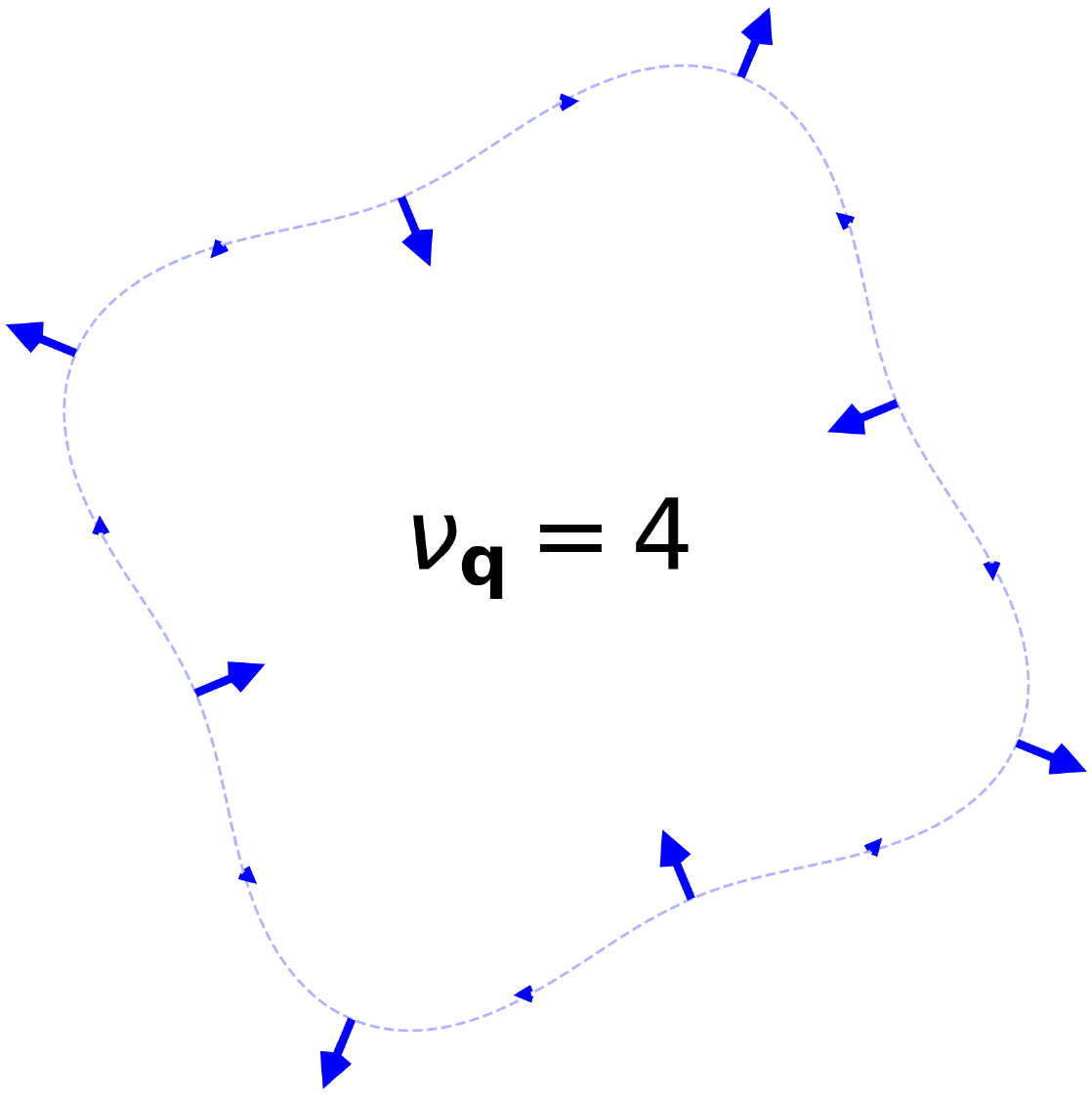}}
\subfigure{\includegraphics[keepaspectratio=true,width=0.15\textwidth]{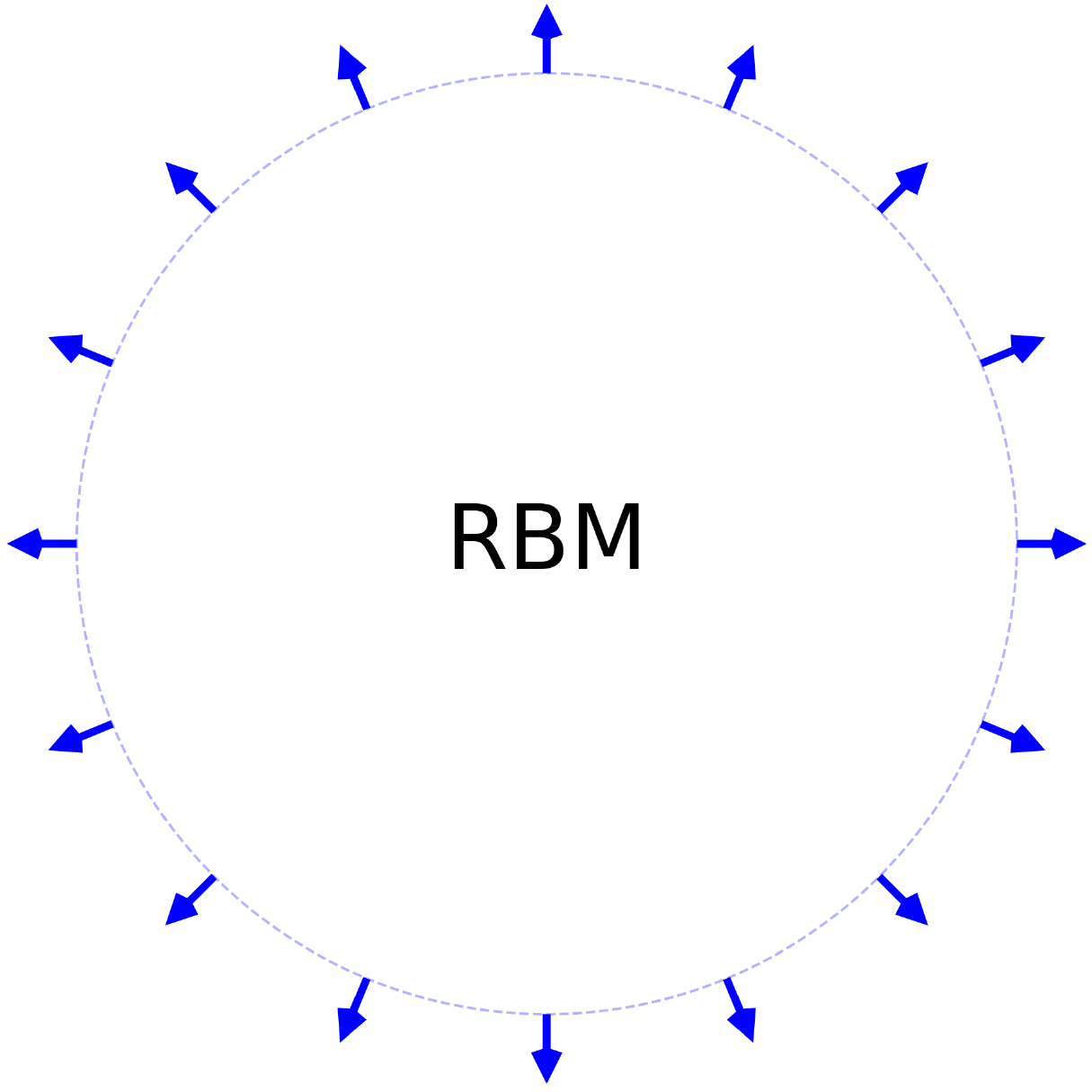}}    
    \caption{Atomic displacements corresponding to the phonon modes of interest that are common to the carbon nanotubes studied, illustrated using the $(16,0)$ nanotube as a representative example. Arrows indicate the directions of atomic perturbations, and their lengths denote the corresponding magnitudes. The torsional mode corresponds to a zero-frequency rigid-body motion. The four ring modes are characterized by the $\nu$-fold symmetry inherent in their vibrational patterns. RBM denotes the radial breathing mode.}
    \label{Fig:modes}
\end{figure}

Next, we study the phonons in carbon nanotubes with varying diameters and chiralities. In particular, we consider five representative nanotubes: zigzag type I $(16,0)$, zigzag type II $(20,0)$, zigzag type III $(18,0)$, armchair $(8,8)$, and chiral $(10,5)$. To visualize the phonon modes, we determine the atomic perturbations corresponding to a given phonon wavevector using the relation:
\begin{align}
\Delta \mathbf{R}_{I'}^{\bq} = \Re \!\left[e^{i \bq \cdot (\tilde{\mathbf{R}}_{I'}-\tilde{\mathbf{R}}_{I})}\, \mathbf{Q}_{\zeta_{I'}\theta}\, \mathbf{Q}_{\mu_{I'}\varphi}\, \frac{[\mathbf{v}_{\bq}]_{I}}{\sqrt{M_I}} \right] \,,
\end{align}
where $[\mathbf{v}_{\bq}]_{I}$ represents the components of the phonon mode that are associated with the $I^{th}$ fundamental atom. The results are summarized in Fig.~\ref{Fig:modes}, which shows the atomic displacements associated with the modes of interest that are found common to the carbon nanotubes studied.

We observe four zero-frequency rigid-body phonon modes, as predicted by the formulation. In particular, the longitudinal translational and torsional modes occur at the wavevector the $\Gamma$-point, with the torsional mode shown in Fig.~\ref{Fig:modes}, while the two transverse translational (flexural) modes occur at the wavevectors $(0,1,\varphi/H)$ and $(0,-1,-\varphi/H)$, with one of these illustrated as the $\nu_{\bq}=1$ mode in Fig.~\ref{Fig:modes}. We also observe the presence of ring modes, in which the atomic vibrations form ring-like patterns, as shown in Fig.~\ref{Fig:modes} for $\nu_{\bq}=1,2,3,4$. These modes are characterized by their underlying symmetry; for example, the $\nu_{\bq}=4$ ring mode exhibits four lobes, reflecting the fourfold rotational symmetry. The $\nu_{\bq}$ ring mode occurs at the wavevector $(0,\nu_{\bq},\nu_{\bq}\varphi/H)$ on the lowest-frequency phonon branch. We also observe the presence of a radial breathing mode, in which the atomic vibrations occur in the radial direction, leading to a uniform expansion and contraction of the nanotube. This mode occurs at the $\Gamma$-point and corresponds to the third-lowest phonon branch.

\begin{figure}[htbp]
        \centering
        \includegraphics[keepaspectratio=true,width=0.4\textwidth]{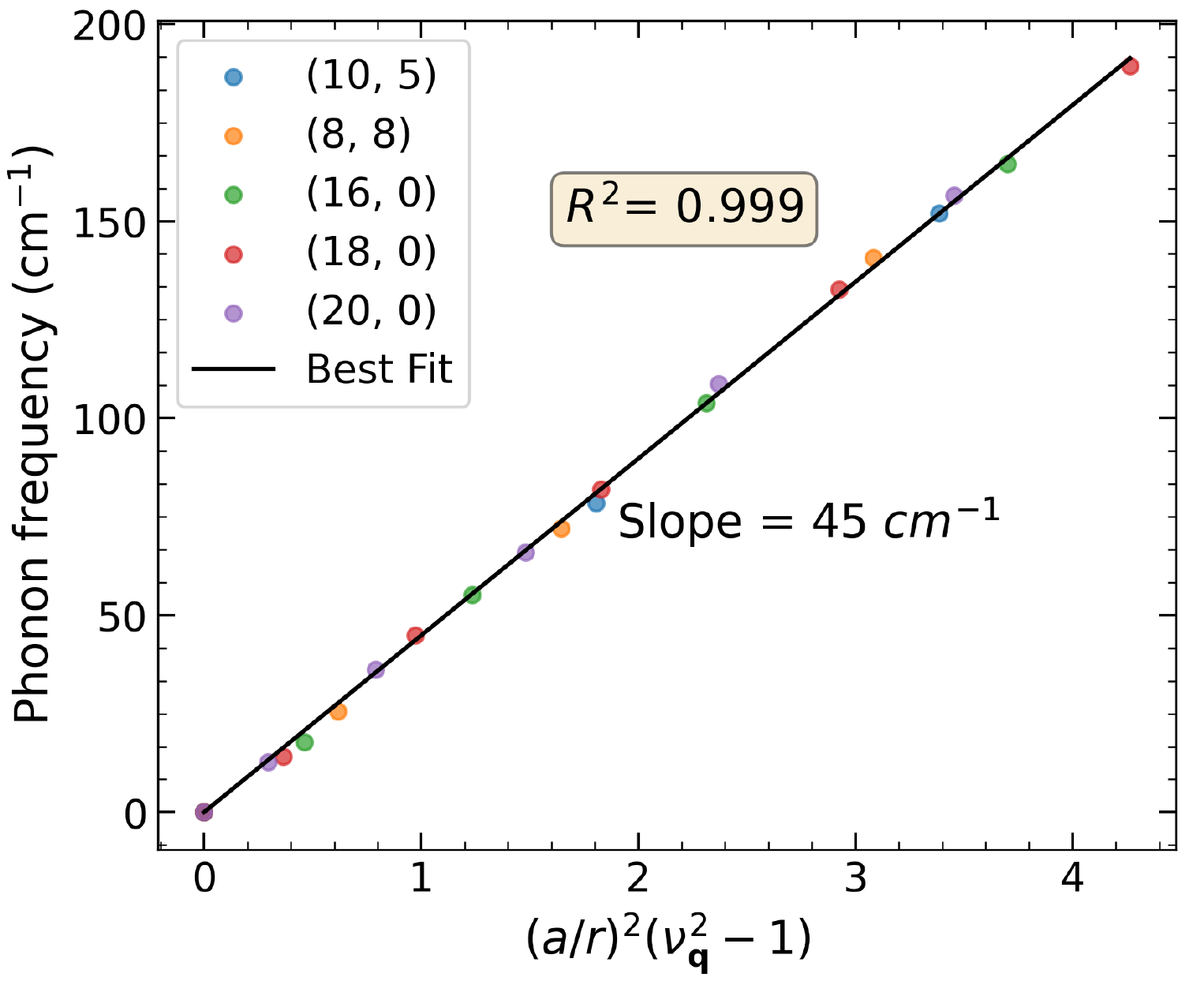}
        \caption{Variation of the ring mode phonon frequency in carbon nanotubes as a function of the radius $r$ and $\nu_{\bq}$.}
        \label{fig:ring_fit}
    \end{figure}

We find from the results that the phonon frequencies of the ring modes primarily depend on the nanotube radius and the ring order of the mode. As shown in Fig.~\ref{fig:ring_fit}, this dependence can be described by the relation:
\begin{equation}
\omega_{\text{RM}}(r,\nu_\bq) \approx 45 \left(\frac{a}{r} \right)^2 (\nu_\bq^2-1) \, \text{cm}^{-1} \,,
\end{equation}
where $a = \sqrt{3}\,a_0$ is the length of graphene's lattice vector and $r$ is the nanotube radius. This scaling law is in very good qualitative agreement with previous studies \cite{sharma2025cyclic, gunlycke2008lattice}. The prefactor of $45~\mathrm{cm}^{-1}$ obtained here is is very good agreement with the value of $46~\mathrm{cm}^{-1}$ reported recently using a cyclic- and helical-symmetry-adapted machine-learned force field \cite{sharma2025cyclic}, but differs noticeably from the value of $54~\mathrm{cm}^{-1}$ obtained earlier using an interatomic potential \cite{gunlycke2008lattice}.

\begin{figure}[htbp]
        \centering
        \includegraphics[keepaspectratio=true,width=0.4\textwidth]{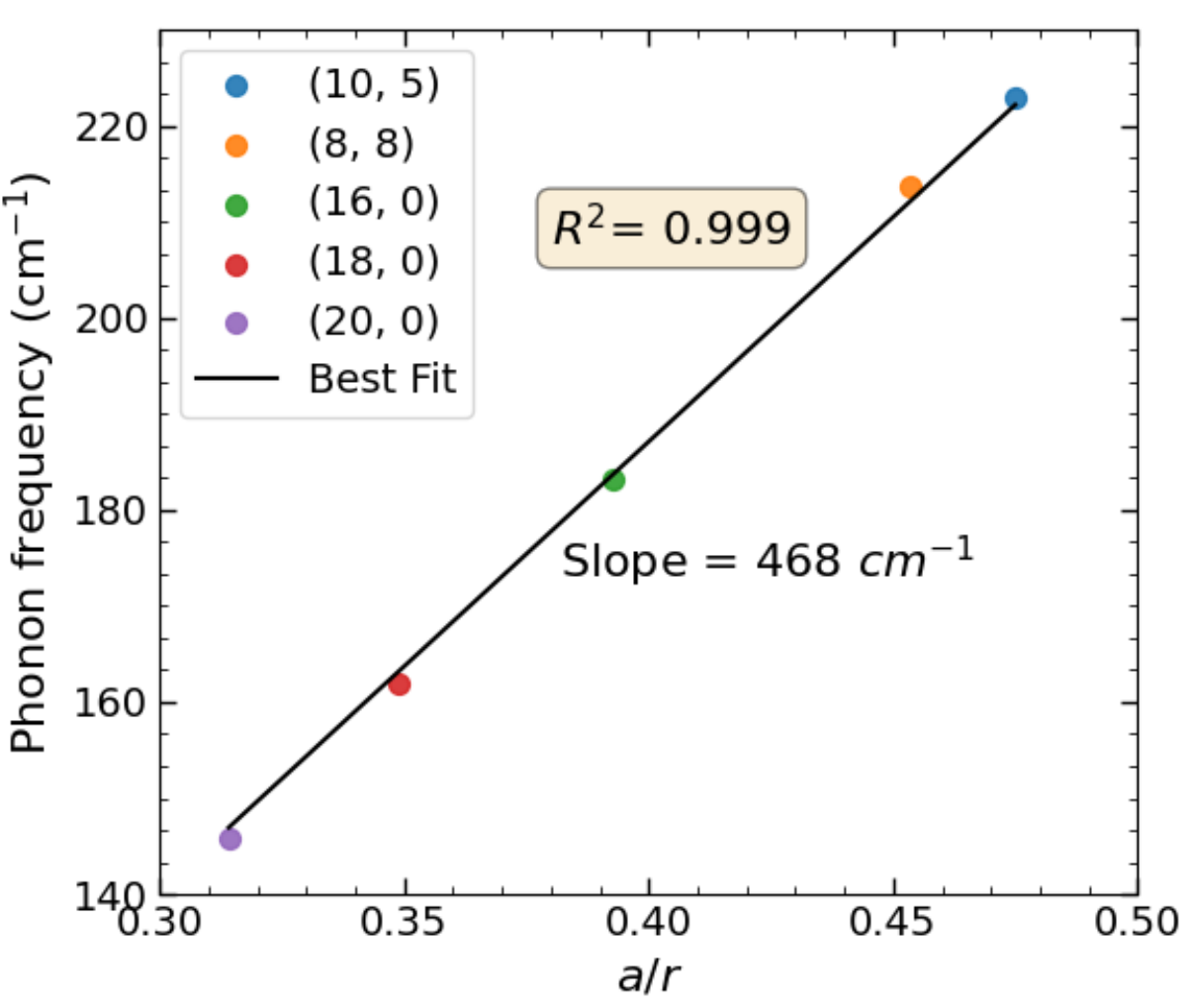}
        \caption{Variation of the radial breathing mode phonon frequency in carbon nanotubes as a function of the radius $r$.}
        \label{fig:RBM_fit}
\end{figure}

Finally, we find from the results that the phonon frequencies of the ring modes primarily depend on the nanotube radius. As shown in Fig.~\ref{fig:RBM_fit}, this dependence can be described by the relation:
\begin{equation}
\omega_{\text{RBM}} \approx 468 \frac{a}{r} \, \text{cm}^{-1} \,.
\end{equation}
This scaling law is also in very good qualitative agreement with the literature \cite{sharma2025cyclic, gunlycke2008lattice}. However, the constant of $468$ cm$^{-1}$ obtained here is noticeably different from the values of $480$ cm$^{-1}$ and $486$ cm$^{-1}$ obtained previously using machine-learned force fields (MLFFs) \cite{sharma2025cyclic} and interatomic potential \cite{gunlycke2008lattice}, respectively.

\section{Concluding remarks} \label{Sec:Conclusions}
In this work, we developed a first-principles framework for the calculation of phonons in nanostructures with cyclic and/or helical symmetry. In particular, we derived cyclic- and helical-symmetry-adapted representations for the dynamical matrix at arbitrary phonon wavevectors within a variationally formulated, symmetry-adapted DFPT framework. In addition, we derived the acoustic sum rules for cylindrical geometries, which include a rigid-body rotational mode in addition to the three translational modes. The resulting formalism was implemented within a high-order finite-difference discretization. Using carbon nanotubes as representative systems, we validated the accuracy of the framework through excellent agreement with periodic plane-wave results. We further applied the framework  to compute the Young’s and shear moduli of carbon nanotubes, as well as the scaling laws governing the dependence of ring and radial breathing mode phonon frequencies on nanotube diameter. The elastic moduli were found to be in agreement with previous DFT and experimental results, while the phonon scaling laws showed qualitative agreement with previous atomistic simulations. \rev{Overall, the symmetry-adapted formalism enables a substantial reduction in computational cost for systems with cyclic and/or helical symmetry. For example, in the case of a (16,0) carbon nanotube, the symmetry-adapted formulation requires only 2 atoms in the fundamental domain, compared to 64 atoms under standard periodic boundary conditions. These advantages become even more pronounced for chiral nanotubes, with increasing diameter, and under applied twist, where the reduction in the number of atoms and the associated computational effort is significantly greater.}

The implementation of the cyclic- and helical-symmetry-adapted phonon formulation  within the large-scale electronic structure code SPARC is expected to significantly reduce computational wall time and enable the study of larger systems, making it a promising direction for future development. Beyond methodological advances, the present framework enables systematic investigations of the effects of mechanical deformations, such as bending and twisting, on phonons and related properties of low-dimensional nanostructures. In addition, it provides a natural foundation for the study of electron--phonon interactions in these systems, thereby opening several compelling avenues for future research.

\section*{ACKNOWLEDGMENTS}
This work draws from the the thesis of A.S. at Georgia Institute of Technology. A.S. gratefully acknowledges the ANRF early career research grant (Grant No. ANRF/ECRG/2024/002362) from Department of Science and Technology India. A.S. and P.S. gratefully acknowledge support from the U.S. National Science Foundation through CAREER Grant No. 155321, during the time A.S. was at Georgia Institute of Technology. This research was also supported by the supercomputing infrastructure provided by Partnership for an Advanced Computing Environment (PACE) through its Phoenix cluster at Georgia Institute of Technology, Atlanta, Georgia.

\section*{Data availability statements}
The data that support the findings of this article are openly available \cite{cyclixphonondataset}.


\begin{thebibliography}{91}%
\makeatletter
\providecommand \@ifxundefined [1]{%
 \@ifx{#1\undefined}
}%
\providecommand \@ifnum [1]{%
 \ifnum #1\expandafter \@firstoftwo
 \else \expandafter \@secondoftwo
 \fi
}%
\providecommand \@ifx [1]{%
 \ifx #1\expandafter \@firstoftwo
 \else \expandafter \@secondoftwo
 \fi
}%
\providecommand \natexlab [1]{#1}%
\providecommand \enquote  [1]{``#1''}%
\providecommand \bibnamefont  [1]{#1}%
\providecommand \bibfnamefont [1]{#1}%
\providecommand \citenamefont [1]{#1}%
\providecommand \href@noop [0]{\@secondoftwo}%
\providecommand \href [0]{\begingroup \@sanitize@url \@href}%
\providecommand \@href[1]{\@@startlink{#1}\@@href}%
\providecommand \@@href[1]{\endgroup#1\@@endlink}%
\providecommand \@sanitize@url [0]{\catcode `\\12\catcode `\$12\catcode
  `\&12\catcode `\#12\catcode `\^12\catcode `\_12\catcode `\%12\relax}%
\providecommand \@@startlink[1]{}%
\providecommand \@@endlink[0]{}%
\providecommand \url  [0]{\begingroup\@sanitize@url \@url }%
\providecommand \@url [1]{\endgroup\@href {#1}{\urlprefix }}%
\providecommand \urlprefix  [0]{URL }%
\providecommand \Eprint [0]{\href }%
\providecommand \doibase [0]{http://dx.doi.org/}%
\providecommand \selectlanguage [0]{\@gobble}%
\providecommand \bibinfo  [0]{\@secondoftwo}%
\providecommand \bibfield  [0]{\@secondoftwo}%
\providecommand \translation [1]{[#1]}%
\providecommand \BibitemOpen [0]{}%
\providecommand \bibitemStop [0]{}%
\providecommand \bibitemNoStop [0]{.\EOS\space}%
\providecommand \EOS [0]{\spacefactor3000\relax}%
\providecommand \BibitemShut  [1]{\csname bibitem#1\endcsname}%
\let\auto@bib@innerbib\@empty
\bibitem [{\citenamefont {Hohenberg}\ and\ \citenamefont
  {Kohn}(1964)}]{Hohenberg}%
  \BibitemOpen
  \bibfield  {author} {\bibinfo {author} {\bibfnamefont {P.}~\bibnamefont
  {Hohenberg}}\ and\ \bibinfo {author} {\bibfnamefont {W.}~\bibnamefont
  {Kohn}},\ }\bibfield  {title} {\enquote {\bibinfo {title} {Inhomogeneous
  electron gas},}\ }\href@noop {} {\bibfield  {journal} {\bibinfo  {journal}
  {Phys. Rev.}\ }\textbf {\bibinfo {volume} {136}},\ \bibinfo {pages}
  {B864--B871} (\bibinfo {year} {1964})}\BibitemShut {NoStop}%
\bibitem [{\citenamefont {Kohn}\ and\ \citenamefont {Sham}(1965)}]{Kohn1965}%
  \BibitemOpen
  \bibfield  {author} {\bibinfo {author} {\bibfnamefont {W.}~\bibnamefont
  {Kohn}}\ and\ \bibinfo {author} {\bibfnamefont {L.~J.}\ \bibnamefont
  {Sham}},\ }\bibfield  {title} {\enquote {\bibinfo {title} {Self-consistent
  equations including exchange and correlation effects},}\ }\href@noop {}
  {\bibfield  {journal} {\bibinfo  {journal} {Phys. Rev.}\ }\textbf {\bibinfo
  {volume} {140}},\ \bibinfo {pages} {A1133--A1138} (\bibinfo {year}
  {1965})}\BibitemShut {NoStop}%
\bibitem [{\citenamefont {Gavini}\ \emph {et~al.}(2023)\citenamefont {Gavini},
  \citenamefont {Baroni}, \citenamefont {Blum}, \citenamefont {Bowler},
  \citenamefont {Buccheri}, \citenamefont {Chelikowsky}, \citenamefont {Das},
  \citenamefont {Dawson}, \citenamefont {Delugas}, \citenamefont {Dogan},
  \citenamefont {Draxl}, \citenamefont {Galli}, \citenamefont {Genovese},
  \citenamefont {Giannozzi}, \citenamefont {Giantomassi}, \citenamefont
  {Gonze}, \citenamefont {Govoni}, \citenamefont {Gygi}, \citenamefont
  {Gulans}, \citenamefont {Herbert}, \citenamefont {Kokott}, \citenamefont
  {Kühne}, \citenamefont {Liou}, \citenamefont {Miyazaki}, \citenamefont
  {Motamarri}, \citenamefont {Nakata}, \citenamefont {Pask}, \citenamefont
  {Plessl}, \citenamefont {Ratcliff}, \citenamefont {Richard}, \citenamefont
  {Rossi}, \citenamefont {Schade}, \citenamefont {Scheffler}, \citenamefont
  {Schütt}, \citenamefont {Suryanarayana}, \citenamefont {Torrent},
  \citenamefont {Truflandier}, \citenamefont {Windus}, \citenamefont {Xu},
  \citenamefont {Yu},\ and\ \citenamefont {Perez}}]{gavini2022roadmap}%
  \BibitemOpen
  \bibfield  {author} {\bibinfo {author} {\bibfnamefont {V.}~\bibnamefont
  {Gavini}}, \bibinfo {author} {\bibfnamefont {S.}~\bibnamefont {Baroni}},
  \bibinfo {author} {\bibfnamefont {V.}~\bibnamefont {Blum}}, \bibinfo {author}
  {\bibfnamefont {D.~R.}\ \bibnamefont {Bowler}}, \bibinfo {author}
  {\bibfnamefont {A.}~\bibnamefont {Buccheri}}, \bibinfo {author}
  {\bibfnamefont {J.~R.}\ \bibnamefont {Chelikowsky}}, \bibinfo {author}
  {\bibfnamefont {S.}~\bibnamefont {Das}}, \bibinfo {author} {\bibfnamefont
  {W.}~\bibnamefont {Dawson}}, \bibinfo {author} {\bibfnamefont
  {P.}~\bibnamefont {Delugas}}, \bibinfo {author} {\bibfnamefont
  {M.}~\bibnamefont {Dogan}}, \bibinfo {author} {\bibfnamefont
  {C.}~\bibnamefont {Draxl}}, \bibinfo {author} {\bibfnamefont
  {G.}~\bibnamefont {Galli}}, \bibinfo {author} {\bibfnamefont
  {L.}~\bibnamefont {Genovese}}, \bibinfo {author} {\bibfnamefont
  {P.}~\bibnamefont {Giannozzi}}, \bibinfo {author} {\bibfnamefont
  {M.}~\bibnamefont {Giantomassi}}, \bibinfo {author} {\bibfnamefont
  {X.}~\bibnamefont {Gonze}}, \bibinfo {author} {\bibfnamefont
  {M.}~\bibnamefont {Govoni}}, \bibinfo {author} {\bibfnamefont
  {F.}~\bibnamefont {Gygi}}, \bibinfo {author} {\bibfnamefont {A.}~\bibnamefont
  {Gulans}}, \bibinfo {author} {\bibfnamefont {J.~M.}\ \bibnamefont {Herbert}},
  \bibinfo {author} {\bibfnamefont {S.}~\bibnamefont {Kokott}}, \bibinfo
  {author} {\bibfnamefont {T.~D.}\ \bibnamefont {Kühne}}, \bibinfo {author}
  {\bibfnamefont {K.-H.}\ \bibnamefont {Liou}}, \bibinfo {author}
  {\bibfnamefont {T.}~\bibnamefont {Miyazaki}}, \bibinfo {author}
  {\bibfnamefont {P.}~\bibnamefont {Motamarri}}, \bibinfo {author}
  {\bibfnamefont {A.}~\bibnamefont {Nakata}}, \bibinfo {author} {\bibfnamefont
  {J.~E.}\ \bibnamefont {Pask}}, \bibinfo {author} {\bibfnamefont
  {C.}~\bibnamefont {Plessl}}, \bibinfo {author} {\bibfnamefont {L.~E.}\
  \bibnamefont {Ratcliff}}, \bibinfo {author} {\bibfnamefont {R.~M.}\
  \bibnamefont {Richard}}, \bibinfo {author} {\bibfnamefont {M.}~\bibnamefont
  {Rossi}}, \bibinfo {author} {\bibfnamefont {R.}~\bibnamefont {Schade}},
  \bibinfo {author} {\bibfnamefont {M.}~\bibnamefont {Scheffler}}, \bibinfo
  {author} {\bibfnamefont {O.}~\bibnamefont {Schütt}}, \bibinfo {author}
  {\bibfnamefont {P.}~\bibnamefont {Suryanarayana}}, \bibinfo {author}
  {\bibfnamefont {M.}~\bibnamefont {Torrent}}, \bibinfo {author} {\bibfnamefont
  {L.}~\bibnamefont {Truflandier}}, \bibinfo {author} {\bibfnamefont {T.~L.}\
  \bibnamefont {Windus}}, \bibinfo {author} {\bibfnamefont {Q.}~\bibnamefont
  {Xu}}, \bibinfo {author} {\bibfnamefont {V.~W.-Z.}\ \bibnamefont {Yu}}, \
  and\ \bibinfo {author} {\bibfnamefont {D.}~\bibnamefont {Perez}},\ }\bibfield
   {title} {\enquote {\bibinfo {title} {Roadmap on electronic structure codes
  in the exascale era},}\ }\href@noop {} {\bibfield  {journal} {\bibinfo
  {journal} {Modelling and Simulation in Materials Science and Engineering}\
  }\textbf {\bibinfo {volume} {31}},\ \bibinfo {pages} {063301} (\bibinfo
  {year} {2023})}\BibitemShut {NoStop}%
\bibitem [{\citenamefont {Xia}\ \emph {et~al.}(2003)\citenamefont {Xia},
  \citenamefont {Yang}, \citenamefont {Sun}, \citenamefont {Wu}, \citenamefont
  {Mayers}, \citenamefont {Gates}, \citenamefont {Yin}, \citenamefont {Kim},\
  and\ \citenamefont {Yan}}]{xia2003one}%
  \BibitemOpen
  \bibfield  {author} {\bibinfo {author} {\bibfnamefont {Y.}~\bibnamefont
  {Xia}}, \bibinfo {author} {\bibfnamefont {P.}~\bibnamefont {Yang}}, \bibinfo
  {author} {\bibfnamefont {Y.}~\bibnamefont {Sun}}, \bibinfo {author}
  {\bibfnamefont {Y.}~\bibnamefont {Wu}}, \bibinfo {author} {\bibfnamefont
  {B.}~\bibnamefont {Mayers}}, \bibinfo {author} {\bibfnamefont
  {B.}~\bibnamefont {Gates}}, \bibinfo {author} {\bibfnamefont
  {Y.}~\bibnamefont {Yin}}, \bibinfo {author} {\bibfnamefont {F.}~\bibnamefont
  {Kim}}, \ and\ \bibinfo {author} {\bibfnamefont {H.}~\bibnamefont {Yan}},\
  }\bibfield  {title} {\enquote {\bibinfo {title} {One-dimensional
  nanostructures: synthesis, characterization, and applications},}\ }\href@noop
  {} {\bibfield  {journal} {\bibinfo  {journal} {Adv. Mater.}\ }\textbf
  {\bibinfo {volume} {15}},\ \bibinfo {pages} {353--389} (\bibinfo {year}
  {2003})}\BibitemShut {NoStop}%
\bibitem [{\citenamefont {James}(2006)}]{james2006objective}%
  \BibitemOpen
  \bibfield  {author} {\bibinfo {author} {\bibfnamefont {R.~D.}\ \bibnamefont
  {James}},\ }\bibfield  {title} {\enquote {\bibinfo {title} {Objective
  structures},}\ }\href@noop {} {\bibfield  {journal} {\bibinfo  {journal} {J.
  Mech. Phys. Solids}\ }\textbf {\bibinfo {volume} {54}},\ \bibinfo {pages}
  {2354--2390} (\bibinfo {year} {2006})}\BibitemShut {NoStop}%
\bibitem [{\citenamefont
  {Allen}(2007{\natexlab{a}})}]{allen2007nanocrystalline1}%
  \BibitemOpen
  \bibfield  {author} {\bibinfo {author} {\bibfnamefont {P.~B.}\ \bibnamefont
  {Allen}},\ }\bibfield  {title} {\enquote {\bibinfo {title} {Nanocrystalline
  nanowires: {I}. structure},}\ }\href@noop {} {\bibfield  {journal} {\bibinfo
  {journal} {Nano Lett.}\ }\textbf {\bibinfo {volume} {7}},\ \bibinfo {pages}
  {6--10} (\bibinfo {year} {2007}{\natexlab{a}})}\BibitemShut {NoStop}%
\bibitem [{\citenamefont {Kit}\ \emph {et~al.}(2011)\citenamefont {Kit},
  \citenamefont {Pastewka},\ and\ \citenamefont {Koskinen}}]{Koskinen2010RPBC}%
  \BibitemOpen
  \bibfield  {author} {\bibinfo {author} {\bibfnamefont {O.~O.}\ \bibnamefont
  {Kit}}, \bibinfo {author} {\bibfnamefont {L.}~\bibnamefont {Pastewka}}, \
  and\ \bibinfo {author} {\bibfnamefont {P.}~\bibnamefont {Koskinen}},\
  }\bibfield  {title} {\enquote {\bibinfo {title} {Revised periodic boundary
  conditions: Fundamentals, electrostatics, and the tight-binding
  approximation},}\ }\href@noop {} {\bibfield  {journal} {\bibinfo  {journal}
  {Phys. Rev. B}\ }\textbf {\bibinfo {volume} {84}},\ \bibinfo {pages} {155431}
  (\bibinfo {year} {2011})}\BibitemShut {NoStop}%
\bibitem [{\citenamefont {Dumitrica}\ and\ \citenamefont
  {James}(2007)}]{DumitricaobjectiveMD}%
  \BibitemOpen
  \bibfield  {author} {\bibinfo {author} {\bibfnamefont {T.}~\bibnamefont
  {Dumitrica}}\ and\ \bibinfo {author} {\bibfnamefont {R.~D.}\ \bibnamefont
  {James}},\ }\bibfield  {title} {\enquote {\bibinfo {title} {Objective
  molecular dynamics},}\ }\href@noop {} {\bibfield  {journal} {\bibinfo
  {journal} {J. Mech. Phys. Solids}\ }\textbf {\bibinfo {volume} {55}},\
  \bibinfo {pages} {2206--2236} (\bibinfo {year} {2007})}\BibitemShut {NoStop}%
\bibitem [{\citenamefont {Dayal}\ and\ \citenamefont
  {James}(2010)}]{DayalnonequilibriumMD}%
  \BibitemOpen
  \bibfield  {author} {\bibinfo {author} {\bibfnamefont {K.}~\bibnamefont
  {Dayal}}\ and\ \bibinfo {author} {\bibfnamefont {R.~D.}\ \bibnamefont
  {James}},\ }\bibfield  {title} {\enquote {\bibinfo {title} {Nonequilibrium
  molecular dynamics for bulk materials and nanostructures},}\ }\href@noop {}
  {\bibfield  {journal} {\bibinfo  {journal} {J. Mech. Phys. Solids}\ }\textbf
  {\bibinfo {volume} {58}},\ \bibinfo {pages} {145--163} (\bibinfo {year}
  {2010})}\BibitemShut {NoStop}%
\bibitem [{\citenamefont {Aghaei}\ and\ \citenamefont
  {Dayal}(2011)}]{aghaei2011symmetry}%
  \BibitemOpen
  \bibfield  {author} {\bibinfo {author} {\bibfnamefont {A.}~\bibnamefont
  {Aghaei}}\ and\ \bibinfo {author} {\bibfnamefont {K.}~\bibnamefont {Dayal}},\
  }\bibfield  {title} {\enquote {\bibinfo {title} {Symmetry-adapted
  non-equilibrium molecular dynamics of chiral carbon nanotubes under tensile
  loading},}\ }\href@noop {} {\bibfield  {journal} {\bibinfo  {journal} {J.
  Appl. Phys.}\ }\textbf {\bibinfo {volume} {109}},\ \bibinfo {pages} {123501}
  (\bibinfo {year} {2011})}\BibitemShut {NoStop}%
\bibitem [{\citenamefont {Aghaei}\ \emph {et~al.}(2013)\citenamefont {Aghaei},
  \citenamefont {Dayal},\ and\ \citenamefont {Elliott}}]{aghaei2013symmetry}%
  \BibitemOpen
  \bibfield  {author} {\bibinfo {author} {\bibfnamefont {A.}~\bibnamefont
  {Aghaei}}, \bibinfo {author} {\bibfnamefont {K.}~\bibnamefont {Dayal}}, \
  and\ \bibinfo {author} {\bibfnamefont {R.~S.}\ \bibnamefont {Elliott}},\
  }\bibfield  {title} {\enquote {\bibinfo {title} {Symmetry-adapted phonon
  analysis of nanotubes},}\ }\href@noop {} {\bibfield  {journal} {\bibinfo
  {journal} {J. Mech. Phys. Solids}\ }\textbf {\bibinfo {volume} {61}},\
  \bibinfo {pages} {557--578} (\bibinfo {year} {2013})}\BibitemShut {NoStop}%
\bibitem [{\citenamefont {White}\ \emph {et~al.}(1993)\citenamefont {White},
  \citenamefont {Robertson},\ and\ \citenamefont
  {Mintmire}}]{mintmire1993symmetries}%
  \BibitemOpen
  \bibfield  {author} {\bibinfo {author} {\bibfnamefont {C.~T.}\ \bibnamefont
  {White}}, \bibinfo {author} {\bibfnamefont {D.~H.}\ \bibnamefont
  {Robertson}}, \ and\ \bibinfo {author} {\bibfnamefont {J.~W.}\ \bibnamefont
  {Mintmire}},\ }\bibfield  {title} {\enquote {\bibinfo {title} {Helical and
  rotational symmetries of nanoscale graphitic tubules},}\ }\href@noop {}
  {\bibfield  {journal} {\bibinfo  {journal} {Phys. Rev. B}\ }\textbf {\bibinfo
  {volume} {47}},\ \bibinfo {pages} {5485(R)} (\bibinfo {year}
  {1993})}\BibitemShut {NoStop}%
\bibitem [{\citenamefont
  {Allen}(2007{\natexlab{b}})}]{allen2007nanocrystalline3}%
  \BibitemOpen
  \bibfield  {author} {\bibinfo {author} {\bibfnamefont {P.~B.}\ \bibnamefont
  {Allen}},\ }\bibfield  {title} {\enquote {\bibinfo {title} {Nanocrystalline
  nanowires: {III}. electrons},}\ }\href@noop {} {\bibfield  {journal}
  {\bibinfo  {journal} {Nano Lett.}\ }\textbf {\bibinfo {volume} {7}},\
  \bibinfo {pages} {1220--1223} (\bibinfo {year}
  {2007}{\natexlab{b}})}\BibitemShut {NoStop}%
\bibitem [{\citenamefont
  {Allen}(2007{\natexlab{c}})}]{allen2007nanocrystalline2}%
  \BibitemOpen
  \bibfield  {author} {\bibinfo {author} {\bibfnamefont {P.~B.}\ \bibnamefont
  {Allen}},\ }\bibfield  {title} {\enquote {\bibinfo {title} {Nanocrystalline
  nanowires: 2. phonons},}\ }\href@noop {} {\bibfield  {journal} {\bibinfo
  {journal} {Nano Lett.}\ }\textbf {\bibinfo {volume} {7}},\ \bibinfo {pages}
  {11--14} (\bibinfo {year} {2007}{\natexlab{c}})}\BibitemShut {NoStop}%
\bibitem [{\citenamefont {Popov}(2004)}]{popov2004carbon}%
  \BibitemOpen
  \bibfield  {author} {\bibinfo {author} {\bibfnamefont {V.~N.}\ \bibnamefont
  {Popov}},\ }\bibfield  {title} {\enquote {\bibinfo {title} {Carbon nanotubes:
  properties and application},}\ }\href@noop {} {\bibfield  {journal} {\bibinfo
   {journal} {{Mater. Sci. Eng. R Rep.}}\ }\textbf {\bibinfo {volume} {43}},\
  \bibinfo {pages} {61--102} (\bibinfo {year} {2004})}\BibitemShut {NoStop}%
\bibitem [{\citenamefont {Popov}\ and\ \citenamefont
  {Lambin}(2006)}]{popov2006radius}%
  \BibitemOpen
  \bibfield  {author} {\bibinfo {author} {\bibfnamefont {V.~N.}\ \bibnamefont
  {Popov}}\ and\ \bibinfo {author} {\bibfnamefont {P.}~\bibnamefont {Lambin}},\
  }\bibfield  {title} {\enquote {\bibinfo {title} {Radius and chirality
  dependence of the radial breathing mode and the {G}-band phonon modes of
  single-walled carbon nanotubes},}\ }\href@noop {} {\bibfield  {journal}
  {\bibinfo  {journal} {Phys. Rev. B Condens. Matter}\ }\textbf {\bibinfo
  {volume} {73}},\ \bibinfo {pages} {085407} (\bibinfo {year}
  {2006})}\BibitemShut {NoStop}%
\bibitem [{\citenamefont {Gunlycke}\ \emph {et~al.}(2008)\citenamefont
  {Gunlycke}, \citenamefont {Lawler},\ and\ \citenamefont
  {White}}]{gunlycke2008lattice}%
  \BibitemOpen
  \bibfield  {author} {\bibinfo {author} {\bibfnamefont {D.}~\bibnamefont
  {Gunlycke}}, \bibinfo {author} {\bibfnamefont {H.~M.}\ \bibnamefont
  {Lawler}}, \ and\ \bibinfo {author} {\bibfnamefont {C.~T.}\ \bibnamefont
  {White}},\ }\bibfield  {title} {\enquote {\bibinfo {title} {Lattice
  vibrations in single-wall carbon nanotubes},}\ }\href@noop {} {\bibfield
  {journal} {\bibinfo  {journal} {Phys. Rev. B Condens. Matter}\ }\textbf
  {\bibinfo {volume} {77}},\ \bibinfo {pages} {014303} (\bibinfo {year}
  {2008})}\BibitemShut {NoStop}%
\bibitem [{\citenamefont {Zhang}\ \emph
  {et~al.}(2009{\natexlab{a}})\citenamefont {Zhang}, \citenamefont {James},\
  and\ \citenamefont {Dumitrica}}]{Zhang2009CNT}%
  \BibitemOpen
  \bibfield  {author} {\bibinfo {author} {\bibfnamefont {D.-B.}\ \bibnamefont
  {Zhang}}, \bibinfo {author} {\bibfnamefont {R.~D.}\ \bibnamefont {James}}, \
  and\ \bibinfo {author} {\bibfnamefont {T.}~\bibnamefont {Dumitrica}},\
  }\bibfield  {title} {\enquote {\bibinfo {title} {Electromechanical
  characterization of carbon nanotubes in torsion via symmetry adapted
  tight-binding objective molecular dynamics},}\ }\href@noop {} {\bibfield
  {journal} {\bibinfo  {journal} {Phys. Rev. B}\ }\textbf {\bibinfo {volume}
  {80}},\ \bibinfo {pages} {115418} (\bibinfo {year}
  {2009}{\natexlab{a}})}\BibitemShut {NoStop}%
\bibitem [{\citenamefont {Zhang}\ \emph
  {et~al.}(2009{\natexlab{b}})\citenamefont {Zhang}, \citenamefont {James},\
  and\ \citenamefont {Dumitrica}}]{Zhang2009dislocation}%
  \BibitemOpen
  \bibfield  {author} {\bibinfo {author} {\bibfnamefont {D.-B.}\ \bibnamefont
  {Zhang}}, \bibinfo {author} {\bibfnamefont {R.~D.}\ \bibnamefont {James}}, \
  and\ \bibinfo {author} {\bibfnamefont {T.}~\bibnamefont {Dumitrica}},\
  }\bibfield  {title} {\enquote {\bibinfo {title} {Dislocation onset and nearly
  axial glide in carbon nanotubes under torsion},}\ }\href@noop {} {\bibfield
  {journal} {\bibinfo  {journal} {J. Chem. Phys.}\ }\textbf {\bibinfo {volume}
  {130}},\ \bibinfo {pages} {071101} (\bibinfo {year}
  {2009}{\natexlab{b}})}\BibitemShut {NoStop}%
\bibitem [{\citenamefont {Zhang}\ and\ \citenamefont
  {Wei}(2017)}]{zhang2017inhomogeneous}%
  \BibitemOpen
  \bibfield  {author} {\bibinfo {author} {\bibfnamefont {D.-B.}\ \bibnamefont
  {Zhang}}\ and\ \bibinfo {author} {\bibfnamefont {S.-H.}\ \bibnamefont
  {Wei}},\ }\bibfield  {title} {\enquote {\bibinfo {title} {Inhomogeneous
  strain-induced half-metallicity in bent zigzag graphene nanoribbons},}\
  }\href@noop {} {\bibfield  {journal} {\bibinfo  {journal} {npj Comput.
  Mater.}\ }\textbf {\bibinfo {volume} {3}},\ \bibinfo {pages} {1--5} (\bibinfo
  {year} {2017})}\BibitemShut {NoStop}%
\bibitem [{\citenamefont {Sharma}\ \emph {et~al.}(2025)\citenamefont {Sharma},
  \citenamefont {Kumar},\ and\ \citenamefont
  {Suryanarayana}}]{sharma2025cyclic}%
  \BibitemOpen
  \bibfield  {author} {\bibinfo {author} {\bibfnamefont {A.}~\bibnamefont
  {Sharma}}, \bibinfo {author} {\bibfnamefont {S.}~\bibnamefont {Kumar}}, \
  and\ \bibinfo {author} {\bibfnamefont {P.}~\bibnamefont {Suryanarayana}},\
  }\bibfield  {title} {\enquote {\bibinfo {title} {Cyclic and helical
  symmetry-informed machine learned force fields: Application to lattice
  vibrations in carbon nanotubes},}\ }\href@noop {} {\bibfield  {journal}
  {\bibinfo  {journal} {J. Mech. Phys. Solids}\ }\textbf {\bibinfo {volume}
  {194}},\ \bibinfo {pages} {105927} (\bibinfo {year} {2025})}\BibitemShut
  {NoStop}%
\bibitem [{\citenamefont {Saito}\ \emph {et~al.}(1992)\citenamefont {Saito},
  \citenamefont {Fujita}, \citenamefont {Dresselhaus},\ and\ \citenamefont
  {Dresselhaus}}]{saito1992electronic}%
  \BibitemOpen
  \bibfield  {author} {\bibinfo {author} {\bibfnamefont {R.}~\bibnamefont
  {Saito}}, \bibinfo {author} {\bibfnamefont {M.}~\bibnamefont {Fujita}},
  \bibinfo {author} {\bibfnamefont {G.}~\bibnamefont {Dresselhaus}}, \ and\
  \bibinfo {author} {\bibfnamefont {M.~S.}\ \bibnamefont {Dresselhaus}},\
  }\bibfield  {title} {\enquote {\bibinfo {title} {Electronic structure of
  chiral graphene tubules},}\ }\href@noop {} {\bibfield  {journal} {\bibinfo
  {journal} {App. Phys. Lett.}\ }\textbf {\bibinfo {volume} {60}},\ \bibinfo
  {pages} {2204--2206} (\bibinfo {year} {1992})}\BibitemShut {NoStop}%
\bibitem [{\citenamefont {Ono}\ and\ \citenamefont
  {Hirose}(2005{\natexlab{a}})}]{OnoHir2005}%
  \BibitemOpen
  \bibfield  {author} {\bibinfo {author} {\bibfnamefont {T.}~\bibnamefont
  {Ono}}\ and\ \bibinfo {author} {\bibfnamefont {K.}~\bibnamefont {Hirose}},\
  }\bibfield  {title} {\enquote {\bibinfo {title} {Real-space
  electronic-structure calculations with a time-saving double-grid
  technique},}\ }\href@noop {} {\bibfield  {journal} {\bibinfo  {journal}
  {Phys. Rev. B}\ }\textbf {\bibinfo {volume} {72}},\ \bibinfo {pages} {085115}
  (\bibinfo {year} {2005}{\natexlab{a}})}\BibitemShut {NoStop}%
\bibitem [{\citenamefont {Ono}\ and\ \citenamefont
  {Hirose}(2005{\natexlab{b}})}]{OnoHir2005gold}%
  \BibitemOpen
  \bibfield  {author} {\bibinfo {author} {\bibfnamefont {T.}~\bibnamefont
  {Ono}}\ and\ \bibinfo {author} {\bibfnamefont {K.}~\bibnamefont {Hirose}},\
  }\bibfield  {title} {\enquote {\bibinfo {title} {First-principles study of
  electron-conduction properties of helical gold nanowires},}\ }\href@noop {}
  {\bibfield  {journal} {\bibinfo  {journal} {Phys. Rev. Lett.}\ }\textbf
  {\bibinfo {volume} {94}},\ \bibinfo {pages} {206806} (\bibinfo {year}
  {2005}{\natexlab{b}})}\BibitemShut {NoStop}%
\bibitem [{\citenamefont {Banerjee}\ and\ \citenamefont
  {Suryanarayana}(2016)}]{Banerjee2016cyclic}%
  \BibitemOpen
  \bibfield  {author} {\bibinfo {author} {\bibfnamefont {A.~S.}\ \bibnamefont
  {Banerjee}}\ and\ \bibinfo {author} {\bibfnamefont {P.}~\bibnamefont
  {Suryanarayana}},\ }\bibfield  {title} {\enquote {\bibinfo {title} {Cyclic
  density functional theory: A route to the first principles simulation of
  bending in nanostructures},}\ }\href@noop {} {\bibfield  {journal} {\bibinfo
  {journal} {J. Mech. Phys. Solids}\ }\textbf {\bibinfo {volume} {96}}
  (\bibinfo {year} {2016})}\BibitemShut {NoStop}%
\bibitem [{\citenamefont {Ghosh}\ \emph {et~al.}(2019)\citenamefont {Ghosh},
  \citenamefont {Banerjee},\ and\ \citenamefont
  {Suryanarayana}}]{ghosh2019symmetry}%
  \BibitemOpen
  \bibfield  {author} {\bibinfo {author} {\bibfnamefont {S.}~\bibnamefont
  {Ghosh}}, \bibinfo {author} {\bibfnamefont {A.~S.}\ \bibnamefont {Banerjee}},
  \ and\ \bibinfo {author} {\bibfnamefont {P.}~\bibnamefont {Suryanarayana}},\
  }\bibfield  {title} {\enquote {\bibinfo {title} {Symmetry-adapted real-space
  density functional theory for cylindrical geometries: Application to large
  group-{IV} nanotubes},}\ }\href@noop {} {\bibfield  {journal} {\bibinfo
  {journal} {Phys. Rev. B}\ }\textbf {\bibinfo {volume} {100}},\ \bibinfo
  {pages} {125143} (\bibinfo {year} {2019})}\BibitemShut {NoStop}%
\bibitem [{\citenamefont {Banerjee}(2021)}]{banerjee2021ab}%
  \BibitemOpen
  \bibfield  {author} {\bibinfo {author} {\bibfnamefont {A.~S.}\ \bibnamefont
  {Banerjee}},\ }\bibfield  {title} {\enquote {\bibinfo {title} {Ab initio
  framework for systems with helical symmetry: theory, numerical implementation
  and applications to torsional deformations in nanostructures},}\ }\href@noop
  {} {\bibfield  {journal} {\bibinfo  {journal} {J. Mech. Phys. Solids}\
  }\textbf {\bibinfo {volume} {154}},\ \bibinfo {pages} {104515} (\bibinfo
  {year} {2021})}\BibitemShut {NoStop}%
\bibitem [{\citenamefont {Sharma}\ and\ \citenamefont
  {Suryanarayana}(2021)}]{sharma2021real}%
  \BibitemOpen
  \bibfield  {author} {\bibinfo {author} {\bibfnamefont {A.}~\bibnamefont
  {Sharma}}\ and\ \bibinfo {author} {\bibfnamefont {P.}~\bibnamefont
  {Suryanarayana}},\ }\bibfield  {title} {\enquote {\bibinfo {title}
  {Real-space density functional theory adapted to cyclic and helical symmetry:
  Application to torsional deformation of carbon nanotubes},}\ }\href@noop {}
  {\bibfield  {journal} {\bibinfo  {journal} {Phys. Rev. B.}\ }\textbf
  {\bibinfo {volume} {103}},\ \bibinfo {pages} {035101} (\bibinfo {year}
  {2021})}\BibitemShut {NoStop}%
\bibitem [{\citenamefont {Kumar}\ and\ \citenamefont
  {Suryanarayana}(2022)}]{Kumar_2022}%
  \BibitemOpen
  \bibfield  {author} {\bibinfo {author} {\bibfnamefont {S.}~\bibnamefont
  {Kumar}}\ and\ \bibinfo {author} {\bibfnamefont {P.}~\bibnamefont
  {Suryanarayana}},\ }\bibfield  {title} {\enquote {\bibinfo {title} {On the
  bending of rectangular atomic monolayers along different directions: an ab
  initio study},}\ }\href@noop {} {\bibfield  {journal} {\bibinfo  {journal}
  {Nanotechnology}\ }\textbf {\bibinfo {volume} {34}},\ \bibinfo {pages}
  {085701} (\bibinfo {year} {2022})}\BibitemShut {NoStop}%
\bibitem [{\citenamefont {Kumar}\ and\ \citenamefont
  {Suryanarayana}(2020)}]{Kumar_2020}%
  \BibitemOpen
  \bibfield  {author} {\bibinfo {author} {\bibfnamefont {S.}~\bibnamefont
  {Kumar}}\ and\ \bibinfo {author} {\bibfnamefont {P.}~\bibnamefont
  {Suryanarayana}},\ }\bibfield  {title} {\enquote {\bibinfo {title} {Bending
  moduli for forty-four select atomic monolayers from first principles},}\
  }\href@noop {} {\bibfield  {journal} {\bibinfo  {journal} {Nanotechnology}\
  }\textbf {\bibinfo {volume} {31}},\ \bibinfo {pages} {43LT01} (\bibinfo
  {year} {2020})}\BibitemShut {NoStop}%
\bibitem [{\citenamefont {Bhardwaj}\ \emph {et~al.}(2021)\citenamefont
  {Bhardwaj}, \citenamefont {Sharma},\ and\ \citenamefont
  {Suryanarayana}}]{Bhardwaj_2021}%
  \BibitemOpen
  \bibfield  {author} {\bibinfo {author} {\bibfnamefont {A.}~\bibnamefont
  {Bhardwaj}}, \bibinfo {author} {\bibfnamefont {A.}~\bibnamefont {Sharma}}, \
  and\ \bibinfo {author} {\bibfnamefont {P.}~\bibnamefont {Suryanarayana}},\
  }\bibfield  {title} {\enquote {\bibinfo {title} {Torsional moduli of
  transition metal dichalcogenide nanotubes from first principles},}\
  }\href@noop {} {\bibfield  {journal} {\bibinfo  {journal} {Nanotechnology}\
  }\textbf {\bibinfo {volume} {32}},\ \bibinfo {pages} {28LT02} (\bibinfo
  {year} {2021})}\BibitemShut {NoStop}%
\bibitem [{\citenamefont {Bhardwaj}\ and\ \citenamefont
  {Suryanarayana}(2022{\natexlab{a}})}]{bhardwaj2022elastic}%
  \BibitemOpen
  \bibfield  {author} {\bibinfo {author} {\bibfnamefont {A.}~\bibnamefont
  {Bhardwaj}}\ and\ \bibinfo {author} {\bibfnamefont {P.}~\bibnamefont
  {Suryanarayana}},\ }\bibfield  {title} {\enquote {\bibinfo {title} {Elastic
  properties of {J}anus transition metal dichalcogenide nanotubes from first
  principles},}\ }\href@noop {} {\bibfield  {journal} {\bibinfo  {journal} {The
  European Physical Journal B}\ }\textbf {\bibinfo {volume} {95}},\ \bibinfo
  {pages} {13} (\bibinfo {year} {2022}{\natexlab{a}})}\BibitemShut {NoStop}%
\bibitem [{\citenamefont {Codony}\ \emph {et~al.}(2021)\citenamefont {Codony},
  \citenamefont {Arias},\ and\ \citenamefont {Suryanarayana}}]{codony_2021}%
  \BibitemOpen
  \bibfield  {author} {\bibinfo {author} {\bibfnamefont {D.}~\bibnamefont
  {Codony}}, \bibinfo {author} {\bibfnamefont {I.}~\bibnamefont {Arias}}, \
  and\ \bibinfo {author} {\bibfnamefont {P.}~\bibnamefont {Suryanarayana}},\
  }\bibfield  {title} {\enquote {\bibinfo {title} {Transversal flexoelectric
  coefficient for nanostructures at finite deformations from first
  principles},}\ }\href@noop {} {\bibfield  {journal} {\bibinfo  {journal}
  {Phys. Rev. Mater.}\ }\textbf {\bibinfo {volume} {5}},\ \bibinfo {pages}
  {L030801} (\bibinfo {year} {2021})}\BibitemShut {NoStop}%
\bibitem [{\citenamefont {Kumar}\ \emph {et~al.}(2021)\citenamefont {Kumar},
  \citenamefont {Codony}, \citenamefont {Arias},\ and\ \citenamefont
  {Suryanarayana}}]{kumar_codony_2021}%
  \BibitemOpen
  \bibfield  {author} {\bibinfo {author} {\bibfnamefont {S.}~\bibnamefont
  {Kumar}}, \bibinfo {author} {\bibfnamefont {D.}~\bibnamefont {Codony}},
  \bibinfo {author} {\bibfnamefont {I.}~\bibnamefont {Arias}}, \ and\ \bibinfo
  {author} {\bibfnamefont {P.}~\bibnamefont {Suryanarayana}},\ }\bibfield
  {title} {\enquote {\bibinfo {title} {Flexoelectricity in atomic monolayers
  from first principles},}\ }\href@noop {} {\bibfield  {journal} {\bibinfo
  {journal} {Nanoscale}\ }\textbf {\bibinfo {volume} {13}},\ \bibinfo {pages}
  {1600--1607} (\bibinfo {year} {2021})}\BibitemShut {NoStop}%
\bibitem [{\citenamefont {Bhardwaj}\ and\ \citenamefont
  {Suryanarayana}(2023)}]{bhardwaj2023ab}%
  \BibitemOpen
  \bibfield  {author} {\bibinfo {author} {\bibfnamefont {A.}~\bibnamefont
  {Bhardwaj}}\ and\ \bibinfo {author} {\bibfnamefont {P.}~\bibnamefont
  {Suryanarayana}},\ }\bibfield  {title} {\enquote {\bibinfo {title} {Ab initio
  study on the electromechanical response of {J}anus transition metal dihalide
  nanotubes},}\ }\href@noop {} {\bibfield  {journal} {\bibinfo  {journal} {The
  European Physical Journal B}\ }\textbf {\bibinfo {volume} {96}},\ \bibinfo
  {pages} {36} (\bibinfo {year} {2023})}\BibitemShut {NoStop}%
\bibitem [{\citenamefont {Bhardwaj}\ and\ \citenamefont
  {Suryanarayana}(2022{\natexlab{b}})}]{bhardwaj2022strain}%
  \BibitemOpen
  \bibfield  {author} {\bibinfo {author} {\bibfnamefont {A.}~\bibnamefont
  {Bhardwaj}}\ and\ \bibinfo {author} {\bibfnamefont {P.}~\bibnamefont
  {Suryanarayana}},\ }\bibfield  {title} {\enquote {\bibinfo {title} {Strain
  engineering of {J}anus transition metal dichalcogenide nanotubes: an ab
  initio study},}\ }\href@noop {} {\bibfield  {journal} {\bibinfo  {journal}
  {The European Physical Journal B}\ }\textbf {\bibinfo {volume} {95}},\
  \bibinfo {pages} {59} (\bibinfo {year} {2022}{\natexlab{b}})}\BibitemShut
  {NoStop}%
\bibitem [{\citenamefont {Bhardwaj}\ and\ \citenamefont
  {Suryanarayana}(2024)}]{Bhardwaj_2024}%
  \BibitemOpen
  \bibfield  {author} {\bibinfo {author} {\bibfnamefont {A.}~\bibnamefont
  {Bhardwaj}}\ and\ \bibinfo {author} {\bibfnamefont {P.}~\bibnamefont
  {Suryanarayana}},\ }\bibfield  {title} {\enquote {\bibinfo {title} {Strain
  engineering of {Z}eeman and {R}ashba effects in transition metal
  dichalcogenide nanotubes and their {J}anus variants: an ab initio study},}\
  }\href@noop {} {\bibfield  {journal} {\bibinfo  {journal} {Nanotechnology}\
  }\textbf {\bibinfo {volume} {35}},\ \bibinfo {pages} {185701} (\bibinfo
  {year} {2024})}\BibitemShut {NoStop}%
\bibitem [{\citenamefont {Clatterbuck}\ \emph {et~al.}(2003)\citenamefont
  {Clatterbuck}, \citenamefont {Krenn}, \citenamefont {Cohen},\ and\
  \citenamefont {Morris~Jr}}]{clatterbuck2003phonon}%
  \BibitemOpen
  \bibfield  {author} {\bibinfo {author} {\bibfnamefont {D.~M.}\ \bibnamefont
  {Clatterbuck}}, \bibinfo {author} {\bibfnamefont {C.~R.}\ \bibnamefont
  {Krenn}}, \bibinfo {author} {\bibfnamefont {M.~L.}\ \bibnamefont {Cohen}}, \
  and\ \bibinfo {author} {\bibfnamefont {J.~W.}\ \bibnamefont {Morris~Jr}},\
  }\bibfield  {title} {\enquote {\bibinfo {title} {Phonon instabilities and the
  ideal strength of aluminum},}\ }\href@noop {} {\bibfield  {journal} {\bibinfo
   {journal} {Phys. Rev. Lett.}\ }\textbf {\bibinfo {volume} {91}},\ \bibinfo
  {pages} {135501} (\bibinfo {year} {2003})}\BibitemShut {NoStop}%
\bibitem [{\citenamefont {Liu}\ \emph {et~al.}(2007)\citenamefont {Liu},
  \citenamefont {Ming},\ and\ \citenamefont {Li}}]{liu2007ab}%
  \BibitemOpen
  \bibfield  {author} {\bibinfo {author} {\bibfnamefont {F.}~\bibnamefont
  {Liu}}, \bibinfo {author} {\bibfnamefont {P.}~\bibnamefont {Ming}}, \ and\
  \bibinfo {author} {\bibfnamefont {J.}~\bibnamefont {Li}},\ }\bibfield
  {title} {\enquote {\bibinfo {title} {Ab initio calculation of ideal strength
  and phonon instability of graphene under tension},}\ }\href@noop {}
  {\bibfield  {journal} {\bibinfo  {journal} {Phys. Rev. B.}\ }\textbf
  {\bibinfo {volume} {76}},\ \bibinfo {pages} {064120} (\bibinfo {year}
  {2007})}\BibitemShut {NoStop}%
\bibitem [{\citenamefont {Savrasov}\ and\ \citenamefont
  {Savrasov}(1996)}]{savrasov1996electron}%
  \BibitemOpen
  \bibfield  {author} {\bibinfo {author} {\bibfnamefont {S.~Y.}\ \bibnamefont
  {Savrasov}}\ and\ \bibinfo {author} {\bibfnamefont {D.~Y.}\ \bibnamefont
  {Savrasov}},\ }\bibfield  {title} {\enquote {\bibinfo {title}
  {Electron-phonon interactions and related physical properties of metals from
  linear-response theory},}\ }\href@noop {} {\bibfield  {journal} {\bibinfo
  {journal} {Phys. Rev. B.}\ }\textbf {\bibinfo {volume} {54}},\ \bibinfo
  {pages} {16487} (\bibinfo {year} {1996})}\BibitemShut {NoStop}%
\bibitem [{\citenamefont {Wu}\ \emph {et~al.}(2005)\citenamefont {Wu},
  \citenamefont {Chen}, \citenamefont {Struzhkin},\ and\ \citenamefont
  {Cohen}}]{wu2005trends}%
  \BibitemOpen
  \bibfield  {author} {\bibinfo {author} {\bibfnamefont {Z.}~\bibnamefont
  {Wu}}, \bibinfo {author} {\bibfnamefont {X.-J.}\ \bibnamefont {Chen}},
  \bibinfo {author} {\bibfnamefont {V.~V.}\ \bibnamefont {Struzhkin}}, \ and\
  \bibinfo {author} {\bibfnamefont {R.~E.}\ \bibnamefont {Cohen}},\ }\bibfield
  {title} {\enquote {\bibinfo {title} {Trends in elasticity and electronic
  structure of transition-metal nitrides and carbides from first principles},}\
  }\href@noop {} {\bibfield  {journal} {\bibinfo  {journal} {Phys. Rev. B.}\
  }\textbf {\bibinfo {volume} {71}},\ \bibinfo {pages} {214103} (\bibinfo
  {year} {2005})}\BibitemShut {NoStop}%
\bibitem [{\citenamefont {Karki}\ \emph {et~al.}(2000)\citenamefont {Karki},
  \citenamefont {Wentzcovitch}, \citenamefont {De~Gironcoli},\ and\
  \citenamefont {Baroni}}]{karki2000high}%
  \BibitemOpen
  \bibfield  {author} {\bibinfo {author} {\bibfnamefont {B.~B.}\ \bibnamefont
  {Karki}}, \bibinfo {author} {\bibfnamefont {R.~M.}\ \bibnamefont
  {Wentzcovitch}}, \bibinfo {author} {\bibfnamefont {S.}~\bibnamefont
  {De~Gironcoli}}, \ and\ \bibinfo {author} {\bibfnamefont {S.}~\bibnamefont
  {Baroni}},\ }\bibfield  {title} {\enquote {\bibinfo {title} {High-pressure
  lattice dynamics and thermoelasticity of {M}g{O}},}\ }\href@noop {}
  {\bibfield  {journal} {\bibinfo  {journal} {Phys. Rev. B.}\ }\textbf
  {\bibinfo {volume} {61}},\ \bibinfo {pages} {8793} (\bibinfo {year}
  {2000})}\BibitemShut {NoStop}%
\bibitem [{\citenamefont {Lee}\ and\ \citenamefont {Gonze}(1995)}]{lee1995ab}%
  \BibitemOpen
  \bibfield  {author} {\bibinfo {author} {\bibfnamefont {C.}~\bibnamefont
  {Lee}}\ and\ \bibinfo {author} {\bibfnamefont {X.}~\bibnamefont {Gonze}},\
  }\bibfield  {title} {\enquote {\bibinfo {title} {Ab initio calculation of the
  thermodynamic properties and atomic temperature factors of $\mathrm{SiO}_2$
  $\alpha$-quartz and stishovite},}\ }\href@noop {} {\bibfield  {journal}
  {\bibinfo  {journal} {Phys. Rev. B.}\ }\textbf {\bibinfo {volume} {51}},\
  \bibinfo {pages} {8610} (\bibinfo {year} {1995})}\BibitemShut {NoStop}%
\bibitem [{\citenamefont {Nie}\ and\ \citenamefont {Xie}(2007)}]{nie2007ab}%
  \BibitemOpen
  \bibfield  {author} {\bibinfo {author} {\bibfnamefont {Y.}~\bibnamefont
  {Nie}}\ and\ \bibinfo {author} {\bibfnamefont {Y.}~\bibnamefont {Xie}},\
  }\bibfield  {title} {\enquote {\bibinfo {title} {Ab initio thermodynamics of
  the hcp metals {M}g, {T}i, and {Z}r},}\ }\href@noop {} {\bibfield  {journal}
  {\bibinfo  {journal} {Phys. Rev. B.}\ }\textbf {\bibinfo {volume} {75}},\
  \bibinfo {pages} {174117} (\bibinfo {year} {2007})}\BibitemShut {NoStop}%
\bibitem [{\citenamefont {Fleszar}\ and\ \citenamefont
  {Gonze}(1990)}]{fleszar1990first}%
  \BibitemOpen
  \bibfield  {author} {\bibinfo {author} {\bibfnamefont {A.}~\bibnamefont
  {Fleszar}}\ and\ \bibinfo {author} {\bibfnamefont {X.}~\bibnamefont
  {Gonze}},\ }\bibfield  {title} {\enquote {\bibinfo {title} {First-principles
  thermodynamical properties of semiconductors},}\ }\href@noop {} {\bibfield
  {journal} {\bibinfo  {journal} {Phys. Rev. Lett.}\ }\textbf {\bibinfo
  {volume} {64}},\ \bibinfo {pages} {2961} (\bibinfo {year}
  {1990})}\BibitemShut {NoStop}%
\bibitem [{\citenamefont {Togo}\ \emph {et~al.}(2010)\citenamefont {Togo},
  \citenamefont {Chaput}, \citenamefont {Tanaka},\ and\ \citenamefont
  {Hug}}]{togo2010first}%
  \BibitemOpen
  \bibfield  {author} {\bibinfo {author} {\bibfnamefont {A.}~\bibnamefont
  {Togo}}, \bibinfo {author} {\bibfnamefont {L.}~\bibnamefont {Chaput}},
  \bibinfo {author} {\bibfnamefont {I.}~\bibnamefont {Tanaka}}, \ and\ \bibinfo
  {author} {\bibfnamefont {G.}~\bibnamefont {Hug}},\ }\bibfield  {title}
  {\enquote {\bibinfo {title} {First-principles phonon calculations of thermal
  expansion in $\mathrm{Ti}_3 \mathrm{SiC}_2, \ \mathrm{Ti}_3 \mathrm{AlC}_2, \
  \mathrm{and} \ \mathrm{Ti}_3 \mathrm{GeC}_2$},}\ }\href@noop {} {\bibfield
  {journal} {\bibinfo  {journal} {Phys. Rev. B.}\ }\textbf {\bibinfo {volume}
  {81}},\ \bibinfo {pages} {174301} (\bibinfo {year} {2010})}\BibitemShut
  {NoStop}%
\bibitem [{\citenamefont {Yin}\ and\ \citenamefont
  {Cohen}(1982)}]{yin1982calculation}%
  \BibitemOpen
  \bibfield  {author} {\bibinfo {author} {\bibfnamefont {M.~T.}\ \bibnamefont
  {Yin}}\ and\ \bibinfo {author} {\bibfnamefont {M.~L.}\ \bibnamefont
  {Cohen}},\ }\bibfield  {title} {\enquote {\bibinfo {title} {Calculation of
  the lattice dynamical properties of {G}e},}\ }\href@noop {} {\bibfield
  {journal} {\bibinfo  {journal} {Solid State Commun.}\ }\textbf {\bibinfo
  {volume} {43}},\ \bibinfo {pages} {391--393} (\bibinfo {year}
  {1982})}\BibitemShut {NoStop}%
\bibitem [{\citenamefont {Car}\ and\ \citenamefont
  {Parrinello}(1985)}]{car1985unified}%
  \BibitemOpen
  \bibfield  {author} {\bibinfo {author} {\bibfnamefont {R.}~\bibnamefont
  {Car}}\ and\ \bibinfo {author} {\bibfnamefont {M.}~\bibnamefont
  {Parrinello}},\ }\bibfield  {title} {\enquote {\bibinfo {title} {Unified
  approach for molecular dynamics and density-functional theory},}\ }\href@noop
  {} {\bibfield  {journal} {\bibinfo  {journal} {Phys. Rev. Lett.}\ }\textbf
  {\bibinfo {volume} {55}},\ \bibinfo {pages} {2471} (\bibinfo {year}
  {1985})}\BibitemShut {NoStop}%
\bibitem [{\citenamefont {Baroni}\ \emph {et~al.}(2001)\citenamefont {Baroni},
  \citenamefont {De~Gironcoli}, \citenamefont {Dal~Corso},\ and\ \citenamefont
  {Giannozzi}}]{baroni2001phonons}%
  \BibitemOpen
  \bibfield  {author} {\bibinfo {author} {\bibfnamefont {S.}~\bibnamefont
  {Baroni}}, \bibinfo {author} {\bibfnamefont {S.}~\bibnamefont
  {De~Gironcoli}}, \bibinfo {author} {\bibfnamefont {A.}~\bibnamefont
  {Dal~Corso}}, \ and\ \bibinfo {author} {\bibfnamefont {P.}~\bibnamefont
  {Giannozzi}},\ }\bibfield  {title} {\enquote {\bibinfo {title} {Phonons and
  related crystal properties from density-functional perturbation theory},}\
  }\href@noop {} {\bibfield  {journal} {\bibinfo  {journal} {Rev. Mod. Phys.}\
  }\textbf {\bibinfo {volume} {73}},\ \bibinfo {pages} {515} (\bibinfo {year}
  {2001})}\BibitemShut {NoStop}%
\bibitem [{\citenamefont {Baroni}\ \emph {et~al.}(1987)\citenamefont {Baroni},
  \citenamefont {Giannozzi},\ and\ \citenamefont {Testa}}]{baroni1987green}%
  \BibitemOpen
  \bibfield  {author} {\bibinfo {author} {\bibfnamefont {S.}~\bibnamefont
  {Baroni}}, \bibinfo {author} {\bibfnamefont {P.}~\bibnamefont {Giannozzi}}, \
  and\ \bibinfo {author} {\bibfnamefont {A.}~\bibnamefont {Testa}},\ }\bibfield
   {title} {\enquote {\bibinfo {title} {Green’s-function approach to linear
  response in solids},}\ }\href@noop {} {\bibfield  {journal} {\bibinfo
  {journal} {Phys. Rev. Lett.}\ }\textbf {\bibinfo {volume} {58}},\ \bibinfo
  {pages} {1861} (\bibinfo {year} {1987})}\BibitemShut {NoStop}%
\bibitem [{\citenamefont {Gonze}\ and\ \citenamefont
  {Vigneron}(1989)}]{gonze1989density}%
  \BibitemOpen
  \bibfield  {author} {\bibinfo {author} {\bibfnamefont {X.}~\bibnamefont
  {Gonze}}\ and\ \bibinfo {author} {\bibfnamefont {J.-P.}\ \bibnamefont
  {Vigneron}},\ }\bibfield  {title} {\enquote {\bibinfo {title}
  {Density-functional approach to nonlinear-response coefficients of solids},}\
  }\href@noop {} {\bibfield  {journal} {\bibinfo  {journal} {Phys. Rev. B.}\
  }\textbf {\bibinfo {volume} {39}},\ \bibinfo {pages} {13120} (\bibinfo {year}
  {1989})}\BibitemShut {NoStop}%
\bibitem [{\citenamefont {Zein}(1984)}]{zein1984density}%
  \BibitemOpen
  \bibfield  {author} {\bibinfo {author} {\bibfnamefont {N.~E.}\ \bibnamefont
  {Zein}},\ }\bibfield  {title} {\enquote {\bibinfo {title} {On density
  functional calculations of crystal elastic modula and phonon spectra},}\
  }\href@noop {} {\bibfield  {journal} {\bibinfo  {journal} {Fiz. tverd. tela}\
  }\textbf {\bibinfo {volume} {26}},\ \bibinfo {pages} {3028--3034} (\bibinfo
  {year} {1984})}\BibitemShut {NoStop}%
\bibitem [{\citenamefont {Dal~Corso}\ \emph {et~al.}(1993)\citenamefont
  {Dal~Corso}, \citenamefont {Baroni}, \citenamefont {Resta},\ and\
  \citenamefont {de~Gironcoli}}]{dal1993ab}%
  \BibitemOpen
  \bibfield  {author} {\bibinfo {author} {\bibfnamefont {A.}~\bibnamefont
  {Dal~Corso}}, \bibinfo {author} {\bibfnamefont {S.}~\bibnamefont {Baroni}},
  \bibinfo {author} {\bibfnamefont {R.}~\bibnamefont {Resta}}, \ and\ \bibinfo
  {author} {\bibfnamefont {S.}~\bibnamefont {de~Gironcoli}},\ }\bibfield
  {title} {\enquote {\bibinfo {title} {Ab initio calculation of phonon
  dispersions in {II}-{VI} semiconductors},}\ }\href@noop {} {\bibfield
  {journal} {\bibinfo  {journal} {Phys. Rev. B.}\ }\textbf {\bibinfo {volume}
  {47}},\ \bibinfo {pages} {3588} (\bibinfo {year} {1993})}\BibitemShut
  {NoStop}%
\bibitem [{\citenamefont {Dal~Corso}\ \emph {et~al.}(1997)\citenamefont
  {Dal~Corso}, \citenamefont {Pasquarello},\ and\ \citenamefont
  {Baldereschi}}]{dal1997density}%
  \BibitemOpen
  \bibfield  {author} {\bibinfo {author} {\bibfnamefont {A.}~\bibnamefont
  {Dal~Corso}}, \bibinfo {author} {\bibfnamefont {A.}~\bibnamefont
  {Pasquarello}}, \ and\ \bibinfo {author} {\bibfnamefont {A.}~\bibnamefont
  {Baldereschi}},\ }\bibfield  {title} {\enquote {\bibinfo {title}
  {Density-functional perturbation theory for lattice dynamics with ultrasoft
  pseudopotentials},}\ }\href@noop {} {\bibfield  {journal} {\bibinfo
  {journal} {Phys. Rev. B.}\ }\textbf {\bibinfo {volume} {56}},\ \bibinfo
  {pages} {R11369} (\bibinfo {year} {1997})}\BibitemShut {NoStop}%
\bibitem [{\citenamefont {Savrasov}(1992)}]{savrasov1992linear}%
  \BibitemOpen
  \bibfield  {author} {\bibinfo {author} {\bibfnamefont {S.~Y.}\ \bibnamefont
  {Savrasov}},\ }\bibfield  {title} {\enquote {\bibinfo {title} {Linear
  response calculations of lattice dynamics using muffin-tin basis sets},}\
  }\href@noop {} {\bibfield  {journal} {\bibinfo  {journal} {Phys. Rev. Lett.}\
  }\textbf {\bibinfo {volume} {69}},\ \bibinfo {pages} {2819} (\bibinfo {year}
  {1992})}\BibitemShut {NoStop}%
\bibitem [{\citenamefont {Yu}\ and\ \citenamefont
  {Krakauer}(1994)}]{yu1994linear}%
  \BibitemOpen
  \bibfield  {author} {\bibinfo {author} {\bibfnamefont {R.}~\bibnamefont
  {Yu}}\ and\ \bibinfo {author} {\bibfnamefont {H.}~\bibnamefont {Krakauer}},\
  }\bibfield  {title} {\enquote {\bibinfo {title} {Linear-response calculations
  within the linearized augmented plane-wave method},}\ }\href@noop {}
  {\bibfield  {journal} {\bibinfo  {journal} {Phys. Rev. B.}\ }\textbf
  {\bibinfo {volume} {49}},\ \bibinfo {pages} {4467} (\bibinfo {year}
  {1994})}\BibitemShut {NoStop}%
\bibitem [{\citenamefont {Gonze}\ and\ \citenamefont
  {Lee}(1997)}]{gonze1997dynamical}%
  \BibitemOpen
  \bibfield  {author} {\bibinfo {author} {\bibfnamefont {X.}~\bibnamefont
  {Gonze}}\ and\ \bibinfo {author} {\bibfnamefont {C.}~\bibnamefont {Lee}},\
  }\bibfield  {title} {\enquote {\bibinfo {title} {Dynamical matrices, born
  effective charges, dielectric permittivity tensors, and interatomic force
  constants from density-functional perturbation theory},}\ }\href@noop {}
  {\bibfield  {journal} {\bibinfo  {journal} {Phys. Rev. B.}\ }\textbf
  {\bibinfo {volume} {55}},\ \bibinfo {pages} {10355} (\bibinfo {year}
  {1997})}\BibitemShut {NoStop}%
\bibitem [{\citenamefont {Verstraete}\ \emph {et~al.}(2008)\citenamefont
  {Verstraete}, \citenamefont {Torrent}, \citenamefont {Jollet}, \citenamefont
  {Z{\'e}rah},\ and\ \citenamefont {Gonze}}]{verstraete2008density}%
  \BibitemOpen
  \bibfield  {author} {\bibinfo {author} {\bibfnamefont {M.~J.}\ \bibnamefont
  {Verstraete}}, \bibinfo {author} {\bibfnamefont {M.}~\bibnamefont {Torrent}},
  \bibinfo {author} {\bibfnamefont {F.}~\bibnamefont {Jollet}}, \bibinfo
  {author} {\bibfnamefont {G.}~\bibnamefont {Z{\'e}rah}}, \ and\ \bibinfo
  {author} {\bibfnamefont {X.}~\bibnamefont {Gonze}},\ }\bibfield  {title}
  {\enquote {\bibinfo {title} {Density functional perturbation theory with
  spin-orbit coupling: Phonon band structure of lead},}\ }\href@noop {}
  {\bibfield  {journal} {\bibinfo  {journal} {Phys. Rev. B.}\ }\textbf
  {\bibinfo {volume} {78}},\ \bibinfo {pages} {045119} (\bibinfo {year}
  {2008})}\BibitemShut {NoStop}%
\bibitem [{\citenamefont {Refson}\ \emph {et~al.}(2006)\citenamefont {Refson},
  \citenamefont {Tulip},\ and\ \citenamefont {Clark}}]{refson2006variational}%
  \BibitemOpen
  \bibfield  {author} {\bibinfo {author} {\bibfnamefont {K.}~\bibnamefont
  {Refson}}, \bibinfo {author} {\bibfnamefont {P.~R.}\ \bibnamefont {Tulip}}, \
  and\ \bibinfo {author} {\bibfnamefont {S.~J.}\ \bibnamefont {Clark}},\
  }\bibfield  {title} {\enquote {\bibinfo {title} {Variational
  density-functional perturbation theory for dielectrics and lattice
  dynamics},}\ }\href@noop {} {\bibfield  {journal} {\bibinfo  {journal} {Phys.
  Rev. B.}\ }\textbf {\bibinfo {volume} {73}},\ \bibinfo {pages} {155114}
  (\bibinfo {year} {2006})}\BibitemShut {NoStop}%
\bibitem [{\citenamefont {Gonze}\ \emph {et~al.}(2002)\citenamefont {Gonze},
  \citenamefont {Beuken}, \citenamefont {Caracas}, \citenamefont {Detraux},
  \citenamefont {Fuchs}, \citenamefont {Rignanese}, \citenamefont {Sindic},
  \citenamefont {Verstraete}, \citenamefont {Zerah}, \citenamefont {Jollet},
  \citenamefont {Torrent}, \citenamefont {Roy}, \citenamefont {Mikami},
  \citenamefont {Ghosez}, \citenamefont {Raty},\ and\ \citenamefont
  {Allan}}]{ABINIT}%
  \BibitemOpen
  \bibfield  {author} {\bibinfo {author} {\bibfnamefont {X.}~\bibnamefont
  {Gonze}}, \bibinfo {author} {\bibfnamefont {J.-M.}\ \bibnamefont {Beuken}},
  \bibinfo {author} {\bibfnamefont {R.}~\bibnamefont {Caracas}}, \bibinfo
  {author} {\bibfnamefont {F.}~\bibnamefont {Detraux}}, \bibinfo {author}
  {\bibfnamefont {M.}~\bibnamefont {Fuchs}}, \bibinfo {author} {\bibfnamefont
  {G.-M.}\ \bibnamefont {Rignanese}}, \bibinfo {author} {\bibfnamefont
  {L.}~\bibnamefont {Sindic}}, \bibinfo {author} {\bibfnamefont
  {M.}~\bibnamefont {Verstraete}}, \bibinfo {author} {\bibfnamefont
  {G.}~\bibnamefont {Zerah}}, \bibinfo {author} {\bibfnamefont
  {F.}~\bibnamefont {Jollet}}, \bibinfo {author} {\bibfnamefont
  {M.}~\bibnamefont {Torrent}}, \bibinfo {author} {\bibfnamefont
  {A.}~\bibnamefont {Roy}}, \bibinfo {author} {\bibfnamefont {M.}~\bibnamefont
  {Mikami}}, \bibinfo {author} {\bibfnamefont {Ph.}\ \bibnamefont {Ghosez}},
  \bibinfo {author} {\bibfnamefont {J.-Y.}\ \bibnamefont {Raty}}, \ and\
  \bibinfo {author} {\bibfnamefont {D.~C.}\ \bibnamefont {Allan}},\ }\bibfield
  {title} {\enquote {\bibinfo {title} {First-principles computation of material
  properties: the {ABINIT} software project},}\ }\href@noop {} {\bibfield
  {journal} {\bibinfo  {journal} {Comput. Mater. Sci.}\ }\textbf {\bibinfo
  {volume} {25}},\ \bibinfo {pages} {478--492(15)} (\bibinfo {year}
  {2002})}\BibitemShut {NoStop}%
\bibitem [{\citenamefont {Giannozzi}\ \emph {et~al.}(2009)\citenamefont
  {Giannozzi}, \citenamefont {Baroni}, \citenamefont {Bonini}, \citenamefont
  {Calandra}, \citenamefont {Car}, \citenamefont {Cavazzoni}, \citenamefont
  {Ceresoli}, \citenamefont {Chiarotti}, \citenamefont {Cococcioni},
  \citenamefont {Dabo}, \citenamefont {{Dal Corso}}, \citenamefont
  {de~Gironcoli}, \citenamefont {Fabris}, \citenamefont {Fratesi},
  \citenamefont {Gebauer}, \citenamefont {Gerstmann}, \citenamefont
  {Gougoussis}, \citenamefont {Kokalj}, \citenamefont {Lazzeri}, \citenamefont
  {Martin-Samos}, \citenamefont {Marzari}, \citenamefont {Mauri}, \citenamefont
  {Mazzarello}, \citenamefont {Paolini}, \citenamefont {Pasquarello},
  \citenamefont {Paulatto}, \citenamefont {Sbraccia}, \citenamefont {Scandolo},
  \citenamefont {Sclauzero}, \citenamefont {Seitsonen}, \citenamefont
  {Smogunov}, \citenamefont {Umari},\ and\ \citenamefont
  {Wentzcovitch}}]{Espresso}%
  \BibitemOpen
  \bibfield  {author} {\bibinfo {author} {\bibfnamefont {P.}~\bibnamefont
  {Giannozzi}}, \bibinfo {author} {\bibfnamefont {S.}~\bibnamefont {Baroni}},
  \bibinfo {author} {\bibfnamefont {N.}~\bibnamefont {Bonini}}, \bibinfo
  {author} {\bibfnamefont {M.}~\bibnamefont {Calandra}}, \bibinfo {author}
  {\bibfnamefont {R.}~\bibnamefont {Car}}, \bibinfo {author} {\bibfnamefont
  {C.}~\bibnamefont {Cavazzoni}}, \bibinfo {author} {\bibfnamefont
  {D.}~\bibnamefont {Ceresoli}}, \bibinfo {author} {\bibfnamefont {G.~L.}\
  \bibnamefont {Chiarotti}}, \bibinfo {author} {\bibfnamefont {M.}~\bibnamefont
  {Cococcioni}}, \bibinfo {author} {\bibfnamefont {I.}~\bibnamefont {Dabo}},
  \bibinfo {author} {\bibfnamefont {A.}~\bibnamefont {{Dal Corso}}}, \bibinfo
  {author} {\bibfnamefont {S.}~\bibnamefont {de~Gironcoli}}, \bibinfo {author}
  {\bibfnamefont {S.}~\bibnamefont {Fabris}}, \bibinfo {author} {\bibfnamefont
  {G.}~\bibnamefont {Fratesi}}, \bibinfo {author} {\bibfnamefont
  {R.}~\bibnamefont {Gebauer}}, \bibinfo {author} {\bibfnamefont
  {U.}~\bibnamefont {Gerstmann}}, \bibinfo {author} {\bibfnamefont
  {C.}~\bibnamefont {Gougoussis}}, \bibinfo {author} {\bibfnamefont
  {A.}~\bibnamefont {Kokalj}}, \bibinfo {author} {\bibfnamefont
  {M.}~\bibnamefont {Lazzeri}}, \bibinfo {author} {\bibfnamefont
  {L.}~\bibnamefont {Martin-Samos}}, \bibinfo {author} {\bibfnamefont
  {N.}~\bibnamefont {Marzari}}, \bibinfo {author} {\bibfnamefont
  {F.}~\bibnamefont {Mauri}}, \bibinfo {author} {\bibfnamefont
  {R.}~\bibnamefont {Mazzarello}}, \bibinfo {author} {\bibfnamefont
  {S.}~\bibnamefont {Paolini}}, \bibinfo {author} {\bibfnamefont
  {A.}~\bibnamefont {Pasquarello}}, \bibinfo {author} {\bibfnamefont
  {L.}~\bibnamefont {Paulatto}}, \bibinfo {author} {\bibfnamefont
  {C.}~\bibnamefont {Sbraccia}}, \bibinfo {author} {\bibfnamefont
  {S.}~\bibnamefont {Scandolo}}, \bibinfo {author} {\bibfnamefont
  {G.}~\bibnamefont {Sclauzero}}, \bibinfo {author} {\bibfnamefont {A.~P.}\
  \bibnamefont {Seitsonen}}, \bibinfo {author} {\bibfnamefont {A.}~\bibnamefont
  {Smogunov}}, \bibinfo {author} {\bibfnamefont {P.}~\bibnamefont {Umari}}, \
  and\ \bibinfo {author} {\bibfnamefont {R.~M.}\ \bibnamefont {Wentzcovitch}},\
  }\bibfield  {title} {\enquote {\bibinfo {title} {{QUANTUM ESPRESSO}: a
  modular and open-source software project for quantum simulations of
  materials},}\ }\href@noop {} {\bibfield  {journal} {\bibinfo  {journal} {J.
  Phys.: Condens. Matter}\ }\textbf {\bibinfo {volume} {21}},\ \bibinfo {pages}
  {395502} (\bibinfo {year} {2009})}\BibitemShut {NoStop}%
\bibitem [{\citenamefont {Segall}\ \emph {et~al.}(2002)\citenamefont {Segall},
  \citenamefont {Lindan}, \citenamefont {Probert}, \citenamefont {Pickard},
  \citenamefont {Hasnip}, \citenamefont {Clark},\ and\ \citenamefont
  {Payne}}]{CASTEP}%
  \BibitemOpen
  \bibfield  {author} {\bibinfo {author} {\bibfnamefont {M.~D.}\ \bibnamefont
  {Segall}}, \bibinfo {author} {\bibfnamefont {P.~J.~D.}\ \bibnamefont
  {Lindan}}, \bibinfo {author} {\bibfnamefont {M.~J.}\ \bibnamefont {Probert}},
  \bibinfo {author} {\bibfnamefont {C.~J.}\ \bibnamefont {Pickard}}, \bibinfo
  {author} {\bibfnamefont {P.~J.}\ \bibnamefont {Hasnip}}, \bibinfo {author}
  {\bibfnamefont {S.~J.}\ \bibnamefont {Clark}}, \ and\ \bibinfo {author}
  {\bibfnamefont {M.~C.}\ \bibnamefont {Payne}},\ }\bibfield  {title} {\enquote
  {\bibinfo {title} {First-principles simulation: ideas, illustrations and the
  {CASTEP} code},}\ }\href@noop {} {\bibfield  {journal} {\bibinfo  {journal}
  {J. Phys.: Condens. Matter}\ }\textbf {\bibinfo {volume} {14}},\ \bibinfo
  {pages} {2717--2744} (\bibinfo {year} {2002})}\BibitemShut {NoStop}%
\bibitem [{\citenamefont {Rivano}\ \emph {et~al.}(2024)\citenamefont {Rivano},
  \citenamefont {Marzari},\ and\ \citenamefont {Sohier}}]{rivano2024density}%
  \BibitemOpen
  \bibfield  {author} {\bibinfo {author} {\bibfnamefont {Norma}\ \bibnamefont
  {Rivano}}, \bibinfo {author} {\bibfnamefont {Nicola}\ \bibnamefont
  {Marzari}}, \ and\ \bibinfo {author} {\bibfnamefont {Thibault}\ \bibnamefont
  {Sohier}},\ }\bibfield  {title} {\enquote {\bibinfo {title} {Density
  functional perturbation theory for one-dimensional systems: Implementation
  and relevance for phonons and electron-phonon interactions},}\ }\href@noop {}
  {\bibfield  {journal} {\bibinfo  {journal} {Phys. Rev. B}\ }\textbf {\bibinfo
  {volume} {109}},\ \bibinfo {pages} {245426} (\bibinfo {year}
  {2024})}\BibitemShut {NoStop}%
\bibitem [{\citenamefont {Sohier}\ \emph {et~al.}(2017)\citenamefont {Sohier},
  \citenamefont {Calandra},\ and\ \citenamefont {Mauri}}]{sohier2017density}%
  \BibitemOpen
  \bibfield  {author} {\bibinfo {author} {\bibfnamefont {Thibault}\
  \bibnamefont {Sohier}}, \bibinfo {author} {\bibfnamefont {Matteo}\
  \bibnamefont {Calandra}}, \ and\ \bibinfo {author} {\bibfnamefont
  {Francesco}\ \bibnamefont {Mauri}},\ }\bibfield  {title} {\enquote {\bibinfo
  {title} {Density functional perturbation theory for gated two-dimensional
  heterostructures: Theoretical developments and application to flexural
  phonons in graphene},}\ }\href@noop {} {\bibfield  {journal} {\bibinfo
  {journal} {Phys. Rev. B}\ }\textbf {\bibinfo {volume} {96}},\ \bibinfo
  {pages} {075448} (\bibinfo {year} {2017})}\BibitemShut {NoStop}%
\bibitem [{\citenamefont {Shang}\ \emph {et~al.}(2018)\citenamefont {Shang},
  \citenamefont {Raimbault}, \citenamefont {Rinke}, \citenamefont {Scheffler},
  \citenamefont {Rossi},\ and\ \citenamefont {Carbogno}}]{shang2018all}%
  \BibitemOpen
  \bibfield  {author} {\bibinfo {author} {\bibfnamefont {H.}~\bibnamefont
  {Shang}}, \bibinfo {author} {\bibfnamefont {N.}~\bibnamefont {Raimbault}},
  \bibinfo {author} {\bibfnamefont {P.}~\bibnamefont {Rinke}}, \bibinfo
  {author} {\bibfnamefont {M.}~\bibnamefont {Scheffler}}, \bibinfo {author}
  {\bibfnamefont {M.}~\bibnamefont {Rossi}}, \ and\ \bibinfo {author}
  {\bibfnamefont {C.}~\bibnamefont {Carbogno}},\ }\bibfield  {title} {\enquote
  {\bibinfo {title} {All-electron, real-space perturbation theory for
  homogeneous electric fields: theory, implementation, and application within
  {DFT}},}\ }\href@noop {} {\bibfield  {journal} {\bibinfo  {journal} {New J.
  Phys.}\ }\textbf {\bibinfo {volume} {20}},\ \bibinfo {pages} {073040}
  (\bibinfo {year} {2018})}\BibitemShut {NoStop}%
\bibitem [{\citenamefont {Sharma}\ and\ \citenamefont
  {Suryanarayana}(2023)}]{sharma2023calculation}%
  \BibitemOpen
  \bibfield  {author} {\bibinfo {author} {\bibfnamefont {A.}~\bibnamefont
  {Sharma}}\ and\ \bibinfo {author} {\bibfnamefont {P.}~\bibnamefont
  {Suryanarayana}},\ }\bibfield  {title} {\enquote {\bibinfo {title}
  {Calculation of phonons in real-space density functional theory},}\
  }\href@noop {} {\bibfield  {journal} {\bibinfo  {journal} {Phys. Rev. E}\
  }\textbf {\bibinfo {volume} {108}},\ \bibinfo {pages} {045302} (\bibinfo
  {year} {2023})}\BibitemShut {NoStop}%
\bibitem [{\citenamefont {McWeeny}(2002)}]{mcweeny2002symmetry}%
  \BibitemOpen
  \bibfield  {author} {\bibinfo {author} {\bibfnamefont {R.}~\bibnamefont
  {McWeeny}},\ }\href@noop {} {\emph {\bibinfo {title} {Symmetry: An
  introduction to group theory and its applications}}}\ (\bibinfo  {publisher}
  {Courier Corporation},\ \bibinfo {year} {2002})\BibitemShut {NoStop}%
\bibitem [{\citenamefont {Martin}(2004)}]{Martin2004}%
  \BibitemOpen
  \bibfield  {author} {\bibinfo {author} {\bibfnamefont {R.}~\bibnamefont
  {Martin}},\ }\href@noop {} {\emph {\bibinfo {title} {Electronic Structure:
  Basic theory and practical methods}}}\ (\bibinfo  {publisher} {Cambridge
  University Press},\ \bibinfo {year} {2004})\BibitemShut {NoStop}%
\bibitem [{\citenamefont {Kleinman}\ and\ \citenamefont
  {Bylander}(1982)}]{kleinman1982efficacious}%
  \BibitemOpen
  \bibfield  {author} {\bibinfo {author} {\bibfnamefont {L.}~\bibnamefont
  {Kleinman}}\ and\ \bibinfo {author} {\bibfnamefont {D.~M.}\ \bibnamefont
  {Bylander}},\ }\bibfield  {title} {\enquote {\bibinfo {title} {Efficacious
  form for model pseudopotentials},}\ }\href@noop {} {\bibfield  {journal}
  {\bibinfo  {journal} {Phys. Rev. Lett.}\ }\textbf {\bibinfo {volume} {48}},\
  \bibinfo {pages} {1425} (\bibinfo {year} {1982})}\BibitemShut {NoStop}%
\bibitem [{\citenamefont {Suryanarayana}\ \emph {et~al.}(2011)\citenamefont
  {Suryanarayana}, \citenamefont {Bhattacharya},\ and\ \citenamefont
  {Ortiz}}]{suryanarayana2011mesh}%
  \BibitemOpen
  \bibfield  {author} {\bibinfo {author} {\bibfnamefont {Phanish}\ \bibnamefont
  {Suryanarayana}}, \bibinfo {author} {\bibfnamefont {Kaushik}\ \bibnamefont
  {Bhattacharya}}, \ and\ \bibinfo {author} {\bibfnamefont {Michael}\
  \bibnamefont {Ortiz}},\ }\bibfield  {title} {\enquote {\bibinfo {title} {A
  mesh-free convex approximation scheme for kohn--sham density functional
  theory},}\ }\href@noop {} {\bibfield  {journal} {\bibinfo  {journal} {J.
  Comput. Phys.}\ }\textbf {\bibinfo {volume} {230}},\ \bibinfo {pages}
  {5226--5238} (\bibinfo {year} {2011})}\BibitemShut {NoStop}%
\bibitem [{\citenamefont {Suryanarayana}\ \emph {et~al.}(2010)\citenamefont
  {Suryanarayana}, \citenamefont {Gavini}, \citenamefont {Blesgen},
  \citenamefont {Bhattacharya},\ and\ \citenamefont {Ortiz}}]{Phanish2010}%
  \BibitemOpen
  \bibfield  {author} {\bibinfo {author} {\bibfnamefont {Phanish}\ \bibnamefont
  {Suryanarayana}}, \bibinfo {author} {\bibfnamefont {Vikram}\ \bibnamefont
  {Gavini}}, \bibinfo {author} {\bibfnamefont {Thomas}\ \bibnamefont
  {Blesgen}}, \bibinfo {author} {\bibfnamefont {Kaushik}\ \bibnamefont
  {Bhattacharya}}, \ and\ \bibinfo {author} {\bibfnamefont {Michael}\
  \bibnamefont {Ortiz}},\ }\bibfield  {title} {\enquote {\bibinfo {title}
  {Non-periodic finite-element formulation of kohn-sham density functional
  theory},}\ }\href@noop {} {\bibfield  {journal} {\bibinfo  {journal} {J.
  Mech. Phys. Solids}\ }\textbf {\bibinfo {volume} {58}},\ \bibinfo {pages}
  {256 -- 280} (\bibinfo {year} {2010})}\BibitemShut {NoStop}%
\bibitem [{\citenamefont {Dal~Corso}(2001)}]{dal2001density}%
  \BibitemOpen
  \bibfield  {author} {\bibinfo {author} {\bibfnamefont {Andrea}\ \bibnamefont
  {Dal~Corso}},\ }\bibfield  {title} {\enquote {\bibinfo {title}
  {Density-functional perturbation theory with ultrasoft pseudopotentials},}\
  }\href@noop {} {\bibfield  {journal} {\bibinfo  {journal} {Physical Review
  B}\ }\textbf {\bibinfo {volume} {64}},\ \bibinfo {pages} {235118} (\bibinfo
  {year} {2001})}\BibitemShut {NoStop}%
\bibitem [{\citenamefont {Dal~Corso}\ and\ \citenamefont
  {de~Gironcoli}(2000)}]{dal2000ab}%
  \BibitemOpen
  \bibfield  {author} {\bibinfo {author} {\bibfnamefont {Andrea}\ \bibnamefont
  {Dal~Corso}}\ and\ \bibinfo {author} {\bibfnamefont {Stefano}\ \bibnamefont
  {de~Gironcoli}},\ }\bibfield  {title} {\enquote {\bibinfo {title} {Ab initio
  phonon dispersions of fe and ni},}\ }\href@noop {} {\bibfield  {journal}
  {\bibinfo  {journal} {Physical Review B}\ }\textbf {\bibinfo {volume} {62}},\
  \bibinfo {pages} {273} (\bibinfo {year} {2000})}\BibitemShut {NoStop}%
\bibitem [{\citenamefont {Xu}\ \emph {et~al.}(2020)\citenamefont {Xu},
  \citenamefont {Sharma},\ and\ \citenamefont {Suryanarayana}}]{xu2020m}%
  \BibitemOpen
  \bibfield  {author} {\bibinfo {author} {\bibfnamefont {Q.}~\bibnamefont
  {Xu}}, \bibinfo {author} {\bibfnamefont {A.}~\bibnamefont {Sharma}}, \ and\
  \bibinfo {author} {\bibfnamefont {P.}~\bibnamefont {Suryanarayana}},\
  }\bibfield  {title} {\enquote {\bibinfo {title} {{M-SPARC}: Matlab-simulation
  package for ab-initio real-space calculations},}\ }\href@noop {} {\bibfield
  {journal} {\bibinfo  {journal} {SoftwareX}\ }\textbf {\bibinfo {volume}
  {11}},\ \bibinfo {pages} {100423} (\bibinfo {year} {2020})}\BibitemShut
  {NoStop}%
\bibitem [{\citenamefont {Zhang}\ \emph {et~al.}(2023)\citenamefont {Zhang},
  \citenamefont {Jing}, \citenamefont {Kumar},\ and\ \citenamefont
  {Suryanarayana}}]{zhang2023version}%
  \BibitemOpen
  \bibfield  {author} {\bibinfo {author} {\bibfnamefont {B.}~\bibnamefont
  {Zhang}}, \bibinfo {author} {\bibfnamefont {X.}~\bibnamefont {Jing}},
  \bibinfo {author} {\bibfnamefont {S.}~\bibnamefont {Kumar}}, \ and\ \bibinfo
  {author} {\bibfnamefont {P.}~\bibnamefont {Suryanarayana}},\ }\bibfield
  {title} {\enquote {\bibinfo {title} {Version 2.0.0 - {M-SPARC}:
  Matlab-simulation package for ab-initio real-space calculations},}\
  }\href@noop {} {\bibfield  {journal} {\bibinfo  {journal} {SoftwareX}\
  }\textbf {\bibinfo {volume} {21}},\ \bibinfo {pages} {101295} (\bibinfo
  {year} {2023})}\BibitemShut {NoStop}%
\bibitem [{\citenamefont {Xu}\ \emph {et~al.}(2021)\citenamefont {Xu},
  \citenamefont {Sharma}, \citenamefont {Comer}, \citenamefont {Huang},
  \citenamefont {Chow}, \citenamefont {Medford}, \citenamefont {Pask},\ and\
  \citenamefont {Suryanarayana}}]{xu2021sparc}%
  \BibitemOpen
  \bibfield  {author} {\bibinfo {author} {\bibfnamefont {Q.}~\bibnamefont
  {Xu}}, \bibinfo {author} {\bibfnamefont {A.}~\bibnamefont {Sharma}}, \bibinfo
  {author} {\bibfnamefont {B.}~\bibnamefont {Comer}}, \bibinfo {author}
  {\bibfnamefont {H.}~\bibnamefont {Huang}}, \bibinfo {author} {\bibfnamefont
  {E.}~\bibnamefont {Chow}}, \bibinfo {author} {\bibfnamefont {A.~J.}\
  \bibnamefont {Medford}}, \bibinfo {author} {\bibfnamefont {J.~E.}\
  \bibnamefont {Pask}}, \ and\ \bibinfo {author} {\bibfnamefont
  {P.}~\bibnamefont {Suryanarayana}},\ }\bibfield  {title} {\enquote {\bibinfo
  {title} {{SPARC}: Simulation package for ab-initio real-space
  calculations},}\ }\href@noop {} {\bibfield  {journal} {\bibinfo  {journal}
  {SoftwareX}\ }\textbf {\bibinfo {volume} {15}},\ \bibinfo {pages} {100709}
  (\bibinfo {year} {2021})}\BibitemShut {NoStop}%
\bibitem [{\citenamefont {Zhang}\ \emph {et~al.}(2024)\citenamefont {Zhang},
  \citenamefont {Jing}, \citenamefont {Xu}, \citenamefont {Kumar},
  \citenamefont {Sharma}, \citenamefont {Erlandson}, \citenamefont {Sahoo},
  \citenamefont {Chow}, \citenamefont {Medford}, \citenamefont {Pask},\ and\
  \citenamefont {Suryanarayana}}]{zhang2024sparc}%
  \BibitemOpen
  \bibfield  {author} {\bibinfo {author} {\bibfnamefont {B.}~\bibnamefont
  {Zhang}}, \bibinfo {author} {\bibfnamefont {X.}~\bibnamefont {Jing}},
  \bibinfo {author} {\bibfnamefont {Q.}~\bibnamefont {Xu}}, \bibinfo {author}
  {\bibfnamefont {S.}~\bibnamefont {Kumar}}, \bibinfo {author} {\bibfnamefont
  {A.}~\bibnamefont {Sharma}}, \bibinfo {author} {\bibfnamefont
  {L.}~\bibnamefont {Erlandson}}, \bibinfo {author} {\bibfnamefont {S.~J.}\
  \bibnamefont {Sahoo}}, \bibinfo {author} {\bibfnamefont {E.}~\bibnamefont
  {Chow}}, \bibinfo {author} {\bibfnamefont {A.~J.}\ \bibnamefont {Medford}},
  \bibinfo {author} {\bibfnamefont {J.~E.}\ \bibnamefont {Pask}}, \ and\
  \bibinfo {author} {\bibfnamefont {P.}~\bibnamefont {Suryanarayana}},\
  }\bibfield  {title} {\enquote {\bibinfo {title} {{SPARC} v2.0.0: Spin-orbit
  coupling, dispersion interactions, and advanced exchange–correlation
  functionals},}\ }\href@noop {} {\bibfield  {journal} {\bibinfo  {journal}
  {Software Impacts}\ }\textbf {\bibinfo {volume} {20}},\ \bibinfo {pages}
  {100649} (\bibinfo {year} {2024})}\BibitemShut {NoStop}%
\bibitem [{\citenamefont {Monkhorst}\ and\ \citenamefont
  {Pack}(1976)}]{monkhorst1976special}%
  \BibitemOpen
  \bibfield  {author} {\bibinfo {author} {\bibfnamefont {H.~J.}\ \bibnamefont
  {Monkhorst}}\ and\ \bibinfo {author} {\bibfnamefont {J.~D.}\ \bibnamefont
  {Pack}},\ }\bibfield  {title} {\enquote {\bibinfo {title} {Special points for
  brillouin-zone integrations},}\ }\href@noop {} {\bibfield  {journal}
  {\bibinfo  {journal} {Phys. Rev. B}\ }\textbf {\bibinfo {volume} {13}},\
  \bibinfo {pages} {5188} (\bibinfo {year} {1976})}\BibitemShut {NoStop}%
\bibitem [{\citenamefont {Perdew}\ and\ \citenamefont
  {Zunger}(1981)}]{PhysRevB.23.5048}%
  \BibitemOpen
  \bibfield  {author} {\bibinfo {author} {\bibfnamefont {J.~P.}\ \bibnamefont
  {Perdew}}\ and\ \bibinfo {author} {\bibfnamefont {A.}~\bibnamefont
  {Zunger}},\ }\bibfield  {title} {\enquote {\bibinfo {title} {Self-interaction
  correction to density-functional approximations for many-electron systems},}\
  }\href@noop {} {\bibfield  {journal} {\bibinfo  {journal} {Phys. Rev. B}\
  }\textbf {\bibinfo {volume} {23}},\ \bibinfo {pages} {5048--5079} (\bibinfo
  {year} {1981})}\BibitemShut {NoStop}%
\bibitem [{\citenamefont {Hamann}(2013)}]{hamann2013optimized}%
  \BibitemOpen
  \bibfield  {author} {\bibinfo {author} {\bibfnamefont {D.~R.}\ \bibnamefont
  {Hamann}},\ }\bibfield  {title} {\enquote {\bibinfo {title} {Optimized
  norm-conserving vanderbilt pseudopotentials},}\ }\href@noop {} {\bibfield
  {journal} {\bibinfo  {journal} {Phys. Rev. B}\ }\textbf {\bibinfo {volume}
  {88}},\ \bibinfo {pages} {085117} (\bibinfo {year} {2013})}\BibitemShut
  {NoStop}%
\bibitem [{\citenamefont {Pratapa}\ and\ \citenamefont
  {Suryanarayana}(2015)}]{pratapa2015restarted}%
  \BibitemOpen
  \bibfield  {author} {\bibinfo {author} {\bibfnamefont {P.~P.}\ \bibnamefont
  {Pratapa}}\ and\ \bibinfo {author} {\bibfnamefont {P.}~\bibnamefont
  {Suryanarayana}},\ }\bibfield  {title} {\enquote {\bibinfo {title} {Restarted
  {P}ulay mixing for efficient and robust acceleration of fixed-point
  iterations},}\ }\href@noop {} {\bibfield  {journal} {\bibinfo  {journal}
  {Chem. Phys. Lett.}\ }\textbf {\bibinfo {volume} {635}},\ \bibinfo {pages}
  {69--74} (\bibinfo {year} {2015})}\BibitemShut {NoStop}%
\bibitem [{\citenamefont {Banerjee}\ \emph {et~al.}(2016)\citenamefont
  {Banerjee}, \citenamefont {Suryanarayana},\ and\ \citenamefont
  {Pask}}]{banerjee2016PeriodicPulay}%
  \BibitemOpen
  \bibfield  {author} {\bibinfo {author} {\bibfnamefont {A.~S.}\ \bibnamefont
  {Banerjee}}, \bibinfo {author} {\bibfnamefont {P.}~\bibnamefont
  {Suryanarayana}}, \ and\ \bibinfo {author} {\bibfnamefont {J.~E.}\
  \bibnamefont {Pask}},\ }\bibfield  {title} {\enquote {\bibinfo {title}
  {Periodic {P}ulay method for robust and efficient convergence acceleration of
  self-consistent field iterations},}\ }\href@noop {} {\bibfield  {journal}
  {\bibinfo  {journal} {Chem. Phys. Lett.}\ }\textbf {\bibinfo {volume}
  {647}},\ \bibinfo {pages} {31 -- 35} (\bibinfo {year} {2016})}\BibitemShut
  {NoStop}%
\bibitem [{\citenamefont {Van~der Vorst}(1992)}]{van1992bi}%
  \BibitemOpen
  \bibfield  {author} {\bibinfo {author} {\bibfnamefont {H.~A.}\ \bibnamefont
  {Van~der Vorst}},\ }\bibfield  {title} {\enquote {\bibinfo {title}
  {Bi-{CGSTAB}: A fast and smoothly converging variant of {B}i-{CG} for the
  solution of nonsymmetric linear systems},}\ }\href@noop {} {\bibfield
  {journal} {\bibinfo  {journal} {SIAM J. Sci. Stat. Comput.}\ }\textbf
  {\bibinfo {volume} {13}},\ \bibinfo {pages} {631--644} (\bibinfo {year}
  {1992})}\BibitemShut {NoStop}%
\bibitem [{\citenamefont {Suryanarayana}\ \emph {et~al.}(2019)\citenamefont
  {Suryanarayana}, \citenamefont {Pratapa},\ and\ \citenamefont
  {Pask}}]{suryanarayana2019alternating}%
  \BibitemOpen
  \bibfield  {author} {\bibinfo {author} {\bibfnamefont {P.}~\bibnamefont
  {Suryanarayana}}, \bibinfo {author} {\bibfnamefont {P.~P.}\ \bibnamefont
  {Pratapa}}, \ and\ \bibinfo {author} {\bibfnamefont {J.~E.}\ \bibnamefont
  {Pask}},\ }\bibfield  {title} {\enquote {\bibinfo {title} {Alternating
  {A}nderson--{R}ichardson method: An efficient alternative to preconditioned
  {K}rylov methods for large, sparse linear systems},}\ }\href@noop {}
  {\bibfield  {journal} {\bibinfo  {journal} {Comput. Phys. Commun.}\ }\textbf
  {\bibinfo {volume} {234}},\ \bibinfo {pages} {278--285} (\bibinfo {year}
  {2019})}\BibitemShut {NoStop}%
\bibitem [{\citenamefont {Pratapa}\ \emph {et~al.}(2016)\citenamefont
  {Pratapa}, \citenamefont {Suryanarayana},\ and\ \citenamefont
  {Pask}}]{pratapa2016anderson}%
  \BibitemOpen
  \bibfield  {author} {\bibinfo {author} {\bibfnamefont {P.~P.}\ \bibnamefont
  {Pratapa}}, \bibinfo {author} {\bibfnamefont {P.}~\bibnamefont
  {Suryanarayana}}, \ and\ \bibinfo {author} {\bibfnamefont {J.~E.}\
  \bibnamefont {Pask}},\ }\bibfield  {title} {\enquote {\bibinfo {title}
  {{A}nderson acceleration of the {J}acobi iterative method: An efficient
  alternative to {K}rylov methods for large, sparse linear systems},}\
  }\href@noop {} {\bibfield  {journal} {\bibinfo  {journal} {J. Comput. Phys.}\
  }\textbf {\bibinfo {volume} {306}},\ \bibinfo {pages} {43--54} (\bibinfo
  {year} {2016})}\BibitemShut {NoStop}%
\bibitem [{\citenamefont {Sharma}\ and\ \citenamefont
  {Suryanarayana}(2018)}]{sharma2018real}%
  \BibitemOpen
  \bibfield  {author} {\bibinfo {author} {\bibfnamefont {A.}~\bibnamefont
  {Sharma}}\ and\ \bibinfo {author} {\bibfnamefont {P.}~\bibnamefont
  {Suryanarayana}},\ }\bibfield  {title} {\enquote {\bibinfo {title} {On
  real-space density functional theory for non-orthogonal crystal systems:
  {K}ronecker product formulation of the kinetic energy operator},}\
  }\href@noop {} {\bibfield  {journal} {\bibinfo  {journal} {Chem. Phys.
  Lett.}\ }\textbf {\bibinfo {volume} {700}},\ \bibinfo {pages} {156--162}
  (\bibinfo {year} {2018})}\BibitemShut {NoStop}%
\bibitem [{\citenamefont {Barros}\ \emph {et~al.}(2006)\citenamefont {Barros},
  \citenamefont {Jorio}, \citenamefont {Samsonidze}, \citenamefont {Capaz},
  \citenamefont {{Souza Filho}}, \citenamefont {{Mendes Filho}}, \citenamefont
  {Dresselhaus},\ and\ \citenamefont {Dresselhaus}}]{ReviewCNT}%
  \BibitemOpen
  \bibfield  {author} {\bibinfo {author} {\bibfnamefont {E.~B.}\ \bibnamefont
  {Barros}}, \bibinfo {author} {\bibfnamefont {A.}~\bibnamefont {Jorio}},
  \bibinfo {author} {\bibfnamefont {G.~G.}\ \bibnamefont {Samsonidze}},
  \bibinfo {author} {\bibfnamefont {R.~B.}\ \bibnamefont {Capaz}}, \bibinfo
  {author} {\bibfnamefont {A.~G.}\ \bibnamefont {{Souza Filho}}}, \bibinfo
  {author} {\bibfnamefont {J.}~\bibnamefont {{Mendes Filho}}}, \bibinfo
  {author} {\bibfnamefont {G.}~\bibnamefont {Dresselhaus}}, \ and\ \bibinfo
  {author} {\bibfnamefont {M.~S.}\ \bibnamefont {Dresselhaus}},\ }\bibfield
  {title} {\enquote {\bibinfo {title} {Review on the symmetry-related
  properties of carbon nanotubes},}\ }\href@noop {} {\bibfield  {journal}
  {\bibinfo  {journal} {Phys. Rep.}\ }\textbf {\bibinfo {volume} {431}},\
  \bibinfo {pages} {261--302} (\bibinfo {year} {2006})}\BibitemShut {NoStop}%
\bibitem [{\citenamefont {Gonze}\ \emph {et~al.}(2009)\citenamefont {Gonze},
  \citenamefont {Amadon}, \citenamefont {Anglade}, \citenamefont {Beuken},
  \citenamefont {Bottin}, \citenamefont {Boulanger}, \citenamefont {Bruneval},
  \citenamefont {Caliste}, \citenamefont {Caracas}, \citenamefont {Côté},
  \citenamefont {Deutsch}, \citenamefont {Genovese}, \citenamefont {Ghosez},
  \citenamefont {Giantomassi}, \citenamefont {Goedecker}, \citenamefont
  {Hamann}, \citenamefont {Hermet}, \citenamefont {Jollet}, \citenamefont
  {Jomard}, \citenamefont {Leroux}, \citenamefont {Mancini}, \citenamefont
  {Mazevet}, \citenamefont {Oliveira}, \citenamefont {Onida}, \citenamefont
  {Pouillon}, \citenamefont {Rangel}, \citenamefont {Rignanese}, \citenamefont
  {Sangalli}, \citenamefont {Shaltaf}, \citenamefont {Torrent}, \citenamefont
  {Verstraete}, \citenamefont {Zerah},\ and\ \citenamefont
  {Zwanziger}}]{gonze2009abinit}%
  \BibitemOpen
  \bibfield  {author} {\bibinfo {author} {\bibfnamefont {X.}~\bibnamefont
  {Gonze}}, \bibinfo {author} {\bibfnamefont {B.}~\bibnamefont {Amadon}},
  \bibinfo {author} {\bibfnamefont {P.-M.}\ \bibnamefont {Anglade}}, \bibinfo
  {author} {\bibfnamefont {J.-M.}\ \bibnamefont {Beuken}}, \bibinfo {author}
  {\bibfnamefont {F.}~\bibnamefont {Bottin}}, \bibinfo {author} {\bibfnamefont
  {P.}~\bibnamefont {Boulanger}}, \bibinfo {author} {\bibfnamefont
  {F.}~\bibnamefont {Bruneval}}, \bibinfo {author} {\bibfnamefont
  {D.}~\bibnamefont {Caliste}}, \bibinfo {author} {\bibfnamefont
  {R.}~\bibnamefont {Caracas}}, \bibinfo {author} {\bibfnamefont
  {M.}~\bibnamefont {Côté}}, \bibinfo {author} {\bibfnamefont
  {T.}~\bibnamefont {Deutsch}}, \bibinfo {author} {\bibfnamefont
  {L.}~\bibnamefont {Genovese}}, \bibinfo {author} {\bibfnamefont {Ph.}\
  \bibnamefont {Ghosez}}, \bibinfo {author} {\bibfnamefont {M.}~\bibnamefont
  {Giantomassi}}, \bibinfo {author} {\bibfnamefont {S.}~\bibnamefont
  {Goedecker}}, \bibinfo {author} {\bibfnamefont {D.~R.}\ \bibnamefont
  {Hamann}}, \bibinfo {author} {\bibfnamefont {P.}~\bibnamefont {Hermet}},
  \bibinfo {author} {\bibfnamefont {F.}~\bibnamefont {Jollet}}, \bibinfo
  {author} {\bibfnamefont {G.}~\bibnamefont {Jomard}}, \bibinfo {author}
  {\bibfnamefont {S.}~\bibnamefont {Leroux}}, \bibinfo {author} {\bibfnamefont
  {M.}~\bibnamefont {Mancini}}, \bibinfo {author} {\bibfnamefont
  {S.}~\bibnamefont {Mazevet}}, \bibinfo {author} {\bibfnamefont {M.~J.~T.}\
  \bibnamefont {Oliveira}}, \bibinfo {author} {\bibfnamefont {G.}~\bibnamefont
  {Onida}}, \bibinfo {author} {\bibfnamefont {Y.}~\bibnamefont {Pouillon}},
  \bibinfo {author} {\bibfnamefont {T.}~\bibnamefont {Rangel}}, \bibinfo
  {author} {\bibfnamefont {G.-M.}\ \bibnamefont {Rignanese}}, \bibinfo {author}
  {\bibfnamefont {D.}~\bibnamefont {Sangalli}}, \bibinfo {author}
  {\bibfnamefont {R.}~\bibnamefont {Shaltaf}}, \bibinfo {author} {\bibfnamefont
  {M.}~\bibnamefont {Torrent}}, \bibinfo {author} {\bibfnamefont {M.~J.}\
  \bibnamefont {Verstraete}}, \bibinfo {author} {\bibfnamefont
  {G.}~\bibnamefont {Zerah}}, \ and\ \bibinfo {author} {\bibfnamefont {J.~W.}\
  \bibnamefont {Zwanziger}},\ }\bibfield  {title} {\enquote {\bibinfo {title}
  {{ABINIT}: First-principles approach to material and nanosystem
  properties},}\ }\href@noop {} {\bibfield  {journal} {\bibinfo  {journal}
  {Comput. Phys. Commun.}\ }\textbf {\bibinfo {volume} {180}},\ \bibinfo
  {pages} {2582--2615} (\bibinfo {year} {2009})}\BibitemShut {NoStop}%
\bibitem [{\citenamefont {Hall}\ \emph {et~al.}(2006)\citenamefont {Hall},
  \citenamefont {An}, \citenamefont {Liu}, \citenamefont {Vicci}, \citenamefont
  {Falvo}, \citenamefont {Superfine},\ and\ \citenamefont
  {Washburn}}]{Hall2006CNT}%
  \BibitemOpen
  \bibfield  {author} {\bibinfo {author} {\bibfnamefont {A.~R.}\ \bibnamefont
  {Hall}}, \bibinfo {author} {\bibfnamefont {L.}~\bibnamefont {An}}, \bibinfo
  {author} {\bibfnamefont {J.}~\bibnamefont {Liu}}, \bibinfo {author}
  {\bibfnamefont {L.}~\bibnamefont {Vicci}}, \bibinfo {author} {\bibfnamefont
  {M.~R.}\ \bibnamefont {Falvo}}, \bibinfo {author} {\bibfnamefont
  {R.}~\bibnamefont {Superfine}}, \ and\ \bibinfo {author} {\bibfnamefont
  {S.}~\bibnamefont {Washburn}},\ }\bibfield  {title} {\enquote {\bibinfo
  {title} {Experimental measurement of single-wall carbon nanotube torsional
  properties},}\ }\href@noop {} {\bibfield  {journal} {\bibinfo  {journal}
  {Phys. Rev. Lett.}\ }\textbf {\bibinfo {volume} {96}},\ \bibinfo {pages}
  {256102} (\bibinfo {year} {2006})}\BibitemShut {NoStop}%
\bibitem [{\citenamefont {Treacy}\ \emph {et~al.}(1996)\citenamefont {Treacy},
  \citenamefont {Ebbesen},\ and\ \citenamefont
  {Gibson}}]{treacy1996exceptionally}%
  \BibitemOpen
  \bibfield  {author} {\bibinfo {author} {\bibfnamefont {M.~M.~J.}\
  \bibnamefont {Treacy}}, \bibinfo {author} {\bibfnamefont {T.~W.}\
  \bibnamefont {Ebbesen}}, \ and\ \bibinfo {author} {\bibfnamefont {J.~M.}\
  \bibnamefont {Gibson}},\ }\bibfield  {title} {\enquote {\bibinfo {title}
  {Exceptionally high young's modulus observed for individual carbon
  nanotubes},}\ }\href@noop {} {\bibfield  {journal} {\bibinfo  {journal}
  {nature}\ }\textbf {\bibinfo {volume} {381}},\ \bibinfo {pages} {678--680}
  (\bibinfo {year} {1996})}\BibitemShut {NoStop}%
\bibitem [{\citenamefont {Sharma}\ and\ \citenamefont
  {Suryanarayana}(2026)}]{cyclixphonondataset}%
  \BibitemOpen
  \bibfield  {author} {\bibinfo {author} {\bibfnamefont {Abhiraj}\ \bibnamefont
  {Sharma}}\ and\ \bibinfo {author} {\bibfnamefont {Phanish}\ \bibnamefont
  {Suryanarayana}},\ }\href {\doibase https://doi.org/10.5281/zenodo.19131108}
  {\enquote {\bibinfo {title} {Cyclic- and helical-symmetry-adapted phonon
  formalism within density functional perturbation theory},}\ } (\bibinfo
  {year} {2026})\BibitemShut {NoStop}%
\end{thebibliography}

%

\end{document}